\documentclass[longbibliography,aps,reprint,groupedaddress]{revtex4-1}

\usepackage{amsmath,amssymb}
\usepackage{amsfonts}
\usepackage[squaren,cdot]{SIunits}
\usepackage{graphicx}
\usepackage{xcolor}
\usepackage{epstopdf}
\usepackage{blindtext}
\usepackage[latin1]{inputenc}
\usepackage{epstopdf, epsfig}

\usepackage{array}
\newcolumntype{L}[1]{>{\raggedright\let\newline\\\arraybackslash\hspace{0pt}}m{#1}}
\newcolumntype{C}[1]{>{\centering\let\newline\\\arraybackslash\hspace{0pt}}m{#1}}
\newcolumntype{R}[1]{>{\raggedleft\let\newline\\\arraybackslash\hspace{0pt}}m{#1}}

 \newcommand{\sds}{\mbox{\tiny SDS}}
  \newcommand{\doh}{\mbox{\tiny DOH}}

 \newcommand{\e}{\varepsilon}
 \newcommand{\de}{\dot{\varepsilon}}
 
 \newcommand{\vv}[1]{{\bf \hat {#1}}}
 
\DeclareGraphicsExtensions{.pdf,.jpg,.png,.mps,.tif,.eps}

\begin{document}

\title{Local origin of the visco-elasticity of a millimetric elementary foam}
\author{Adrien Bussonni\`ere}
\author{Isabelle Cantat}

\affiliation{Univ Rennes, CNRS, IPR (Institut de Physique de Rennes) - UMR 6251, F- 35000 Rennes.}
\date{\today}

\begin{abstract}
Liquid foam exhibits surprisingly high viscosity, higher than each of its phases. This dissipation enhancement has been rationalized by invoking either a geometrical confinement of the shear in the liquid phase, or the influence of the  interface viscosity. However, a precise localization of the dissipation, and its mechanism, at the bubble scale is still lacking. To this aim, we simultaneously monitored the evolution of the local flow velocity, film thickness and surface tension of a five films assembly, induced by different controlled deformations. These measurements allow us to build local constitutive relations for  this foam elementary brick. We first show that, for our millimetric foam films, the main part of the film has a purely elastic, reversible behavior, thus  ruling out the  interface viscosity to explain the observed dissipation. We then highlight a generic frustration at the menisci,  controlling the  interface transfer between neighbor films and resulting in the localization of a bulk shear flow close to the menisci. A model accounting for surfactant transport in these small sheared regions is developed. It is in good agreement with the experiment, and   demonstrate that most of the dissipation is localized in these domains. The length of these sheared regions, determined by the physico-chemical properties of the solution, sets a transition between a large bubble regime in which the films are mainly stretched and compressed, and a small bubble regime in which they are sheared. Finally, we discuss the parameter range where a model of foam viscosity could be built on the basis of these local results.
\end{abstract}


\maketitle

\section{Introduction}

A foam, made of inviscid gas and Newtonian liquid, has an effective viscosity that may reach thousand times the viscosity of the foaming solution \cite{marze08,krishan10}. Liquid phase confinement is classically  assumed to be at the origin of this spectacular viscosity enhancement, with a local shear rate scaling as the imposed one multiplied by the confinement factor $d/h$, with $d$ the bubble size and $h$ the film thickness \cite{cohenaddad13}. However, how and where the imposed stress is transmitted to the liquid phase remains an open question.  In the absence of any solid part in the foam structure, the degrees of freedom of the liquid phase are numerous, and an imposed external deformation can lead to many different local deformations and flows, which have been listed in the seminal work of Buzza and Cates \cite{buzza95}. 

The problem has been addressed experimentally both at the bubble scale and at the foam sample scale. In the first approach, an assembly of few millimetric films are deformed, either due to the structure relaxations after a triggered T1 event \cite{durand06,biance09,petit15}, or due to the controlled motion of the supporting frame \cite{besson07, costa13b,seiwert13, bussonniere20}.  In most of these studies, the local film tensions are deduced from the film structure geometry, and/or the local film thicknesses measured using absorption or interferometry. The observations are rationalized with models involving film extensions and compressions, associated to a viscoelastic response of the surfactant monolayer, and a relaxation of the interface area variations by the surfactant monolayer transfer from the compressed films to the stretched ones. The velocities of both film interfaces are assumed identical. In contrast, at the sample scale, with typical bubble sizes of the order of 100 $\mu$m, the rheometric measurements are usually modeled using the assumption of bubbles sliding on top of each other and thus shearing the liquid film that separates them, without any interface extension \cite{denkov08}. Depending on the physico-chemistry, different scaling laws are observed, which are difficult to interpret in term of one model or the other \cite{krishan10, costa13}. 

There is thus a clear need for a full characterization of the flows induced in the foam films by an imposed deformation, with a synchronized measure of both the kinematic quantities (local interface extension and extension rate, interface transfer velocity) and of the local tension in films, in order to discriminate between both approaches, involving either film shearing or film extension. This is an unavoidable milestone to fully elucidate the local origin of foam viscosity.

To this end, we built a dedicated set-up that allows us to impose controlled deformations to a five films assembly. In a previous paper, we measured the transfer velocity from one film to its neighbor due to this deformation, as well as other kinematic quantities  \cite{bussonniere20}. The aim here is to relate these kinematic quantities to the film tensions, in order to build a constitutive law for each part of the foam structure, and, eventually, to build the resulting constitutive law for the foam sample.

The main results of this paper are the following: (i) The films are shown to be governed by a perfectly reversible elastic law, with no influence of the  shear or extensional interface viscosities. This proves that the viscoelastic response of the interface is not, as often assumed, at the origin of the foam dissipation; (ii) We measure the relationship between  the  interface transfer velocity, from one  film to its neighbor, and the tension difference between these films. We evidence a generic geometrical frustration at the meniscus: as it prevents the free transfer of the interface,  this frustration is at the origin of the largest part of the dissipation; (iii) We predict that this dominant dissipation is localized in a small part of the films, close to the meniscus, that is sheared. In this domain, a well controlled  scale separation is used to simplify the hydrodynamics and transport equations, which become easily numerically solvable. One important prediction of our model is the scaling law for the length $\ell$ of this sheared part of the film, as a function of the physico-chemical properties of the foaming solution. This length increases when the surfactant solubility decreases and is typically of the order of 100 $\mu$m. Importantly, it defines a frontier between the foams having bubbles smaller than $\ell$, in which the whole film should be sheared, and the foams having bubbles larger than $\ell$, in which the main film deformation should be an extension/compression. This reconciles the various classes of model, based either on extension or on shear, found in the literature.

On these bases, we built a model of foam viscosity for the large bubble regime and/or high surfactant solubility,  {\it e.g.} for a bubble radius larger than $\ell$. It reproduces the variations of the foam loss modulus as the square root of the frequency, and as the inverse of the bubble size, which are observed for a large class of foams \cite{krishan10, costa13}, and predicts the prefactor as a function of well defined, measurable, physico-chemical properties.

The paper is organized as follows. We first introduce the dedicated experimental setup in section \ref{sec:setup} and then describe the measurement of the relevant kinematic quantities in section \ref{sec:kinematic}. In section \ref{sec:force}, we describe the technique used to measure the evolution of the tension of the five different films. In section \ref{sec:const_film} and \ref{sec:const_men} we build the constitutive relationships of the thin film and of the meniscus, respectively. In section \ref{sec:const_exchange}, we unravel and model the dominant dissipation mechanism associated to the film/meniscus exchange and compare our model to the experiments in section \ref{sec:comp_shear}. Finally in section \ref{sec:foam_rev} we discuss the relevance of our findings in a foam context.

%%%%%%%%%%%%%%%%%%%%%%%%%%%%%%%%%%%%%%%%%%%%%%%%%%%%%%%%%%%%%%%%%%%%%%%%%%%%%%%%%%

\section{Experimental set-up and foaming solution}
\label{sec:setup}

The same experimental set-up has been used previously in \cite{bussonniere20}, and the measure of the kinematic properties of the film (velocity, extension) has already been presented in this former article for a restricted range of the control parameters. We recall here the main features and describe the physico-chemical properties of the solution.

\subsection{Mechanical device}

\begin{figure}[htp!]
\begin{center}
\includegraphics[width=0.95\linewidth]{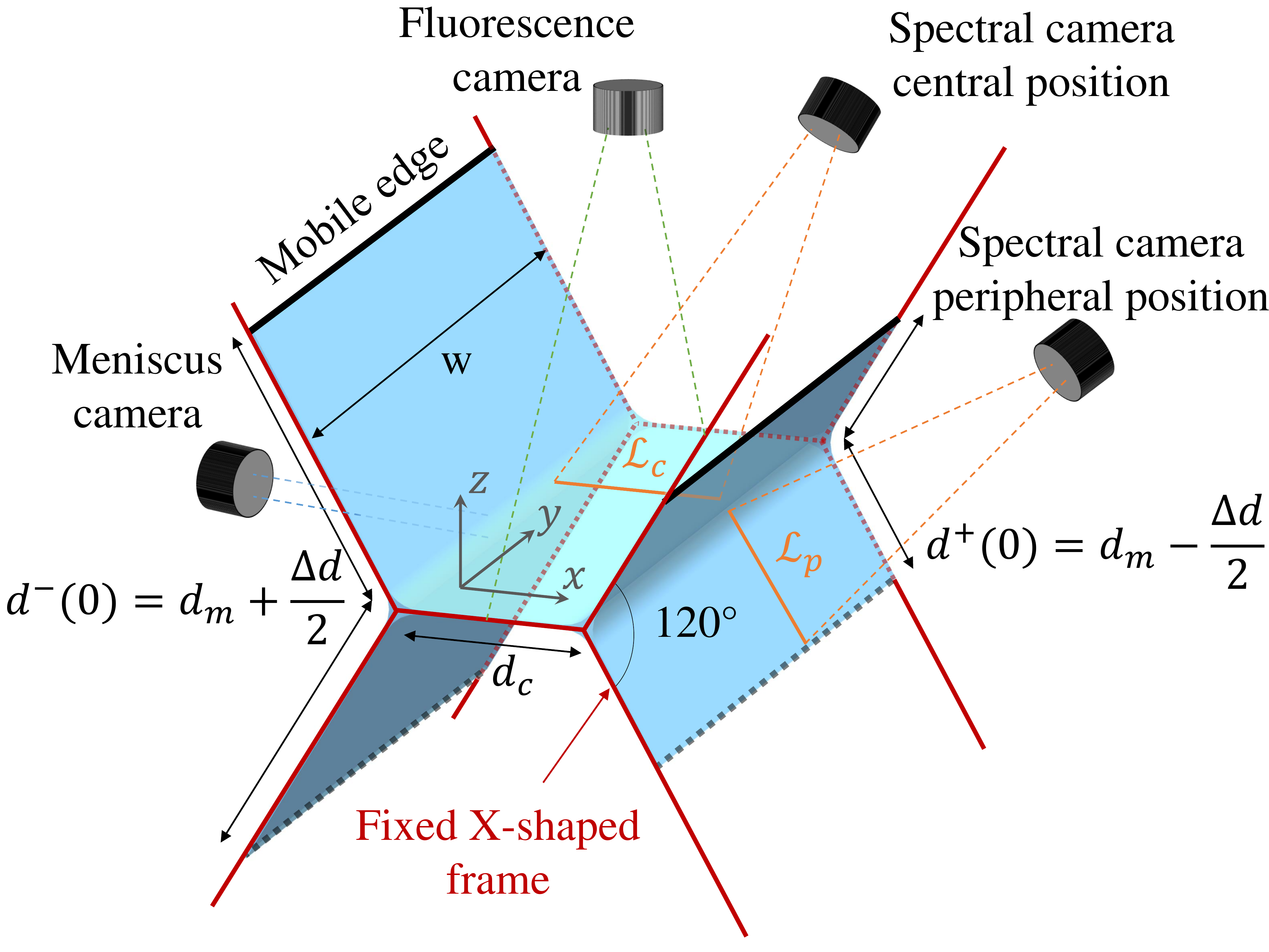}
\caption{Schematic of the experimental setup showing the mechanical device controlling film deformation and the different optical devices used to monitor the film motion. \label{fig:setup}}
\end{center}
\end{figure}

The film assembly is made of five foam films held by two metallic X-shaped pieces as shown in Fig.~\ref{fig:setup}. The central horizontal film has a width $d_c=6$ mm and a length $W=41.5$ mm. The length has been chosen such that $W\gg d_c$ so the middle part of the central film is not influenced by the boundary effects on the solid frame. 
The short edges of the central film form menisci with the metallic frames (the {\it supported menisci}) while its long edges are menisci connected to the four peripheral films at an angle of 120$^o$ (the {\it free menisci}).  The  external edges of the peripheral films are connected to metallic pieces of length $W$ (black pieces in Fig.~\ref{fig:setup}) which can translate along the lateral arms of the X-shape pieces. The mobile edge positions are controlled by four linear piezo motors (PI U-521.23). This geometry ensures that each film remains flat and stay in the same plane whatever the position of the motor, if the films are at mechanical equilibrium.
Unless otherwise specified, an invariance in the $y$ direction will be assumed for all quantities. They are expressed as a function of the curvilinear abscissa $s$, defined for each film as the position along a line in the $(x,z)$ plane. 

As in \cite{bussonniere20}, the deformation studied consists of a compression of the left peripheral films from an initial length $d^-(0)=d_m+\Delta d/2$ to $d_m-\Delta d/2$ and a simultaneous symmetric extension of the right films from $d^+(0)=d_m-\Delta d/2$ to $d_m+\Delta d/2$ at a constant velocity $V$. In this study, we explore the influence of the motor velocity $V$, of the deformation amplitude $\Delta d$ and of the mean position $d_m$.

\subsection{Optical device}
Three cameras are used to monitor the dynamic induced by the film deformations as shown in Fig.~\ref{fig:setup}. 
The meniscus camera records the size and vertical position of one free meniscus. It is magnified with a telecentric lens (Edmund Optics SilverTL 4x) and illuminated with a collimated white LED.
The free menisci position in the $(x,y)$ plane and the  gross thickness variations of the central film are  recorded with the fluorescence camera located on the top of the setup. The  fluorescein added in the foaming solution is excited with a 488 nm laser (Oxxius LBX 200 mW) and its emission is filtered using a band-pass filter. Finally, more precise thickness profiles $h$ are monitored along the line ${\cal{L}}_c$ in the central film or ${\cal{L}}_p$ in a peripheral film with a hyperspectral camera (Resonon Pika L). This camera, described in \cite{bussonniere20}, measures the interfered spectrum intensity $I(\lambda)$ of the light emitted by a halogen lamp and reflected by the thin film. The thickness is extracted using the relation :

\begin{equation}
I(\lambda) \propto 1 - \cos\left(\frac{4\pi h n}{\lambda}\, \left(1-\frac{\sin^2{\theta}}{n^2}\right)^{1/2}\right) \; , 
\end{equation}
where $n=1.33$ is the film refractive index, $\lambda$ the wavelength and $\theta$ the light incident angle. In our set-up, $\theta$  is $45^\circ$ for the central camera position and $58.5^\circ$ for the peripheral one. 

\subsection{Physico-chemical properties of the foaming solution}
\label{sec:phychi}
The foaming solution used is a mixture of sodium dodecyl sulfate ($c_{\sds} =5.6$ g/l), dodecanol ($c_{\doh} =0.05$ g/l), fluorescein ($c_{\mbox{\tiny fluo}}=0.8$ g/l) dissolved in a solution of distilled water-glycerol (15 \% in volume). The equilibrium surface tension was measured at $\gamma_{0}= 33\pm 1$ mN/m using the pendant drop method and the viscosity  is $\eta = 1.5$ mPa.s. Four needles located on the top mobile edges of the peripheral films can be used to supply solution to the film assembly during the entire experiment. Two sets of experiments have been performed, one where the foaming solution is injected at a rate of 0.2 ml/min (0.05 ml/min per needle) and one without injection.

Such mixture of sodium dodecyl sulfate and dodecanol has been extensively studied for its relevance in foam science and has been chosen here to optimize film stability and rheological response. However, its  physico-chemical properties remain difficult to model due to the strong interactions between the anionic surfactant and the non-ionic alcohol which can lead to the formation of a complex \cite{lu95,nguyen19,vollhardt00,kralchevsky03}. Moreover, above the critical micelle concentration (CMC$=2.33$ g/l for pure SDS), DOH molecules can be solubilized in SDS micelles. These interactions lead to co-adsorption processes and mixed diffusion \cite{fang92}.
As the chemistry has not been varied in this study, the potentially complex equation of state of the interface, adsorption and transport laws, taking into account the different species, can not be confronted to our experimental results. We thus choose to keep our thermodynamic model of interface as simple as possible by linearizing the different laws.

Surface tension of pure SDS remains almost constant above the CMC \cite{elworthy66}. The important variations observed in our experiments are therefore assumed to be associated to the dodecanol only. At thermodynamic equilibrium, the surface tension $\gamma_{th}$ is related to the DOH surface excess $\Gamma$ by :
\begin{gather}
    \gamma_{th}=\gamma_0+\left. \frac{\partial \gamma_{th}}{\partial \Gamma}\right|_{\Gamma_0}(\Gamma-\Gamma_0)=\gamma_0-E\frac{\Gamma-\Gamma_0}{\Gamma_0},
    \label{eq:elas_gibbs}
\end{gather}
with $\gamma_0$  the  surface tension of the foaming solution,  $\Gamma_0$ the corresponding surface excess,  and $E=-\left.\frac{\partial \gamma_{th}}{\partial \Gamma}\right|_{\Gamma_0}\Gamma_0$ the so called Gibbs-Marangoni elasticity. This elasticity can be estimated using the Langmuir model of DOH/micellar SDS solution proposed in \cite{fang92} which leads to $E\approx 10$ mN/m (see appendix \ref{app:phys_chem}).

The adsorption of dodecanol at the interface is characterized  by 
\begin{gather}
    \Gamma=\Gamma_0+\left.\frac{\partial \Gamma}{\partial c}\right|_{c_0}(c-c_0)=\Gamma_0+h_{\Gamma}(c-c_0),
    \label{eq:surf_trans}
\end{gather}
where $c$ is the local dodecanol bulk concentration, $c_0$ the initial concentration and $h_{\Gamma}=\left.\frac{\partial \Gamma}{\partial c}\right|_{c_0}$, hereafter called reservoir length. Based on \cite{fang92}, we estimated $h_\Gamma\approx 5.4$ $\mu$m (see appendix \ref{app:phys_chem}). For processes faster than the micellization \cite{patist98}, we also need to consider the equilibrium between the surface excess and the concentration $c^m$ of dodecanol in its monomer form, involving the parameter  $h^m_\Gamma=\left.\frac{\partial \Gamma}{\partial c^m}\right|_{c^m_0}\approx 370$ $\mu$m (see appendix \ref{app:phys_chem}). 

The disjoining pressure as a function of the film thickness is also an important physico-chemical property of the system. However, it is negligible in our study  as the films are always larger than 100 nm.  

Finally, on each interface, we define $\gamma(s)$ as the full interfacial stress which includes the surface tension  $\gamma_{th}(\Gamma)$ as well as the potential contributions  associated to the surface extensional and shear interfacial viscosities, respectively $\eta_s$ and $\kappa_s$. Note that the interfacial stress is thus {\it a priori} of tensorial nature, and $\gamma(s)$ represents its projection in the direction orthogonal to the direction of invariance $y$. 
For the thin films, we also define  the film tension $\sigma \sim 2 \gamma$ that takes into account the contribution of both interfaces and of the film bulk (see section \ref{sec:def_tens}).

\subsection{Control parameters}
In this study we explored the influence of the deformation parameters on the film assembly dynamic by performing around 480 experiments.

 A first experimental campaign was  performed with the fluorescent camera used at a  frame rate of 130 fps,  for $\Delta d$ varying between 2 to 12 mm, $V$ between 1 to 100 mm/s and $d_m$ between 7 to 17 mm. Each set of parameters has been repeated at least 3 times with and without solution injection representing a total of 186 experiments. For some parameter values, the measures have been refined in a second campaign,  by increasing the frame rate to 300 fps, increasing the amount of experiments, and/or using the spectral camera in the peripheral position, instead of the central position only. 

%%%%%%%%%%%%%%%%%%%%%%%%%%%%%%%%%%%%%%%%%%%%%%%%%%%%%%%
\section{Determination of the kinematic quantities}
\label{sec:kinematic}

\begin{figure}[htp!]
\begin{center}
\includegraphics[width=1\linewidth]{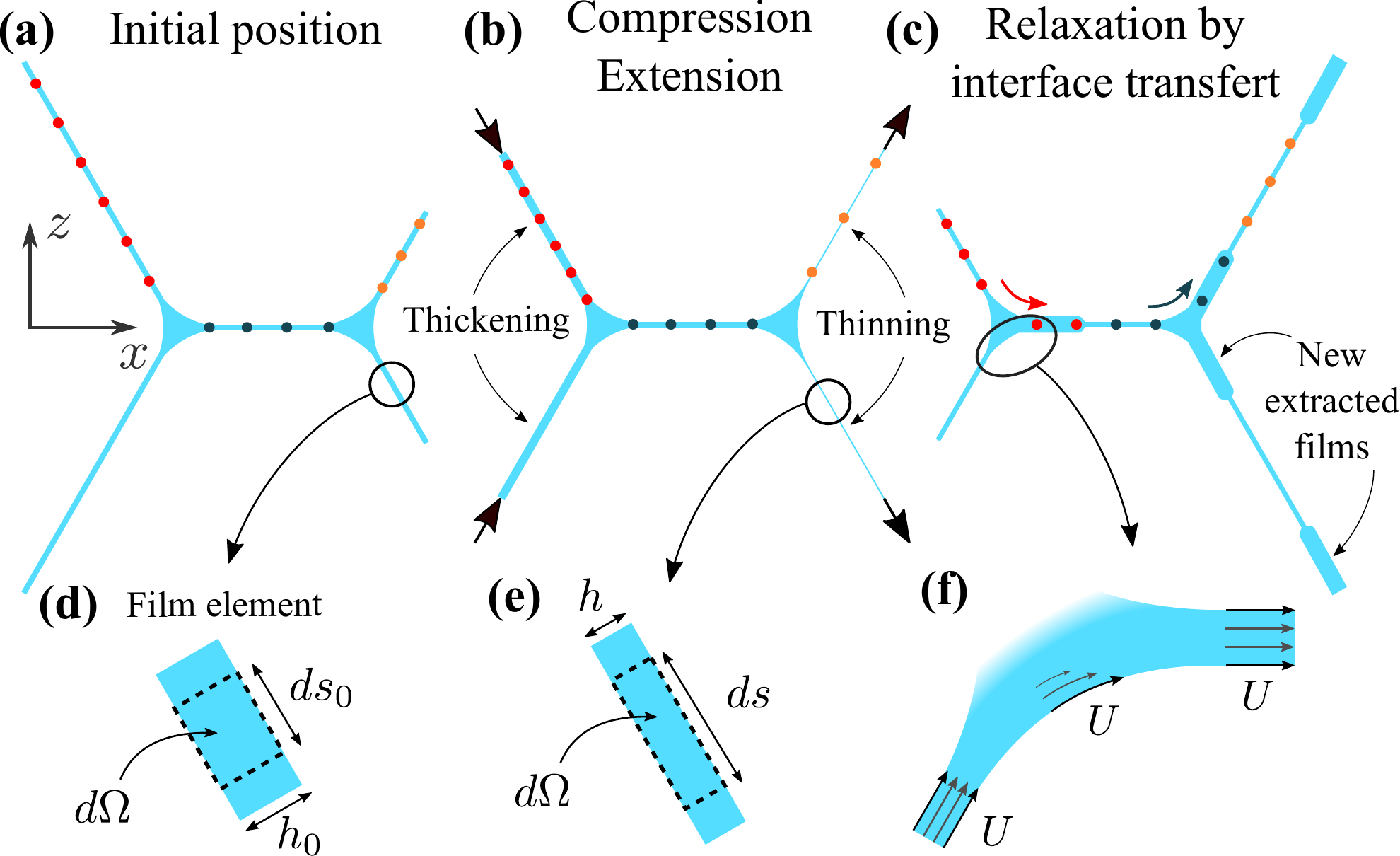}
\caption{(a) to (c) Schematic of the typical dynamic of the foam assembly. The colored dots represent elementary material systems at the interface which are followed along their trajectories, thus illustrating film compression/extension as well as interface transfers from one film to the other. Films thicknesses evolve because of the compression/extension and because of the extraction of thick film from the menisci, associated to the interface transfers. (d)-(e) Zoom on a closed volume film element before and during the deformation. (f) Zoom on a meniscus showing the interface transfer dynamic, at velocity $U$. \label{fig:def}}
\end{center}
\end{figure}

 As shown in our previous study \cite{bussonniere20} and summarized in Fig.~\ref{fig:def}, the typical dynamic is composed of an extension of the peripheral films on the stretched side, and a compression of the peripheral films on the pushed side, at the first instants (Fig.~\ref{fig:def} (b)). The imposed deformation then relaxes through interface transfers between adjacent films (Fig.~\ref{fig:def} (c)). A visible signature of this transfer is the appearance of thick films, extracted from the menisci, in the central and stretched films.

\subsection{Definition of the kinematic quantities}
\label{sec:def_kin}
The extension of the films and the transfer velocity at the menisci are the two relevant kinematic quantities of the problem and they will be related to the film tensions through constitutive relationships in section \ref{sec:const_film} and \ref{sec:const_men}. To properly define and measure the extensions and the transfer velocities we first clarify here different assumptions.

In the films, the relative velocity of the bulk phase with respect to the interfaces is a Poiseuille flow governed by the gravity forces and by the Laplace pressure gradients. As quantitatively discussed in section \ref{sec:dom_def}, these relative velocities are negligible far enough from the menisci, and the velocity can be assumed to be constant across the film (Fig. \ref{fig:def} (f)). In the central part of each film, we can therefore define a {\it film element}  ${\cal S}$  as an elementary material system of volume $d\Omega = h dS = h dy ds $ (see Fig. \ref{fig:def} (d)-(e)) spanning the film from one interface to the other. It is a closed system  which can be followed along its trajectory and which is entirely determined by the shape and position of its interface. The invariance in the $y$ direction imposes that $dy$ is constant. However stretching or compression modifies $ds$. 
In such a film element, the film extension, or equivalently the interface extension, can be defined as 
 \begin{equation}
 \varepsilon(t)=\frac{ds(t)}{ds_0} -1 = \frac{h_0}{h(t)}-1 \, ,
 \label{eq:defeps}
 \end{equation}
with $h_0$ and $ds_0$ the initial characteristic of the film element, before deformation. The second equality is deduced from the volume conservation of the system, which imposes $h(t) ds(t) dy  = h_0 ds_0 dy$. With this definition $\varepsilon > 0$ for an extension and $\varepsilon  < 0$ for a compression.

The transfer velocity is a dynamical property associated to each free meniscus. In \cite{bussonniere20} we experimentally checked that, for the imposed deformation, when some film is extracted at one side of a free meniscus, a similar amount of film is absorbed  on the other side, at the same rate, as schematized in Fig. \ref{fig:def} (f). The surfactant monolayer slides on the meniscus interface, from one film to its neighbor, with negligible deformation. This allows us to define the transfer velocity $U$ as, indifferently, the velocity of the film entering the meniscus at one side, the velocity of the film extracted on the other side, or the velocity of the surfactant monolayer at the meniscus interface connecting both films. Experimentally, $U$ is measured in the central film. The model of section \ref{sec:const_exchange} goes beyond this first order description and provides a prediction for the interface velocity difference between both sides of the free menisci, thus refining this first definition of the transfer velocity.

 \subsection{Measure of the transfer velocity}
\label{sec:meastrans}

\begin{figure}[htp!]
\begin{center}
\includegraphics[width=.8\linewidth]{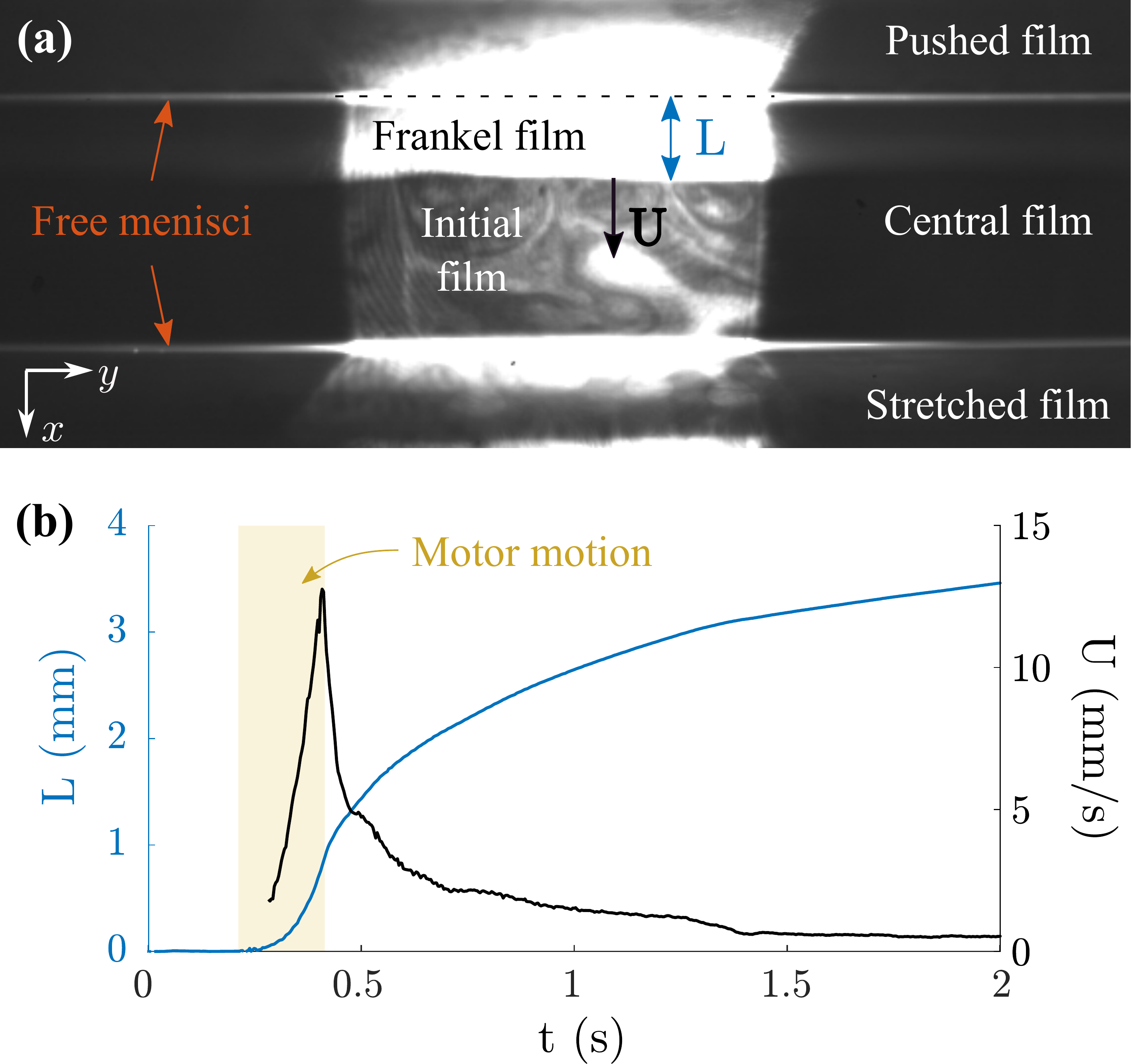}
\caption{(a) Image of the fluorescence camera (top view) after the deformation showing the Frankel film extraction. This Frankel film is invariant in the $y$ direction and the black parts on both sides are due to the fact  that the illuminated domain size is smaller than $W$. (b) Typical Frankel film length $L$ (blue) and velocity $U= dL/dt$ (black) evolution with time for $\Delta d=10$ mm, $V=50$ mm/s and $d_m=12$ mm. Yellow shaded area indicates when the motors move.
\label{fig:frankel}}
\end{center}
\end{figure}
 
An image of the central film is shown in figure \ref{fig:frankel} (a). In this film, the relaxation of the peripheral film deformation by interface transfer results in an extraction of a new film from the free meniscus on the compressed side (see Fig.~\ref{fig:frankel} (a)) and in  a film entry in the free meniscus on the stretched side. The film extraction is governed by the Frankel's theory \cite{mysels,bussonniere20} and the new film, called hereafter Frankel's film, is thicker than the remaining part of the film, which is  the film initially present (denoted {\it initial film} hereafter). The Frankel's film therefore appears bright on the fluorescence camera, with a very well defined frontier at a distance $L(t)$ from the pushed meniscus. The extraction begins as soon as the motors start and accelerates until the motors stop. The velocity is maximum at this time (Fig.~\ref{fig:frankel} (b)) and then the extraction slows down over a characteristic timescale of 1 second. 

This motion occurs without compression nor extension of the central film,  which simply translates in the $x$ direction \cite{bussonniere20}. The central film dynamics  is thus fully resolved by tracking the position $L(t)$ of the Frankel's film frontier, with respect to the pushed meniscus position. For experiments recorded at a high frame rate (300 fps), the central film velocity is computed by smoothing the time derivative of $L$. For longer experiments with slower frame rate (130 fps), this velocity is extracted by first  fitting the evolution of $L$ with a four order polynomial during motor motion and a logarithm function after motor stops. This uniform central film velocity is our experimental definition of the transfer velocity $U$, which happens to be identical at both free menisci, for the deformation we impose.

\subsection{Measure of the film extension}
\label{sec:thickmeas}

The fact that the films may be absorbed by or extracted from the menisci implies that each individual film can not be considered as a closed material system. Consequently, the distance $d$ between the menisci on both sides of a film does not provide a measure of its extension $\varepsilon$.

\begin{figure}[htp!]
\begin{center}
\includegraphics[width=0.95\linewidth]{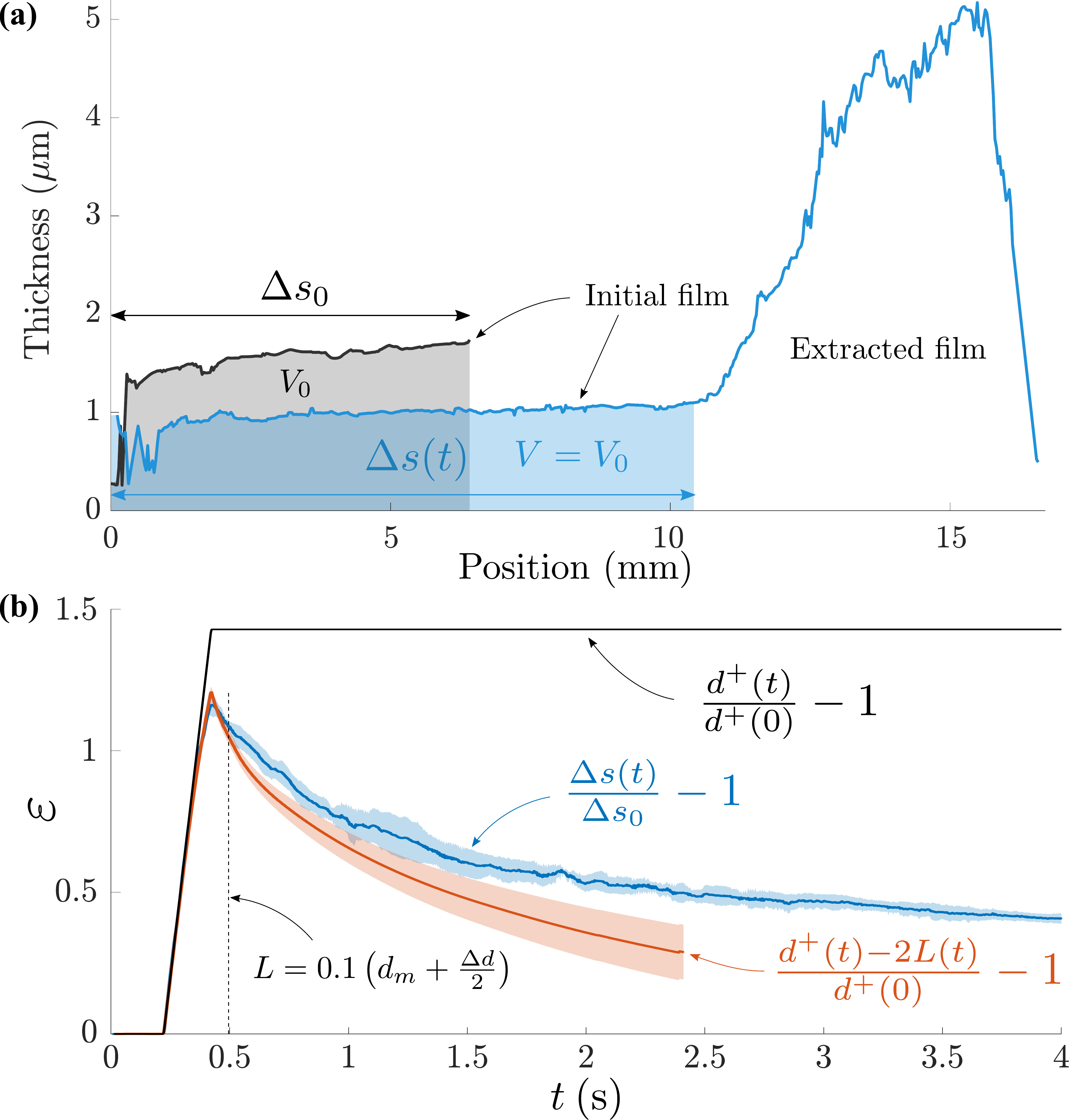}
\caption{(a) Thickness profiles of the stretched bottom film at the initial time (black) and after the deformation (blue), as measured with the spectral camera between the right free meniscus (at the origin position) and the meniscus on the bottom right moving edge. The shaded areas correspond to the initial volume of the film (per unit length). (b) Extension as a function of time. Blue : film extension based on the thickness  (eq. \eqref{eq:eps_thick}) ; Red : approximated film extension obtained from  the transfer length $L$ (eq.  \eqref{eq:eps_L}) ; Black : geometrical extension imposed by the motor positions. Averages (solid line) and standard deviations (shaded area) are calculated over 5 experiments for the thickness definition and 50 for the transfer length definition.  The control parameters are  $\Delta d=10$ mm, $V=50$ mm/s and $d_m=12$ mm. The dashed line delimits the validity range of the extension based on the transfer length. \label{fig:extension}}
\end{center}
\end{figure}

A first method to determine $\varepsilon$ is based on thickness measurements.
A film thickness profile in the stretched film, measured with the spectral camera, is shown in Fig.~\ref{fig:extension} (a). After deformation we can see a thin part, corresponding to the  initial film, in contact with a thicker part, corresponding to newly extracted Frankel's film, with a sharp transition between both. As shown in \cite{bussonniere20}, a Frankel's film is extracted both at the free meniscus (at the origin position in Fig.~\ref{fig:extension} (a)) and at the supported meniscus, on the bottom right moving edge. However, gravity imposes a stratification of the non-horizontal films, and both Frankel's films merge at the film bottom \cite{shabalina19}, thus explaining the film profile. The key point here is that the initial film is a well identified material system, which does not leave nor enter the film during the experiment, and which is well separated from the Frankel's film by a measurable frontier. 

To follow this material system, we proceed as follows: the volume $V_0$ (per unit length in the $y$ direction) of the initial film is calculated by integrating the thickness profile at $t=0$ over the total length $d^+(0)=\Delta s_0$ of the film. During the dynamic, its length $\Delta s(t)$ is deduced from the volume conservation : the thickness profile is integrated from the free meniscus at $s=0$  to the position $\Delta s(t)$ at which the integral equals $V_0$. Note that consistently $\Delta s(t)$ coincides with the position of the thickness transition, which is however known with a smaller precision.

As discussed in section \ref{sec:elas_exp}, the extension is uniform in the film, thus allowing us to integrate eq. \eqref{eq:defeps} over the whole initial film to obtain
 \begin{equation}
 \varepsilon(t)=\frac{\Delta s(t)}{\Delta s_0} -1 \, ,
 \label{eq:eps_thick}
 \end{equation}
which is plotted in Fig. \ref{fig:extension} (b) as a function of time. 

On the compressed side, the initial film is absorbed by the meniscus and the previous method unfortunately fails. Moreover, marginal regeneration plumes form at the bottom meniscus, move upward and merge with the film, draining the compressed film much faster than the other ones and making the extension measure based on eq. \eqref{eq:defeps} impossible. In that case, the actual size of the monolayer initially covering the film is estimated as $d^-(t) + L_1(t) + L_2(t)$,  with $d^-(t)$ the imposed film length at time $t$ and $L_1$ and $L_2$ the monolayer areas (per unit length in the $y$ direction) that have been lost by the compressed film respectively at the free meniscus and at the supported meniscus. As previously discussed, it is shown in \cite{bussonniere20} that $L_1 \sim L(t)$; at short time, we also observed that $L_2 \sim L(t)$.  The same assumptions can be made for the stretched film 
to take into account the gain of area on both film sides. The extension can finally be estimated by, using the appropriate sign for each case,  
\begin{equation}
    \varepsilon(t)=\frac{d^{\pm}(t) \mp 2 L(t)}{d^\pm(0)} -1.
    \label{eq:eps_L}
\end{equation}

The values of $\varepsilon$ in the stretched films, obtained using both definitions (eqs. \eqref{eq:eps_thick} and \eqref{eq:eps_L}), are plotted in Fig.~\ref{fig:extension} (b) for one series. As expected an excellent agreement is obtained at short time, but the two curves become different at longer times. Based on this comparison, we define a cut off length $L_c=  0.1 (d_m+\Delta d/2)$, represented by the dashed line in figure \ref{fig:extension} (b): for $ L(t) < L_c$, the extension can be calculated using \eqref{eq:eps_L}. Then $L_1$ and $L_2$ begin to significantly differ from $L$ and eq. \eqref{eq:eps_L} becomes invalid. 

In the following,  the extension is computed for one parameter set using eq. \eqref{eq:eps_thick},  in the stretched film and  for the whole time range. For the other cases, extension and compression are computed with eq. \eqref{eq:eps_L}, at short times only, for $L(t) < 0.1(d_m+\Delta d/2)$. As this measure is much faster, it  allows us to scan a large set of deformation parameters.
 
%%%%%%%%%%%%%%%%%%%%%%%%%%%%%%%%%%%%%%%%%%%%%%%%%%%%%%%%%%%%%%
\section{Determination of the film tensions}
\label{sec:force}

The set-up is designed so that, as long as the film structure is at equilibrium, the two free menisci stay at a constant position whatever the motor position. A meniscus motion is therefore the signature of some  dynamical forces \cite{besson07}. 
We demonstrate in this section that the dominant forces are the tension differences between the films, which  can therefore be modeled  by a minimal surface of vanishing mean curvature during the dynamics.  The position and shape  of the free menisci, that we have  extracted over time, can thus be used to measure the  film tensions.

\subsection{Meniscus motion}

\begin{figure}[htp!]
\begin{center}
\includegraphics[width=0.9\linewidth]{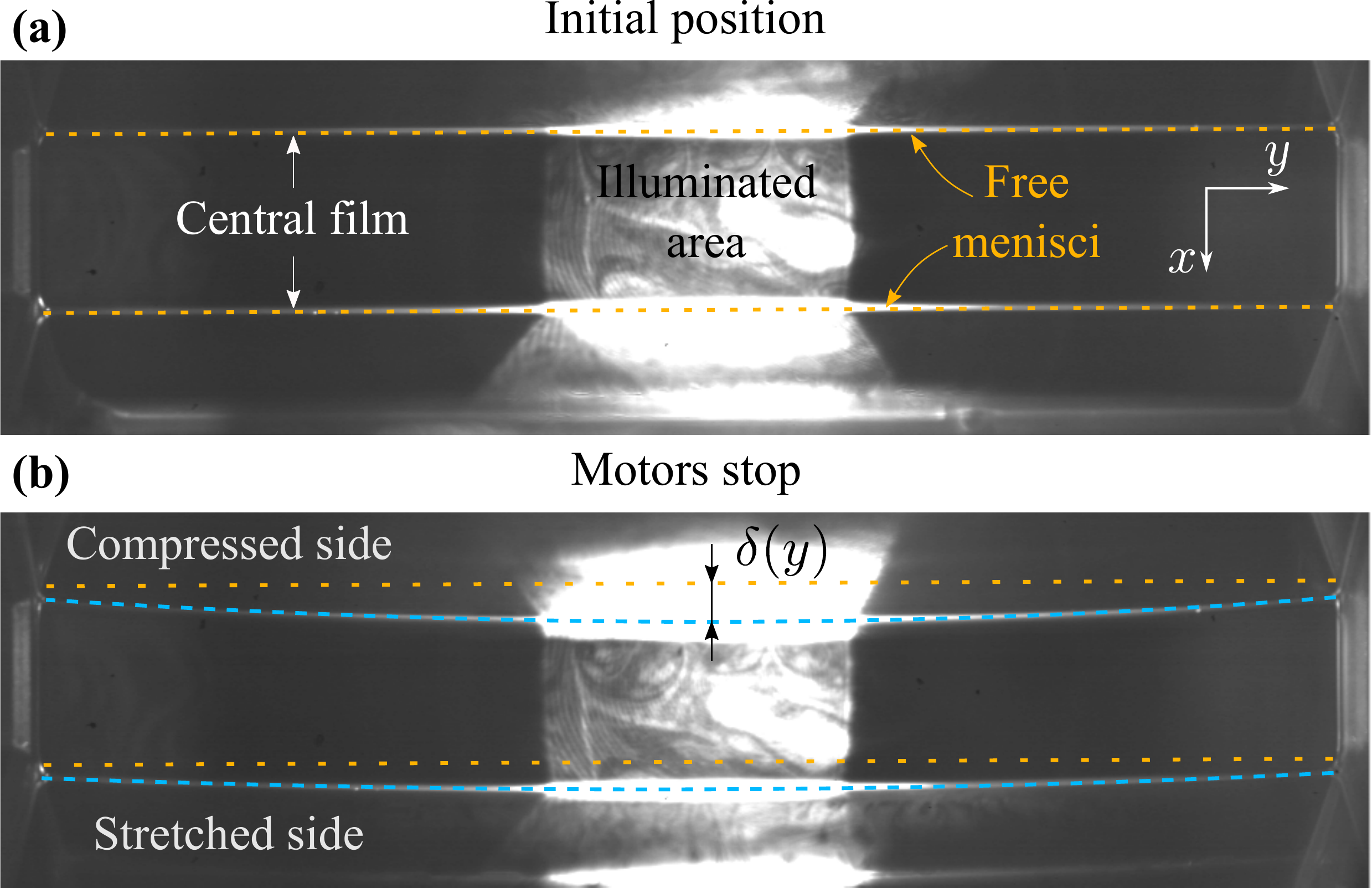}
\caption{(a) Fluorescence image of the central film at the initial state showing the position of the free menisci (yellow dashed line). (b) Fluorescence image at the end of the motor motion. New positions of the menisci are highlighted by dashed blue lines.
\label{fig:pos_menisci}}
\end{center}
\end{figure}

During the dynamics, both free menisci delimiting the central film move in the $(x,y)$ plane toward the stretched side. As shown in Fig.~\ref{fig:pos_menisci}, the meniscus ends slide on the solid frame and the whole meniscus  curves in the direction of motion. The displacement $\delta^\pm(y)$ of each free meniscus (the symbols $-$ and $+$ refer to the compressed and stretched  sides, respectively) can be fitted at each time by a second order polynomial, from which we deduce the sliding motion (the constant term $\delta_1^\pm$) and the meniscus curvature (from the quadratic term $ \delta_2^\pm(y) $):
\begin{equation}
\delta^\pm(y)= \delta_1^\pm +  \delta_2^\pm(y) = \delta_1^\pm +  \delta_2^\pm(0) \left (1- \frac{2y}{W}\right)^2    \, . 
\label{fit_men}
\end{equation}
In this expression  $y=0$ is chosen in the middle of the film. The motion in the $z$ direction is measured with the meniscus camera (see Fig.~\ref{fig:setup}) and is negligible. 

\subsection{Estimation of the tangential forces and film tension definition}
\label{sec:def_tens}

To estimate the value of the external forces acting on the films, we use the following orders of magnitude, corresponding to our observations: the film in-plane velocity scales as  the transfer velocity $U \sim 10^{-2}$ m/s, the film normal velocity scales as the meniscus velocity $U_m\sim 10^{-3}$ m/s, the film extension is $|\e| \sim 1$ and the fastest deformation time scale is $T \sim  10^{-1}$ s, corresponding to an extension rate $\de \sim 10 $ s$^{-1}$. Finally, we anticipate that the film tension differences between the different films $\Delta \sigma$, that is  deduced from the meniscus shape in section \ref{sec:tens_exp},  are of the order of $10^{-3}$ N/m. 

\begin{figure}[htp!]
\begin{center}
\includegraphics[width=0.65\linewidth]{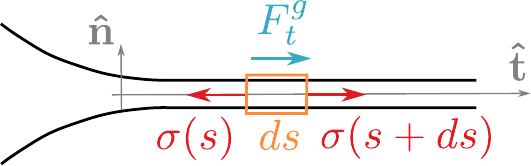}
\caption{Schematic of a thin film with the tangential forces. The variable $s$ in used along the direction $\hat{t}$ and the variable $\zeta$ along $\hat{n}$. 
\label{fig:tangent}}
\end{center}
\end{figure}

As shown in Fig. \ref{fig:tangent}, we use on each thin film the normal and tangential unit vectors $\vv{n}$ and $ \vv{t}$, along the thin film profile (in the $(x,z)$ plane),  associated to the spatial variables $\zeta$ and $s$, respectively. 
We define the film tension $\sigma(s) \vv{t}$ as the action of the film at an abscissa larger than $s$ on the film at an abscissa smaller than $s$. This quantity takes into account the interfacial stress $\gamma$ on both interfaces, and the contribution of the pressure in the liquid bulk, governed by the Laplace pressure (the latter term being negligible in the central part of each film).
The tangential force balance on the piece of film located between $s$ and $s+ ds$ is 
\begin{equation}
\rho \left (  \frac{ \partial (h \langle u \rangle)}{\partial t} + \frac{\partial  (h \langle u^2 \rangle) }{\partial s} \right) = \frac{\partial \sigma}{\partial s} + \rho g_s h + F_t^g  \, , 
\label{eq:forc_bal_film}
\end{equation}
with $\rho$ the solution density, $g_s$ the gravity component along $ \vv{t}$, $\langle u \rangle$ the tangential velocity averaged in the normal direction, scaling as $U$, and $F_t^g$ the tangential stress due to the gas phase at both interfaces. 

The first  inertial term scales as  $\rho h U/T  \sim 10^{-4}$ Pa,  the weight as $  \rho g_s h \sim 10^{-2}$ Pa (for the peripheral films) and the air-borne stress  as $F_t^g \sim \eta_g U/\delta_{Bl}$, with $\delta_{Bl}$ the thickness of the Blasius visco-inertial boundary layer, $\eta_g \sim 10^{-5}$ Pa$\cdot$s the gas shear viscosity and $\rho_g \sim 1$kg/m$^3$ the gas density \cite{rutgers96}. The value of $\delta_{Bl}$ is of the order of $\sqrt{\eta_g T/\rho_g} \sim  10 ^{-3}$ m (or $\sqrt{\eta_g d/(U \rho_g)}$ of similar order) and thus $F_t^g \sim 10^{-4}$ Pa. 
The convection term arises from the fact that we considered an open system and scales as $\rho h U^2 /d \sim 10^{-5}$ Pa. 

In the horizontal film, it results from these orders of magnitude that the Marangoni term $\partial \sigma/\partial s$, also appearing in eq. \eqref{eq:forc_bal_film}, can not be larger than $10^{-4}$ Pa. Its variation between both ends of the film is thus below $10^{-6}$ N/m, which is much smaller than $\Delta \sigma$. 
The surface tension variation induced by the gravity in the peripheral film simply balances its weight and is easily determined as $\Delta \sigma^{grav} \sim \rho g_s h d \sim 10^{-4}$ N/m, which is negligible too (and could be easily taken into account if needed).

One important consequence is that, in the parameter range we explored, the film tension is necessarily uniform on each thin film, whatever its physico-chemical properties \cite{durand06}. We thus define the film tension $\sigma^-$ in the two compressed peripheral films (top and bottom films are identical by symmetry, as gravity is negligible), $\sigma^+$ in the two stretched peripheral films, and $\sigma^c$ in the central film. 

The contribution of both interfaces and of the film bulk to this film tension will be discussed in section \ref{sec:const_exchange}. Note that, as tensions and extensions are related to each others, the tension uniformity validates the assumption of uniform extension in each given film made in section \ref{sec:thickmeas}.

\subsection{Estimation of the normal forces}

\begin{figure}[htp!]
\begin{center}
\includegraphics[width=0.45\linewidth]{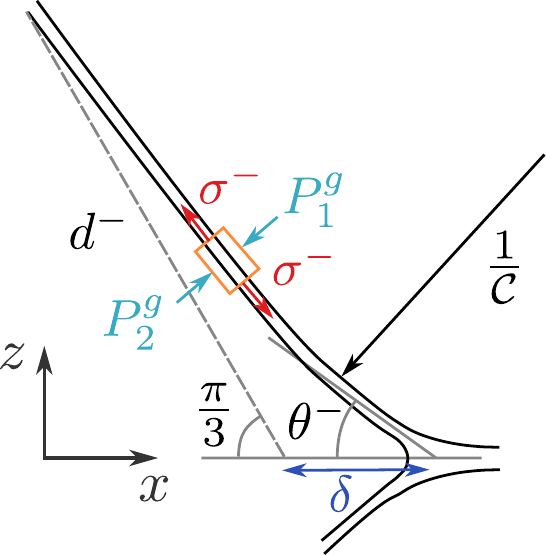}
\caption{Schematic of the pushed film with the normal forces applied.
\label{fig:normal}}
\end{center}
\end{figure}

We now consider the normal motion of the peripheral films in order to show that they keep a negligible mean curvature during the deformation. Disregarding gravity effects, the $(x,y)$ plane is a symmetry plane, so the central film remains flat and stay in the $(x,y)$ plane. As shown in Fig. \ref{fig:normal}, the normal velocity of a piece of peripheral film is of the order of the meniscus velocity  $U_m$. The Newton law in the normal direction applied to this system involves thus an inertial  term (per unit film area) scaling as  $I_f= h \rho U_m/T \sim 10^{-5} Pa$. The normal forces  are the gas pressures on both sides and the Laplace pressure, {\it i.e.} the normal component of the film tension contribution  \cite{salkin16}. The convection term is, as for the tangential projection, negligible. 

In our set-up, the gas constitutes a continuous phase and the pressure variations are thus of dynamical origin only. The inertial gas pressure scales as $P^{\mbox{g}}= \rho_g U_m^2 \sim 10^{-6}$ Pa $ \ll I_f$. The gas phase can thus be assumed to be at rest. In the following,  the atmospheric pressure is chosen as pressure reference and all the pressures defined in the liquid phase are the actual pressure minus this uniform atmospheric pressure.

The force balance thus only involves the film inertia $I_f$ and the Laplace pressure $\sigma {\cal C} $.  This  provides a scaling law  for the film  mean  curvature ${\cal C}$:
\begin{equation}
 {\cal C} \sim \frac{h \rho U_m}{\sigma  T} \sim 10^{-3} \mbox{ m}^{-1}.
\label{eq:dyn_curv}
\end{equation}

This mean curvature is much smaller than the observed curvature in the $(x,y)$ plane, of the order of 20 m$^{-1}$ (see Fig. \ref{fig:pos_menisci}) and is therefore negligible. The peripheral films remain thus minimal surfaces of vanishing mean curvature, even during the motor motion, and their entire shape can be deduced from the position of their boundaries, \textit{i.e.} from the position and shape of the free menisci.

\subsection{Determination of the angles between the films}

\begin{figure}[htp!]
\begin{center}
\includegraphics[width=1\linewidth]{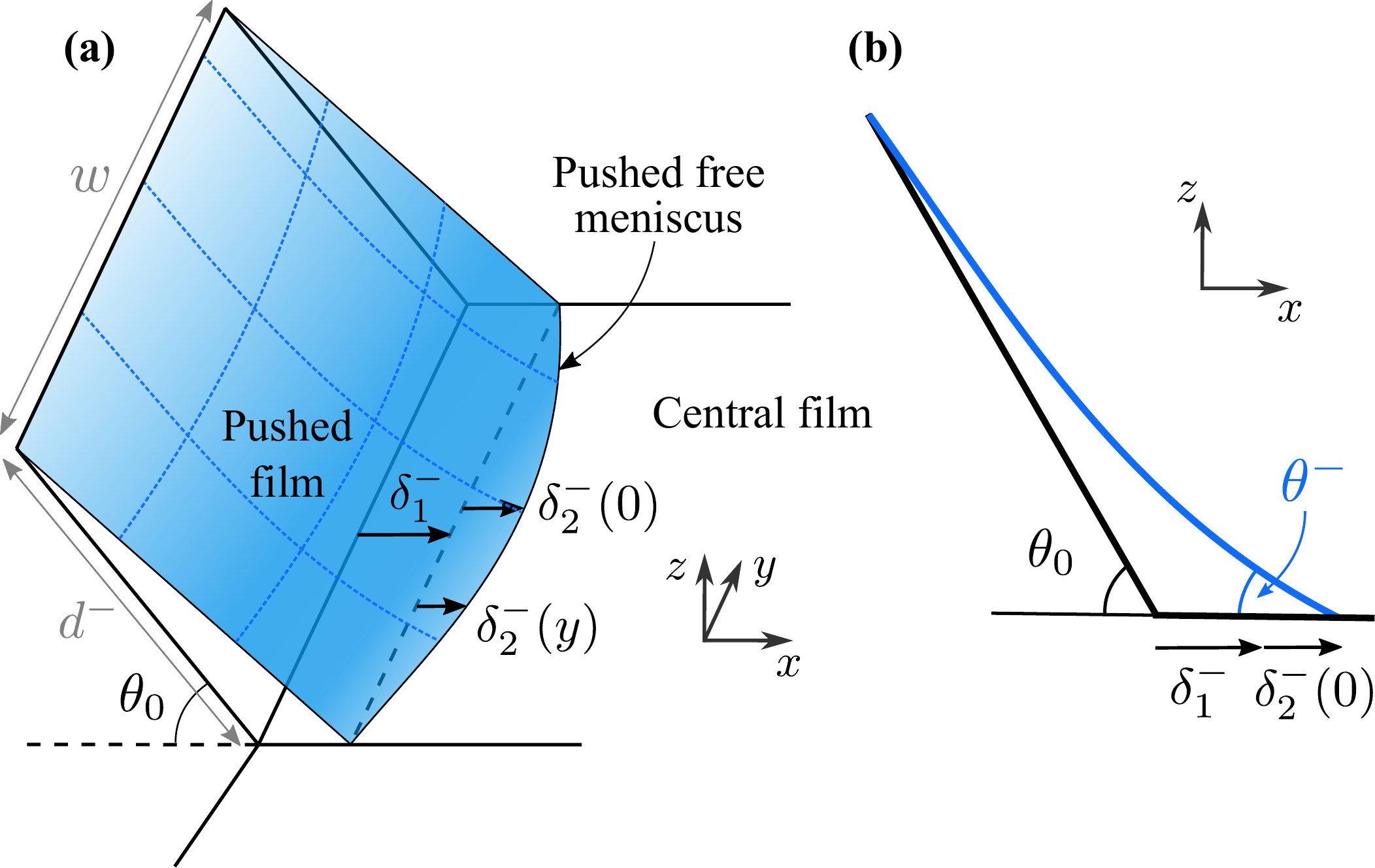}
\caption{(a) 3D schematic of the pushed film shape ensuring zero mean curvature. (b) Cutting scheme of the pushed film at the middle of the film.
\label{fig:film_shape}}
\end{center}
\end{figure}

Figure \ref{fig:film_shape} shows the scheme of the pushed film shape ensuring vanishing mean curvature and geometrical constrains. The relevant geometrical quantity is the angle $2 \theta^-$ (resp.  $2 \theta^+$) between the tangent vectors of the top and bottom pushed (resp. stretched) film, measured at the free meniscus position, in the $y=0$ plane ({\it i. e.  } in the middle of the film). Its expression as a function of the free meniscus shape in the $(x,y)$ plane  (Fig. \ref{fig:film_shape}) is derived in Appendix \ref{app:angles} and is given by :
\begin{multline}
\theta^{\pm}=\tan^{-1}\left(\frac{d^\pm\sin\theta_0}{d^\pm\cos\theta_0+\delta_1^{\pm}}\right)\\
-\frac{\delta_2^{\pm}(0)}{w}\frac{\pi\sin\theta_0}{\tanh\left( \frac{\pi d^\pm}{w} \right)}.
\label{eq:theta}
\end{multline}
The first term on the right-hand side is due to the meniscus sliding displacement ($\delta_1$) and the second term is a correction induced by the meniscus curvature ($\delta_2$), as defined in eq. \eqref{fit_men}. The initial equilibrium angle is $\theta_0=\pi/3$.

\subsection{Determination of the film tensions}
\label{sec:tens_exp}

The film tension differences between the adjacent films can now be obtained from the force balance on the free menisci. Along the $x$ direction we have, for the compressed ($-$)  and stretched ($+$) sides, 
\begin{equation}
 \rho \,  \alpha_{men}  \,  r_m^2 \frac{d^2x^\pm_m}{dt^2} =  \pm \left (  2 \sigma^\pm \cos \theta^\pm - \sigma_c \right) \; , 
\label{eq:menisc_bal}
\end{equation}
with $\alpha_{men} r_m^2$ the section area of the meniscus.
The meniscus inertia scales as $  \rho r_m^2 U_m/T \sim 10^{-6} $ N/m and is much smaller than $\Delta \sigma$. We thus obtained the sought relationship :
\begin{equation}
\sigma^\pm  = \frac{\sigma_c}{2 \cos \theta^\pm},
\label{eq:delta_sig}
\end{equation}
with  $\theta^-$ and  $\theta^+$ expressed as a function of the meniscus displacement in eq. \eqref{eq:theta}. 

\begin{figure}[htp!]
\begin{center}
\includegraphics[width=0.8\linewidth]{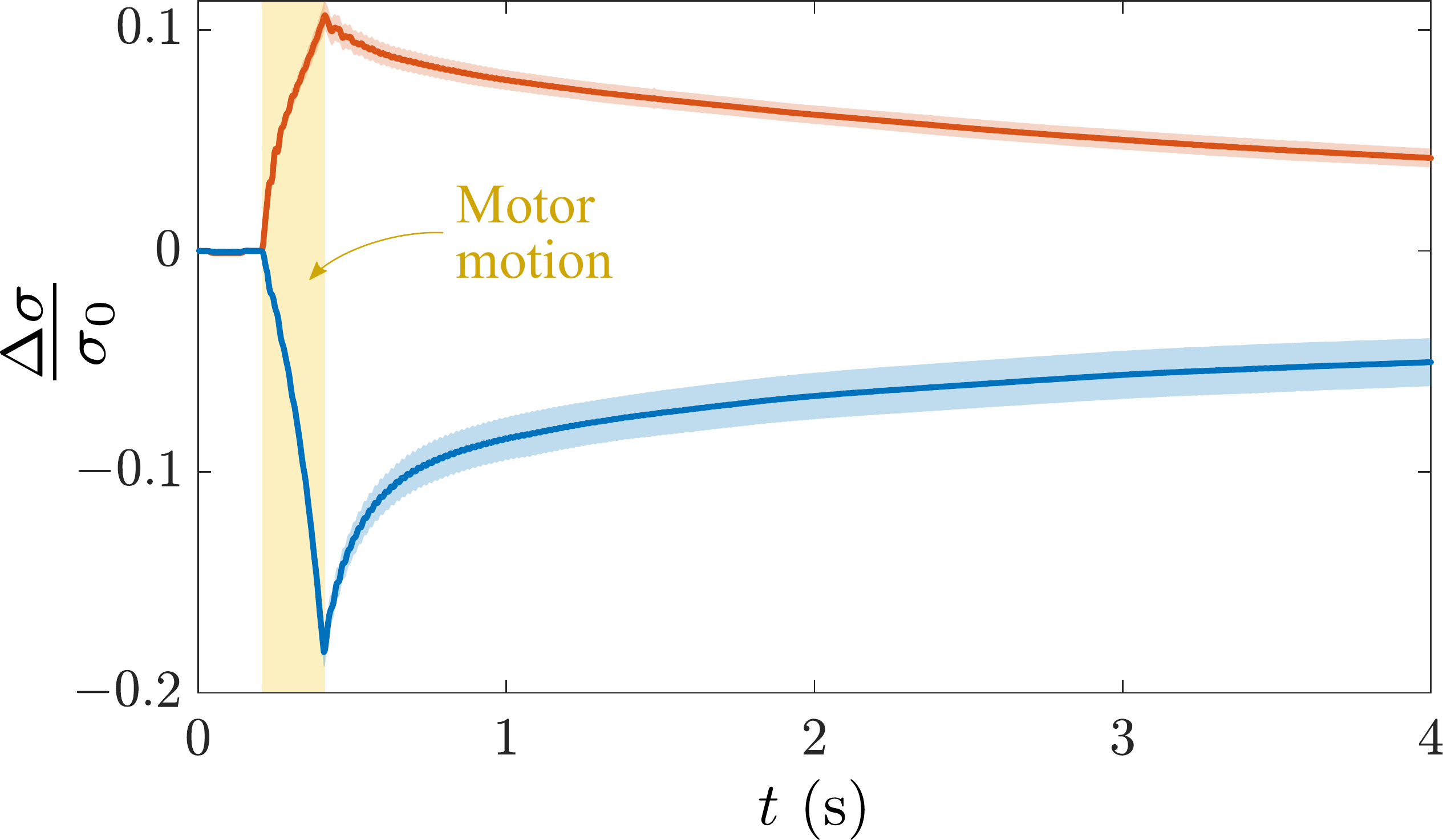}
\caption{Typical time evolution of the film tension variation of the stretched (in red) and compressed (in blue) side for $\Delta d$=10 mm, $V$=50 mm/s and $d_m$=12 mm. Solid lines (resp. shaded areas) represents the average (resp. the standard deviation) over 50 experiments. \label{fig:ex_tension}}
\end{center}
\end{figure}

We show in the next section that any variation of film tension is associated to a film extension $\varepsilon$. As the central film is never stretched or compressed we can assume  $\sigma_c = \sigma_0 = 2\gamma_0$. The film tension variations in the stretched and compressed peripheral films, with respect to their equilibrium values, are thus given by: 
\begin{equation}
\frac{\Delta \sigma^\pm}{ \sigma_0} = \frac{1}{2 \cos \theta^\pm}-1. 
\label{eq:delta_sig2}
\end{equation}

An example of film tension variation is shown in Fig.~\ref{fig:ex_tension} as a function of time. We find, as anticipated, an order of magnitude of a few mN/m for $\Delta \sigma$.  As soon as the motors start, the film tension begins to deviate from its equilibrium value.  It reaches a maximum/minimum when the motors stop and then relaxes.

Note that, by symmetry,  the final state reached by the system is identical to its initial state. The total amount of dissipated energy ${\cal D}$ (per unit length) is therefore the total work provided by the motor to the system: 
\begin{equation}
{\cal D} = V \int_0^{t_m}  2 (\sigma^+  - \sigma^-) dt \, , 
\label{eq:distot}
\end{equation}
with $V$ the motor velocity and $[0-t_m]$ the duration of the motor motion. The tension difference between the stretched and compressed films is thus a direct signature of the system dissipation.
%%%%%%%%%%%%%%%%%%%%%%%%%%%%%%%%%%%%%%%%%%%%%%%%%%%%%%%%%%%%%%

\section{Constitutive relation for the film}
\label{sec:const_film}

In the previous sections we determined the tension and the extension of the different films, which now allows us to build the film experimental constitutive relation, {\it i.e.} the relationship between the two quantities.  

\subsection{Experimental results}
\label{sec:elas_exp}

\begin{figure}[htp!]
\begin{center}
\includegraphics[width=.85\linewidth]{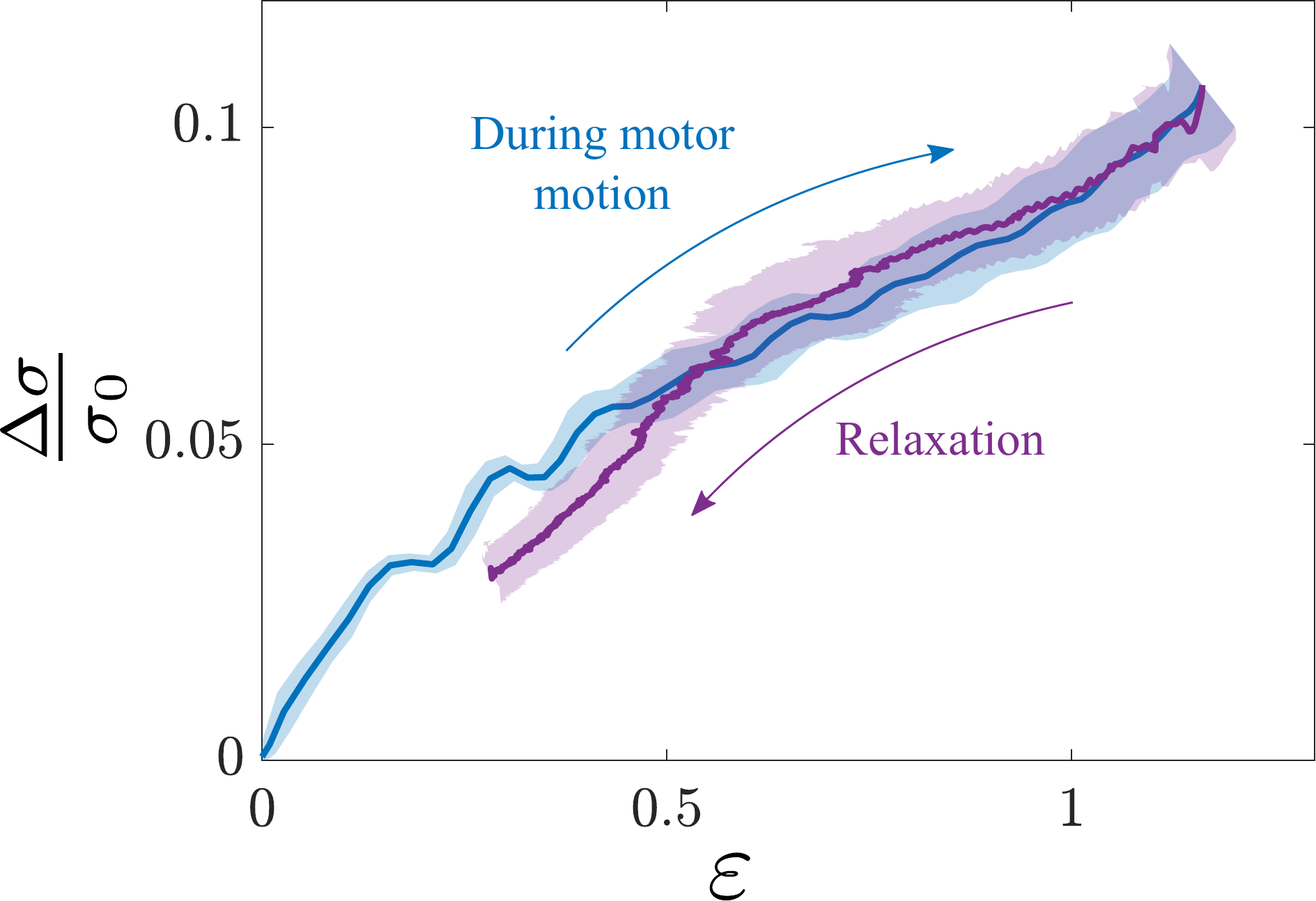}
\caption{Film tension relative variation as a function of $\varepsilon$ deduced  from the film thickness (using eq. \eqref{eq:eps_thick}) during the motor motion (blue) and after motor stops (purple) for $\Delta d$=10 mm, $V$=50 mm/s and $d_m$=12 mm. The solid lines (shaded areas) represent the averages (standard deviations).
\label{fig:elasLT}}
\end{center}
\end{figure}

We first plot, in Fig. \ref{fig:elasLT}, the tension variation in term of the film extension for the experiments where the extension has been deduced from the thickness measurements in stretched films. This  allows us  to monitor the relationship over the entire experiment, during 4 seconds. The film is first stretched (blue data) and then relaxes toward its initial length (purple data). The most noticeable result of the paper is that the two parts of the curves are perfectly superimposed, thus proving unambiguously the purely elastic behavior of the film. 

To investigate further the role of the extension rate, we varied the motor motion parameters in a large range. The amount of data was too large to use the definition of eq. \eqref{eq:eps_thick} of the extension (which requires manual check during the data processing) and we used the definition of eq. \eqref{eq:eps_L} instead, for the stretched and compressed films, at short times ({\it i.e.} during motor motion and just after).  

At each time, for each experiment and each film, we measure the data set ($\e, \de, \sigma$). All 
data points are then considered together, whatever the values of the control parameters. They are binned by extension rate $|\de|$, and averaged.
The bins have been chosen to show the whole range of extension rate while keeping a significant number of points in each bin.  Most of the points are associated to a small $|\de|$, but there is still 8500 data points for the $[5\ ;\ 25]$ s$^{-1}$ bin.

\begin{figure}[htp!]
\begin{center}
\includegraphics[width=1\linewidth]{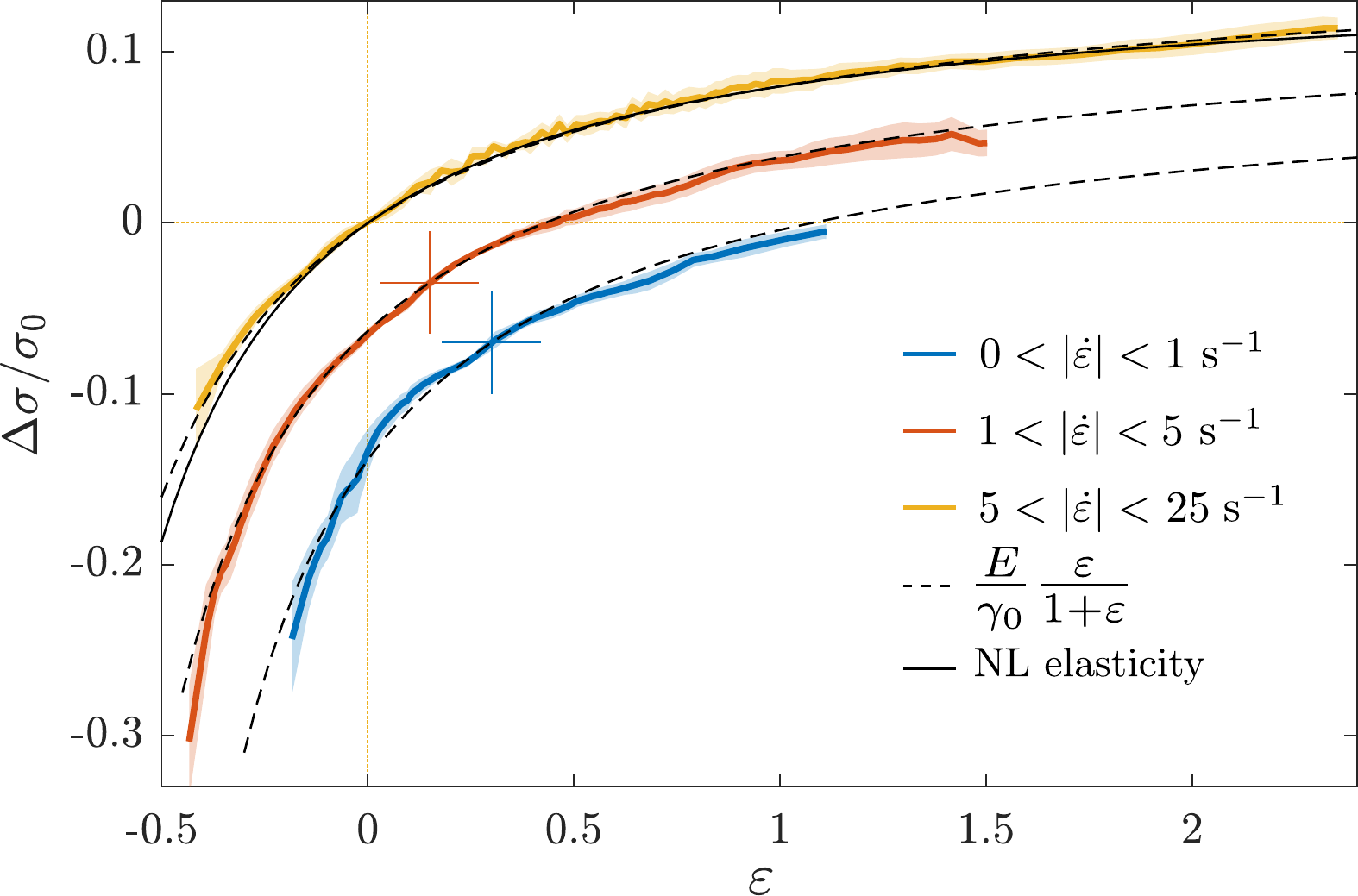}
\caption{Film tension relative variation as a function of the local film extension $\varepsilon$. Each color corresponds to the average over all the data having an extension rate in a given range. The 3 curves are perfectly superimposed and the red and blue curves have been shifted for the sake of visibility: the corresponding axis origins are the cross of the same color. The error bar (shaded area) is the standard deviation and is less than $10\%$ in average. The dashed lines are 3 times the same curve: the theoretical prediction of eq. \eqref{eq:elastheo}, with the relative elasticity $E/\gamma_0=0.16$. The solid line corresponds to the non-linear model using Langmuir adsorption of reference \cite{fang92}, derived in appendix \ref{app:NL_elas}, with an initial DOH concentration of $0.6 \,  c_0$.
\label{fig:elasticity}}
\end{center}
\end{figure}

The obtained results are shown in Fig. \ref{fig:elasticity}. Negative $\e$ correspond to a compression and positive $\e$ to an extension. Note that to provide a better readability of the data, we shifted the $x$ and $y$ axis of the two lowest extension rates. The three curves would otherwise be perfectly superimposed, as indicated by the theoretical law (dashed black lines) plotted on each graph, which is each time the same curve. For the investigated parameters, {\it i. e} $|\de|$ in the range  $[0 \; ; \; 25]$ s$^{-1}$ and  $\e$ in the range $[-0.5 \; ; \; 2]$,  the film tension is thus a function of the extension only. The relationship between both quantities is discussed in the next section, on the basis of the classical models.

\subsection{Gibbs-Marangoni elasticity}
\label{sec:gibbsmar}
The full interfacial stress $\gamma$ of an interface involves the thermodynamic definition of the surface tension $\gamma_{th}$ which depends solely on the local surfactant interfacial excess ($\Gamma$) and the intrinsic surface extensional and shear viscosities, respectively $\kappa_s$ and $\eta_s$. In our $y$-invariant geometry, similar to a Langmuir trough geometry, this stress is \cite{edwards,stone10}:
\begin{equation}
\gamma = \gamma_{th}(\Gamma(s)) + (\eta_s + \kappa_s) \,  \dot{\varepsilon}  \, ,
\label{eq:interfacial_stress}
\end{equation}
The local surface excess might deviate from its initial equilibrium value and depends on the surfactant transport processes. 
In the general case, the surfactant  advection-diffusion and the exchanges between  the bulk and the interface result in an elastic and an apparent viscous behavior due to, respectively, the in-phase and out-of-phase (delayed) response of the surface excess with the deformation.

Here, the diffusion time scales of the surfactants in the directions parallel or transverse to a thin film scale as $\tau^p \sim d^2/D \sim 10^6$ s and $\tau^t \sim h^2/D \sim 10^{-2}$ s, respectively, whereas the experimental time scale is of the order of 1 s. This scale separation allows us to assume that (i) there is no diffusive transport along the film; (ii) at a given location $s$ in the film, the equilibrium between the  bulk concentration and the interface excess  is immediately reached. From these assumptions, and following \cite{couder89}, we can deduce the relation between the surface excess and the film extension. 

As established in section \ref{sec:force}, the film tension $\sigma$ is homogeneous in each film. However, some important dynamical processes, discussed in section \ref{sec:const_exchange}, occur close to the meniscus and lead to variations of the interface tension $\gamma$ on both film interfaces, while keeping the resulting film tension constant. Here we focus on the central part of the films, where the bulk pressure is the reference pressure,    both interface tensions are identical, and  the velocity field across the film is homogeneous (see section \ref{sec:dom_def} for more details). In this domain we thus simply have $\sigma = 2 \gamma$.  
Moreover, the interface $dS(t)$ of a film element $d\Omega = h(t) dS(t)$  is always in contact with the same liquid bulk. At our experimental time scale $\tau^t \ll T \ll \tau^p$, we can thus  assume that (i) the film element ${\cal S}$  is a closed system (both for the liquid phase and for the surfactants); (ii) the bulk concentration $c$ has a homogeneous value $c_0+ \Delta c$ in $d\Omega$ and is at equilibrium with the interface concentration, so that  $\Gamma=\Gamma_0+h_\Gamma\Delta c$ (eq. \eqref{eq:surf_trans}). The surfactant mass conservation leads to \cite{couder89}:
\begin{equation}
\Gamma= \Gamma_0 \frac{1 + h_0/(2 h_\Gamma)}{1+\varepsilon +  h_0/(2 h_\Gamma)}\ .
\label{eq:gamma_couder}
\end{equation}

The surface tension $\gamma_{th}$ is related to the local surface concentration through eq. \eqref{eq:elas_gibbs} and, using eq. \eqref{eq:gamma_couder}, the interfacial stress in eq. \eqref{eq:interfacial_stress} becomes :
\begin{equation}
\gamma = \gamma_0 + E\ \frac{\varepsilon}{1+\varepsilon +  h_0/(2 h_\Gamma)}+(\eta_s + \kappa_s)\ \de.
\end{equation}

Finally, the relative film tension variation is predicted to be:
\begin{equation}
\frac{\Delta\sigma}{\sigma_0}
=\frac{\sigma^\pm-\sigma_c}{\sigma_c} 
= \frac{E}{\gamma_0}\ \frac{\varepsilon}{1+\varepsilon +  h_0/(2 h_\Gamma)}+\frac{(\eta_s + \kappa_s)}{\gamma_0}\ \dot{\varepsilon}\ .
\label{eq:elastheo}
\end{equation}

This prediction is plotted in Fig. \ref{fig:elasticity}, using  $\kappa_s+\eta_s=0$, $h_0/h_\Gamma=0$ and $E/\gamma_0=0.16$ as fitting parameters. The agreement with experimental data is excellent for the whole range of deformation and deformation rate explored, and the constitutive relation for the films is thus 
\begin{equation}
\sigma= \sigma_0 + 2 E \,  \frac{\varepsilon}{1+\varepsilon}\ .
\label{eq:elastheo2}
\end{equation}

A first important consequence of this agreement with the  experiments  is that the films do not exhibit any measurable viscous behavior neither intrinsic nor effective. The potential viscous contribution is actually hidden by the experimental error estimated around 0.5 mN/m. The viscous term is thus below  0.5 mN/m for extension rate reaching $\dot{\varepsilon}\approx 10$ s$^{-1}$  which  provides the upper limit for the surface viscosities $\kappa_s+\eta_s\le 5.10^{-5}$ kg$\cdot$s$^{-1}$. This result  is consistent  with different measurements reported in the literature \cite{wantke03,drenckhan07,zell14}. 
Therefore, the  dissipation observed in our experiments cannot be attributed to the viscosity of the interface.

A good fit of the experimental results by the equation \eqref{eq:elastheo} is obtained for a large range of $h_\Gamma$ ($5\ \mu$m $<h_{\Gamma}<\infty$) while the relative elasticity remains in a narrow range $0.16<E/\gamma_0<0.18$ corresponding to an elasticity $E\approx 5-6$ mN/m. This indicates that the film tension variation arises from insoluble surfactants (important $h_\Gamma$) and validates our assumption that the dodecanol is at the origin of the observed tension variations. SDS molecules are mainly passive to the deformation due to the high bulk concentration as well as the fast adsorption/desorption dynamic of the order of 1 ms $\ll$ T \cite{chang92}.

The elasticity $E$ extracted from the experiments is closed to the one estimated by using   the physico-chemical model of \cite{fang92} ($E^{th}= 10.6$ mN/m, see Appendix \ref{app:phys_chem}). However, $E$ is slightly  lower than this estimated value. This deviation may be due to the fact that the dodecanol is depleted during the film formation \cite{couder89}. Such a DOH depletion would indeed results in a decrease of the elasticity, in agreement with the experimental trend.

A refined model based on the non-linear Langmuir adsorption (see appendix \ref{app:NL_elas}) gives also a good fit of the experiment if the initial DOH concentration is assumed to be $0.6 \, c_0$, as shown by the solid line in figure \ref{fig:elasticity}. At the present, it is not possible to discriminate between the different models of adsorption isotherms, nor to determine $h_\Gamma$. In the following, for the sake of simplicity, we choose the simplest linear model with $E/\gamma_0=0.16$ obtained in the limit $h_\Gamma\rightarrow\infty$. Note that this choice introduces only a small error on $E$, as $h_\Gamma$ is much larger than $h_0$ and thus  has little influence on the fit.

%%%%%%%%%%%%%%%%%%%%%%%%%%%%%%%%%%%%%%%%%%%%%%%%%%%%%%%%%%%%%%

\section{Flow properties in the meniscus and around}
\label{sec:const_men}
As shown in the previous section, the thin films far from menisci confer a pure elastic behavior to the foam assembly. In this section, we show that the viscous, dissipative, behavior arises from a generic  geometrical frustration at the meniscus.

\subsection{Experimental relationship between the transfer velocity and the tensions}

\begin{figure}[htp!]
\begin{center}
\includegraphics[width=.8\linewidth]{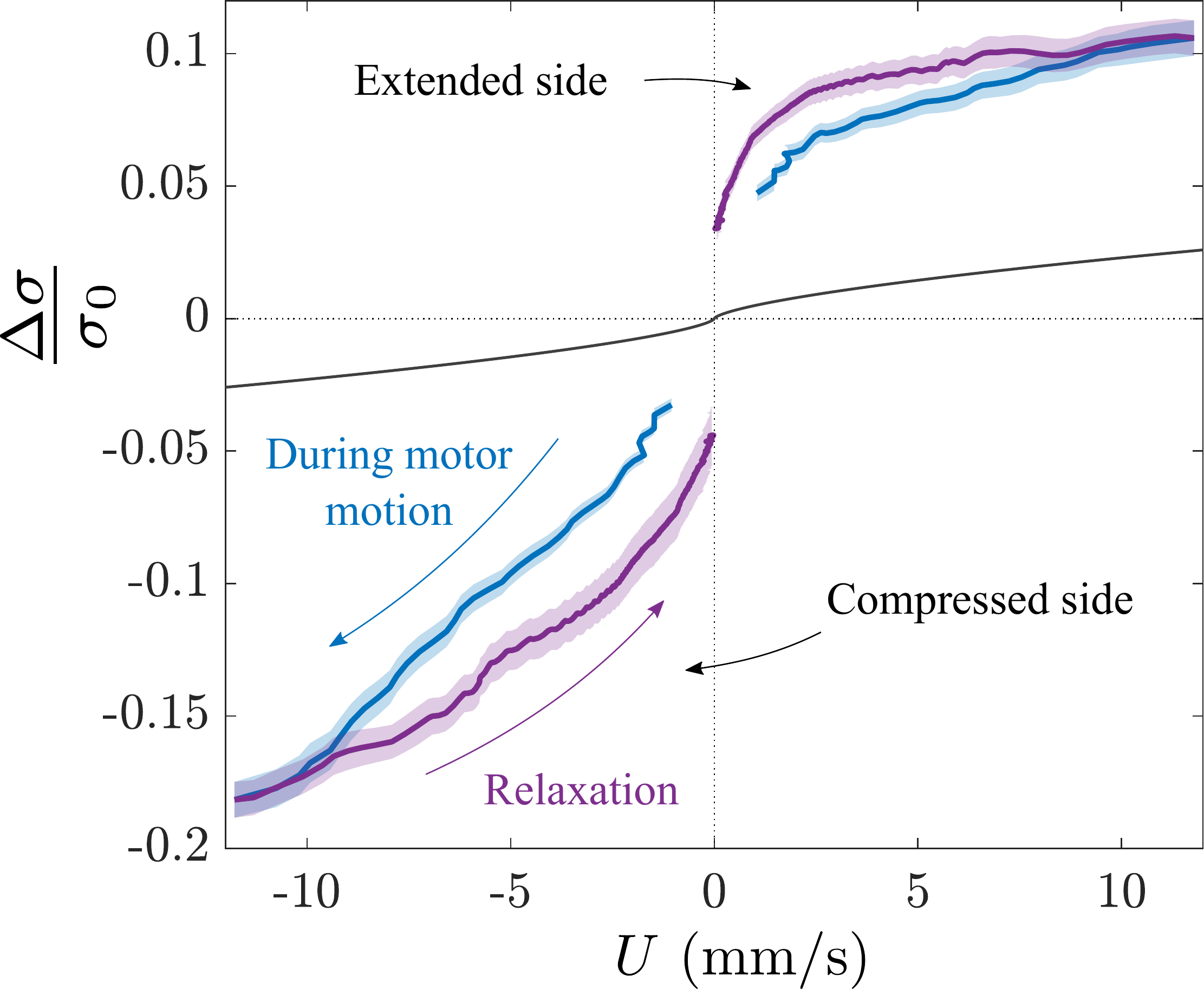}
\caption{Relative film tension variation in term of the transfer velocity $U$ for $\Delta d$=10 mm, $V$=50 mm/s and $d_m$=12 mm. Solid lines (shaded areas) represent the averages (standard deviations) over 50 experiments, respectively in blue and purple for the short times (during the motor motion) and the long times (after the motors stop). The black line is the theoretical tension relative variation associated to a Frankel film extraction at velocity $U$, given by  eq. \eqref{eq:dgpull}.\label{fig:tension_velocity_men}}
\end{center}
\end{figure}

A surface tension difference between films arises from the extension/compression of the peripheral films at short time, and this tension difference tends to relax through interface transfer from one film to its neighbor at later times. Figure \ref{fig:tension_velocity_men} shows the experimental relationship between the tension difference and the transfer velocity, i.e. the observed viscous response of the film assembly. By convention, the velocity $U$ is positive on the stretched side, and negative on the compressed one. 

The surface tension evolution is asymmetric between the compressed and stretched  sides: on the stretched side, surface tension rises rapidly with the transfer velocity and seems to reach a plateau at $\Delta\sigma/\sigma_0\approx 0.1$ ; on the compressed side, tension keeps decreasing at higher velocity. Despite a significant difference between the curves obtained  during the motor motion and afterwards,  they  remains qualitatively the same, for both sides.

The tension difference between adjacent films and the transfer velocity are well correlated for our whole parameter set (see section \ref{sec:comp_shear}), and, in a first approximation, the relationship between $\Delta \sigma$ and $U$ plays the role of a constitutive relation for the meniscus. However, a non negligible deviation is observed between experiments with different parameters and, except in some limits that we identify in section \ref{sec:const_exchange}, other dynamical parameters should be taken into account to fully describe the experimental data. 

As the interface transfer is the process allowing to relax the elastic energy stored in the peripheral films, it is a dissipative process. The only dissipative features in the system are the viscous and diffusive transports,  the first contribution scaling as the square of the velocity gradients, and the second as the square of the concentration gradients. 
The dissipative processes in the central part of the peripheral films have been shown to be negligible in the section \ref{sec:const_film}: the interface and bulk viscosities do not contribute to the dynamics, and the diffusion in the films is either to fast or to slow to induce a significant dissipation. The dissipation is thus localized in the menisci or in their  vicinity, as shown below.

\subsection{Meniscus frustration - Domain definitions}
\label{sec:dom_def}

The prediction of the relationship between the velocity transfer and the tension difference between adjacent films first requires  to  analyze where the tension gradients are located. This is performed using specific approximations in the different domains defined in this section. 

\begin{figure}[htp!]
\begin{center}
\includegraphics[width=1\linewidth]{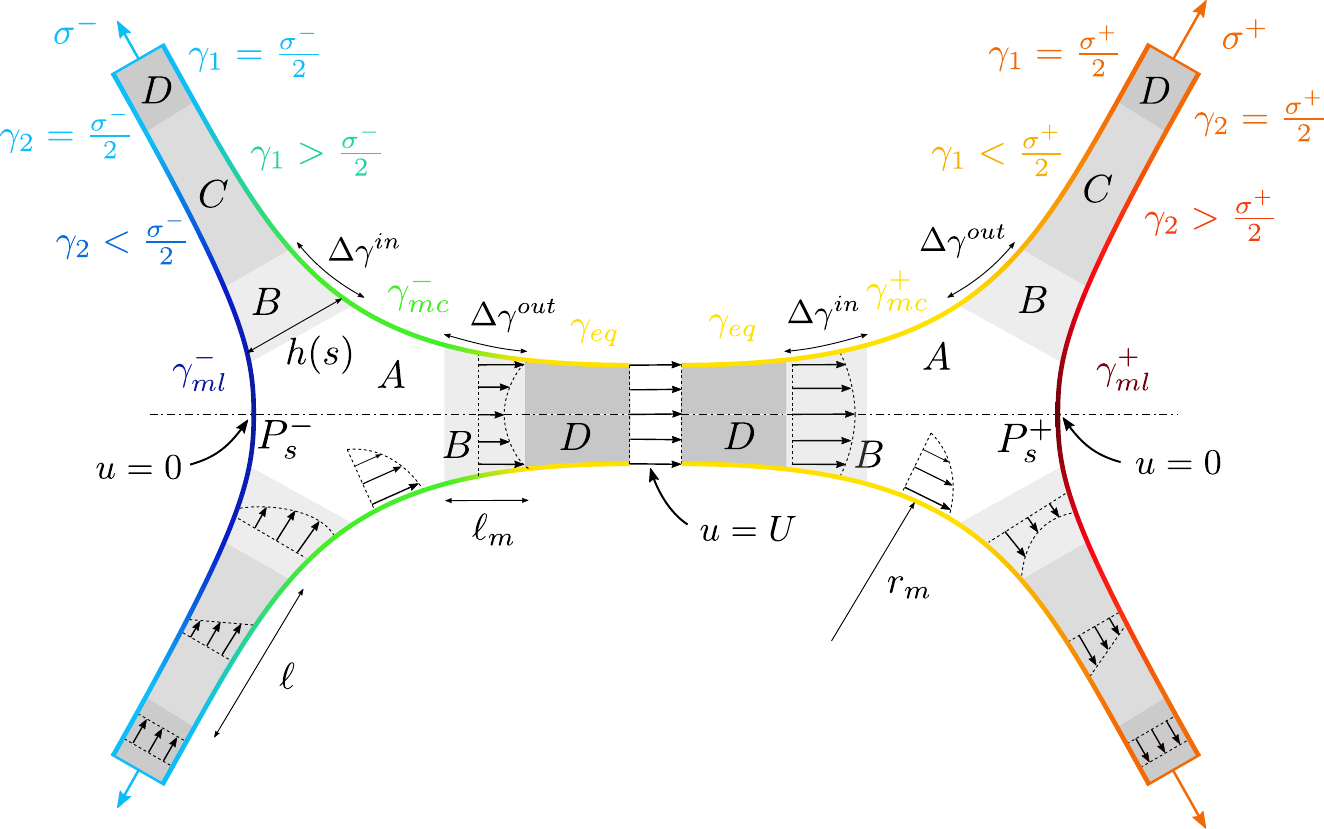}
\caption{Schematic of the velocity profiles and of the different domains. The interface color codes for  the surface tension: the lower value is $\gamma_{ml}^-$ in dark blue and the higher value is  $\gamma_{ml}^+$ in dark red.
\label{fig:men_frustration}}
\end{center}
\end{figure}

The surface tension variation along an interface is related to the bulk velocity $v(s,\zeta)$ beneath it trough the continuity of the tangential stress. We showed in section \ref{sec:def_tens} that air drag is negligible and the stress continuity thus  simplifies into the Marangoni relation
\begin{gather}
    \frac{\partial \gamma}{\partial s}=\pm \eta\frac{\partial v}{\partial \zeta}.
    \label{eq:marangoni}
\end{gather}
respectively for the interface at $\zeta>0$ and for the one at $\zeta<0$ (see the notation convention in Fig. \ref{fig:tangent}). The variation of the surface tension is thus coupled to the flow profile which depends on the liquid confinement. Fig. \ref{fig:men_frustration} represents the different domains and their corresponding flow profiles. Note that the schematic is not to scale for clarity and that $d_c \gg r_m\gg h$. 

The domain A is usually called the static meniscus, in which the curvature remains close to the equilibrium one. Elsewhere, the liquid is confined in thin films characterized by a thickness profile $h(s)$, in which  $\partial_s h \ll 1$ so that  the classical lubrication approximations apply. One important  consequence is that the pressure in the films does not depend on $\zeta$ and is only controlled by the Laplace pressure:
\begin{equation}
P(s)= \frac{\gamma_0}{2} \partial_{ss}h \, , 
\label{eq:lapl}
\end{equation} 
with  $\partial_{ss}h/2$ the curvature of each interface. The reference surface tension $\gamma_0$ is used in this expression as tension variations would lead to higher order corrections. The film tension can thus be expressed as :
\begin{equation}
    \sigma(s) = \gamma_1(s) + \gamma_2(s) - \frac{\gamma_0}{2} h(s) \partial_{ss}h
\label{eq:tens_fi}
\end{equation}
with $\gamma_1$ and $\gamma_2$ the tensions on both  film interfaces. 

The domain B, usually called the dynamical meniscus, of extension $\ell_m$, is defined as the part of the films in which the Laplace pressure is non-negligible. It connects the static meniscus at low pressure to the films  at  reference pressure. A Poiseuille flow results from the Laplace pressure gradient, which controls the volume exchanges between the films and the meniscus. 

The domain D  is the central part of the film, where the only degree of freedom  is a stretching / compression deformation. The  tensions verify $\gamma_1=\gamma_2=\sigma/2$ and the velocity field is a plug flow. The {\it film elements} used in the previous section can only be defined in this domain.   

The novelty of our approach is to define the domain C,  of length $\ell$, between the domains B and D, in which the Laplace pressure is negligible, but the tensions on both film sides are different. The tensions $ \gamma_1(s)$ and  $\gamma_2(s)$ are  equal by symmetry in the central film but may indeed differ in the peripheral films. 

These domains are called  the sheared film in the following.
They arise from a mismatch of surface velocity appearing on the peripheral films in the vicinity of the meniscus, due to a geometrical frustration. On the stretched side for example, the interface coming from the central film slips almost freely over the meniscus whereas, on the other interface, the  velocity must vanish on the symmetry plane (at the point $P_s^+$ in Fig. \ref{fig:men_frustration}). This results in the shearing of the thin film close to the meniscus  and in a tension difference between both interfaces. 
This domain C is far enough from the meniscus for the Laplace pressure to be negligible, but close enough from it so that the boundary condition difference on both interfaces is not screened. Such behavior has already been observed and quantified for a meniscus in contact with a solid wall \cite{cantat11, reichert19}, and was conjectured in \cite{petitphd} for a free meniscus.

Note that this geometrical frustration is not specific to our deformation, and is a generic feature for any meniscus connected to three (or any odd number) films : it is not possible to impose a uniform velocity on each of the three meniscus interfaces without getting a velocity difference between both interfaces in at least one film. 

The relative size of each domain depends on the physico-chemical properties of the solution. From our experimental observations, we assume that the sheared film length is much larger than the dynamical meniscus length, and much  smaller than the peripheral film size, thus leading to the condition $d^{\pm}  \gg \ell \gg \ell_m$ which  allows us to separate the regions $B$, $C$ and $D$. These conditions will be discussed and verified in section \ref{sec:const_exchange}. The approximations relevant for each domain are discussed below.

\subsection{Tension in the static menisci - Domain A}
\label{sec:dom_A}
On each of the three interfaces of the static meniscus, the surface tension variation is given by the flow profile in the bulk through eq. \eqref{eq:marangoni}. The length $\kappa$ over which the bulk velocity $v$ varies  in the normal direction  is {\it a priori} unknown, and can be much smaller than the meniscus size $r_m$, thus potentially leading to high velocity gradients. For $\kappa < r_m$, the boundary layer theory imposes :
\begin{gather}
    \kappa^{st}=\sqrt{\frac{\eta r_m}{\rho U}} \quad \mbox{or} \quad  \kappa^{tr}=\sqrt{\eta T/\rho} \; , 
\end{gather}
respectively for the steady and transient cases. Both lengths are of the order of 0.1 mm, which is comparable to the meniscus size. Consequently, the bulk flow is a recirculation  extending over the whole meniscus and  $r_m$ is  the relevant length scale for velocity gradient as well as for tension variation along the interface. The corresponding interfacial stress difference between the point in contact with the peripheral film and the point in contact with the central film  scales as $ \Delta \gamma \sim \eta U \sim 10^{-5}$ N/m which is much smaller than the tension difference observed in our experiments ($\Delta\sigma\sim 10^{-3}$ N/m).

We can thus conclude that the meniscus has a uniform tension on each of its 3 interfaces, $\gamma_{ml}^{\pm}$ on the lateral sides, and $\gamma_{mc}^{\pm}$ on the interfaces connected to the central film.

\subsection{Tension  in the dynamical menisci  - Domain B}
\label{sec:dom_B}
This part of the film has been extensively studied for films having the same velocity and the  same tension on both interfaces. For incompressible interfaces moving  at the velocity $U$ toward the thin film, the asymptotic thickness is given by the Frankel's law  \cite{mysels}:
\begin{equation}
h^{Fr}= 2.66 \,  r_m  \left(\frac{\eta U}{\gamma_0} \right)^{2/3}.
\label{eq:frankel}
\end{equation}
The associated surface tension difference between the film and the meniscus is
\begin{equation}
\Delta \gamma^{out}= 3.84 \gamma_0  \left(\frac{\eta U}{\gamma_0} \right)^{2/3}   
\label{eq:dgpull}
\end{equation}

Some corrections have been obtained for elastic interfaces \cite{seiwert14,champougny15}, and are negligible if 
\begin{equation}
\frac{E}{\gamma_0} \gg \left(\frac{\eta U}{\gamma_0} \right)^{2/3}  \, 
\label{eq:corr_elas}
\end{equation}
which is always the case in our experiments. 

When the film is pushed toward the meniscus at the velocity $U$, the situation is more complicated. A steady solution also exists, and leads to 
\begin{equation}
\Delta \gamma^{in} = \gamma_0 (3 Ca)^{2/3} \left(2.55 \alpha^{1/3}  -  2.68  \right) \, . 
\label{eq:gammam}
\end{equation}
with $\alpha = (h_\infty/r_m) (3 Ca)^{-2/3}$ and $Ca=\eta U/\gamma_0$ the capillary number \cite{mysels}. In this case, the steady solution not only depends on the velocity but also of the asymptotic thickness in the film $h_\infty$. This solution has been observed  in the Landau Levich geometry, showing a quantitative agreement between the theoretical and experimental film profiles \cite{denkov06, cantat13}. However, we recently evidenced that this solution is unstable for suspended film, and that the invariance in the $y$ direction is spontaneously broken. We show in \cite{gros20} that the tension difference $\Delta \gamma^{in}$ between the film and the meniscus is positive even when $U$ is oriented toward the meniscus, and that $\Delta \gamma^{in} \ll \Delta \gamma^{out}$ at a given capillary number. The tension jump associated with a film motion toward the meniscus will thus be neglected. 

Consequently the tension difference between the peripheral films and the central one arising from the dynamical meniscus  is given by eq \eqref{eq:dgpull}. This viscous response of the dynamical menisci is plotted in figure \ref{fig:tension_velocity_men}, and it clearly appears that this contribution is not large enough to explain our experimental results: a given transfer velocity $U$ requires a higher tension difference than the one predicted by Mysels's theory. 

 Note that, in our case, the interface velocities on each side will be shown to be different. We show in appendix \ref{app:frankel_u} that the prediction of  eq. \eqref{eq:dgpull} still holds if the velocity $U$  is replaced by the mean velocity $(U_1+U_2)/2$, $U_1$ and $U_2$ being the velocities on both interfaces in the dynamical meniscus. As this mean velocity is lower than the transfer velocity  measured in the central film, this reinforce the conclusion that the observed tension difference between adjacent films can not be explained by this contribution only. 
The tension variations in our foam assembly, and equivalently its  dissipation,  must originate from the domains C where thin films are sheared. The prediction of this flow and of the induced dissipation is the aim of the next section.  

%%%%%%%%%%%%%%%%%%%%%%%%%%%%%%%%%%%%%%%%%%%%%%%%%%%%

\section{Constitutive relation for the meniscus}
\label{sec:const_exchange}

It results from the previous analysis that the main dissipation should be localized in the {\it sheared  films}, in the peripheral films, close to the free menisci.
In this section, the surfactant and liquid transports are modeled in order to predict the relationship between the transfer velocity and the tension difference between adjacent films. This relationship, coupling a velocity and a force, rationalized the effective viscosity of the system and constitutes the constitutive relation for the meniscus. 

\subsection{Equations set}

The key fact at the origin of the dissipation is the dead-end role played by the lateral side of the free meniscus. In the top left film in Fig. \ref{fig:men_frustration} for example, 
the top interface can slide over the meniscus and be transferred to the central film, whereas the lateral one encounters the interface coming from the bottom left peripheral film. 
The meniscus can not instantaneously absorb the surfactant flux and the lateral interface must slow down when reaching the meniscus. This  breaks the symmetry between both interfaces and shears the film.

\begin{figure}[htp!]
\begin{center}
\includegraphics[width=0.95\linewidth]{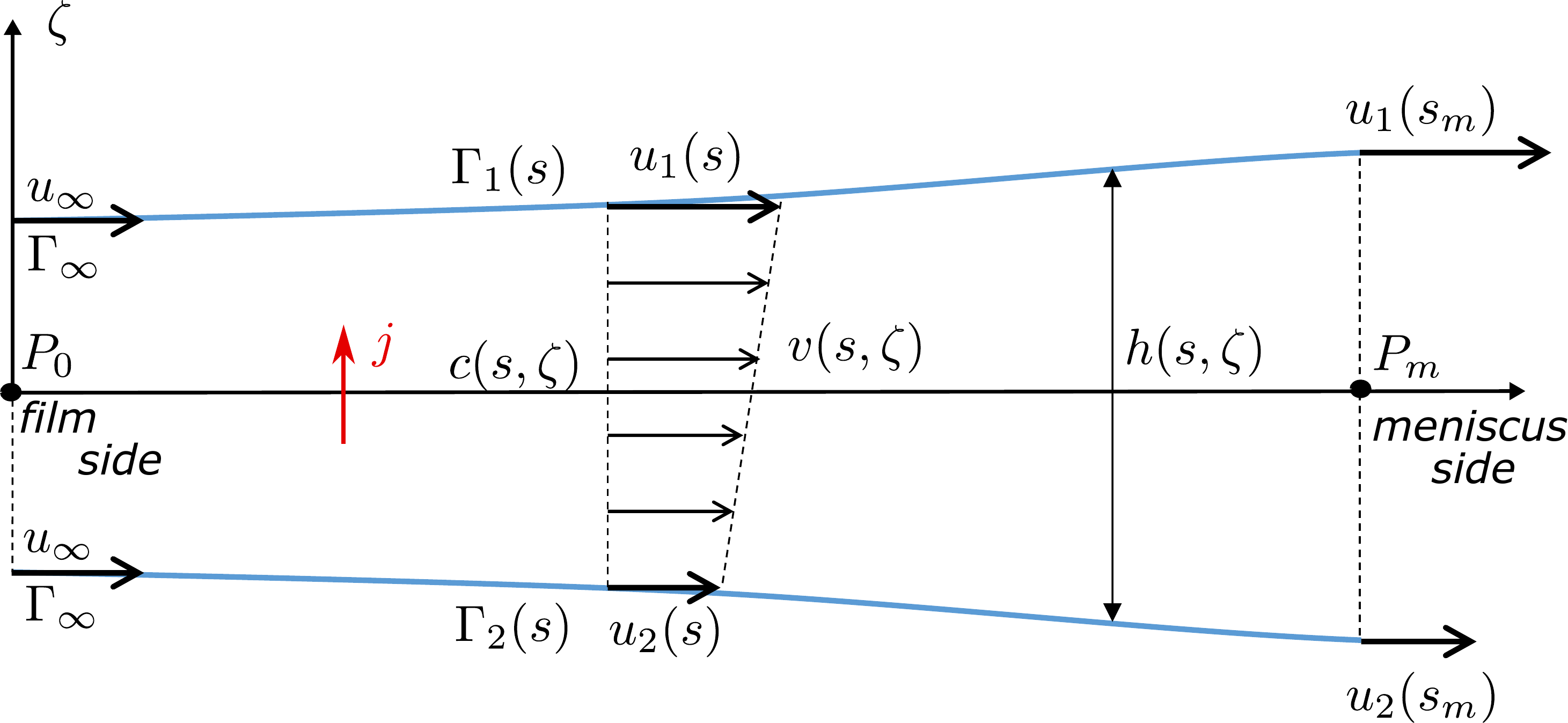}
\caption{Scheme of the sheared film (domain C in Fig. \ref{fig:men_frustration})  and notations used in the text. The peripheral film is on the left side of the figure ($s<0$) and the free meniscus on the right side ($s>s_m$). The  interface (2) is connected the interface of the symmetric peripheral film, whereas the  interface (1) is connected to the central film. The case $u_\infty>0$ represented here corresponds to the compression case.
\label{fig:Schema_film_dyn}}
\end{center}
\end{figure}

The model quantifies this mechanism by solving the coupled Stokes and surfactant transport equations in the appropriate limits. To this aim, we consider the piece of peripheral film shown in Fig. \ref{fig:Schema_film_dyn}. The abscissa $s$ and the film tangent $\vv{t}$ are oriented from the peripheral film to the free meniscus, and the indices 1 and 2 indicate respectively the interface connected to the central film and the interface connected to the other peripheral film. The normal to the film is $\vv{n}$, oriented from the interface 2 to the interface 1 and the corresponding variable is $\zeta$, with an origin in the middle of the film.  
The film thickness is $h(s)$, the bulk velocity is $v(s,\zeta) \,  \vv{t}$, the interface velocities are $u_1(s)$ and $u_2(s)$, and the surface coverages are $\Gamma_1(s)$ and $\Gamma_2(s)$. The notation $\partial_x$ indicates the partial derivative with respect to any variable $x$. 

For the sake of simplicity, and in an attempt to build a relationship between the transfer velocity and the film tension at a given time, independently of the film history, we assume that the liquid and surfactant transports are stationary. This requires that the transient regime is shorter than the experimental time scale.

We start from the lubrication theory and neglect the Laplace pressure. The velocity field is therefore governed by $ \partial_{\zeta \zeta} v=0$, and the velocity profile is equal to, with $u_m=(u_1+u_2)/2$ and $\Delta^f u= u_1-u_2$:
\begin{equation}
    v(s,\zeta)=\frac{\Delta^f u}{h} \zeta +u_m  \, .
    \label{eq:val_v}
\end{equation}
The flow rate at the position $s$ is :
\begin{gather*}
    Q=\int_{-h/2}^{h/2} v \,  d \zeta=u_m h,
\end{gather*}
and, from the mass conservation, we find:
\begin{gather}
    h_{\infty}u_\infty=\frac{h}{2}(u_1+u_2) \, , 
    \label{eq:cons_mass}
\end{gather}
with  $h_{\infty}$ and $u_\infty$ the thickness and velocity in the central part of the peripheral film, where both interfaces are identical. 

The shear flow in the film imposes a viscous stress at the interface, coupled to  a surface tension gradient by the Marangoni law:
\begin{equation}
    \eta\frac{\Delta^f u}{h}=\partial_s\gamma_1 \quad \mbox{and} \quad
    \partial_s\gamma_2=-\partial_s\gamma_1.
    \label{eq:val_dga}
\end{equation}
The surface tension is related to surface coverage $\Gamma$ using eq. \eqref{eq:elas_gibbs} leading to 
\begin{gather}
    -\frac{E}{\Gamma_0} \partial_s\Gamma_1=\eta\frac{\Delta^f u}{h}  \quad \mbox{and} \quad
    \partial_s\Gamma_2=-\partial_s\Gamma_1 \, .\label{eq:marangoni2}
\end{gather}

As already discussed in section \ref{sec:gibbsmar}, surfactant diffusion along the film is slower than the convection, and surfactant diffusion across the film is faster than the convection. The convection diffusion equation thus simplifies into $\partial_{\zeta \zeta}c = 0 $. We assume a fast adsorption process (no adsorption barrier) so that the equilibrium relation between the interface and the bulk, eq.  \eqref{eq:surf_trans}, can be used. The boundary conditions at the interfaces are thus $c_i=c_0+(\Gamma_i-\Gamma_0)/h_\Gamma$ (with i=1 or 2) and the bulk concentration is
\begin{gather}
    c(s,\zeta)=c_0+\frac{\Gamma_1-\Gamma_2}{h h_{\Gamma}} \zeta +\frac{\Gamma_1+\Gamma_2}{2 h_{\Gamma}}-\frac{\Gamma_0}{h_\Gamma}.
\end{gather}

Using this profile and neglecting surface diffusion, the surfactant conservation on each interface gives, with $j$ the diffusive flux coming from the bulk to interface 1:

\begin{align}
 &   \partial_s\left(\Gamma_1 u_1\right)=j= -D\frac{\Gamma_1-\Gamma_2}{hh_{\Gamma}}\label{eq:conv_surf_1}\\
&    \partial_s\left(\Gamma_1 u_1\right)=-\partial_s\left(\Gamma_2 u_2\right)\label{eq:conv_surf_2}.
\end{align}

The equations \eqref{eq:marangoni2} and \eqref{eq:conv_surf_2} imply that the two quantities $\Gamma_1+\Gamma_2$ and $\Gamma_1u_1+\Gamma_2u_2$ are conserved along the film so 
\begin{gather}
  \Gamma_1+\Gamma_2=2\, \Gamma_{\infty} \, ,  \label{eq:ga1ga2}\\
   \Gamma_1u_1+\Gamma_2u_2=2\,\Gamma_{\infty}u_\infty \, , \label{eq:uga1uga2}
\end{gather}
leading to 
\begin{gather}
    \Gamma_2=2\, \Gamma_{\infty}-\Gamma_1 \, , \label{eq:gamma2}\\
   u_2=\frac{2\,\Gamma_{\infty}u_{\infty}}{2\,\Gamma_{\infty}-\Gamma_1}-\frac{u_1\Gamma_1}{2\Gamma_{\infty}-\Gamma_1}\, . \label{eq:u2}
\end{gather}

The whole dynamics is finally controlled by a set of two coupled differential equations, deduced respectively from the Marangoni law and from the surfactant mass balance at interface 1:
\begin{gather}
    \partial_s \Gamma_1=-\frac{\eta\Gamma_0}{E}\frac{\Delta^f u}{h} \; , \label{eq:reduced_sys_1}\\
    \partial_s\left(\Gamma_1 u_1\right)=-2D\frac{\Gamma_1-\Gamma_{\infty}}{hh_{\Gamma}}\label{eq:reduced_sys_2},
\end{gather}
where
\begin{gather}
    \Delta^f u=u_1-u_2
    =\left(u_1-u_{\infty}\right)\frac{2\Gamma_{\infty}}{2\Gamma_{\infty}-\Gamma_1} \, , \label{eq:deltau}\\
    h=\frac{2h_{\infty}u_{\infty}}{u_1+u_2}
    =\frac{h_{\infty}u_{\infty}(2\Gamma_{\infty}-\Gamma_1)}{\Gamma_{\infty}u_{\infty}+u_1(\Gamma_{\infty}-\Gamma_1)}\; .  \label{eq:h}
\end{gather}

\subsection{Boundary conditions}
\label{sec:boundcond}

The model applies only in the {\it sheared  film} defined in section \ref{sec:dom_def} and depicted as the domain C in Fig. \ref{fig:men_frustration}. 
The problem is thus solved  between the point $P_0$, chosen as the abscissa origin  $s=0$,  at the frontier between this domain and the central part of the peripheral film (domain D), and the point $P_m$ at $s=s_m$ at its frontier with the dynamical meniscus (domain B). 

 By definition, the conditions at $s<0 $ are $u_1 = u_{\infty}$ and $\Gamma_1= \Gamma_{\infty}$  imposed in the central part of the peripheral film (see Fig. \ref{fig:Schema_film_dyn}). If the peripheral film is compressed $\Gamma_{\infty} >  \Gamma_0$   and    $u_{\infty}>0 $; the signs are opposite  if the film is stretched. 
 
 These boundary conditions are sufficient to solve the  system (\ref{eq:reduced_sys_1}-\ref{eq:reduced_sys_2}).
However, the aim of the resolution is to determine the relationship between the surface coverage $\Gamma_{\infty}$  (related to  the tension in the peripheral film) and the film velocity $u_\infty$ (related to its transfer velocity). In the following, $\Gamma_{\infty}$ will thus be considered as our control parameter, and $u_\infty$ as an unknown quantity. As expected, the  problem should thus be closed with additional conditions, at the meniscus. These conditions quantify the dead-end role of the meniscus for the interface 2 and thus provides the sought relationship between  $u_\infty$ and $\Gamma_{\infty}$.

For large values of $s$, the meniscus is reached and the assumption of vanishing Laplace pressure  fails. The boundary conditions must therefore be imposed at the frontier $P_m$, and not in the central film (for interface 1) nor in the symmetry plane $z=0$ (for interface 2), at the  point $P_s$  of the meniscus lateral interface, where the conditions are well defined. We thus need to make additional assumptions.

For interface 1, the tension in the central film is the equilibrium tension, and does not vary much along the static meniscus interface, nor along the dynamical meniscus interface, as shown in sections \ref{sec:dom_A} and  \ref{sec:dom_B}. We therefore impose the condition $\Gamma_1(s_m) = \Gamma_0$, which determines $s_m$. 

On interface 2,   the  velocity vanishes at the point $P_s$ by symmetry. This information must be used to build the  condition at the required point $P_m$. 
The surfactant mass balance made on the piece of interface between $P_m$ and $P_s$ imposes
\begin{equation}
 \Gamma_2(s_m) u_2(s_m) + j_m = 0 \, , 
 \label{eq:mass_bal}
\end{equation}
with $j_m$ the amount of surfactant adsorbed from the bulk along the meniscus interface, per unit time, between $P_m$ and $P_s$. This quantity is difficult to predict and its modeling would require a better control of the solution transport along the axis of the meniscus ({\it i. e.} in the $y$ direction).  In our model, we use the simplest phenomenological relationship: 
\begin{equation}
 j_m = - \frac{r_m}{\tau} \left( \Gamma_2(s_m) - \Gamma_0  \right)  \; , 
 \label{eq:flux_men}
\end{equation}
with $\tau$ the characteristic adsorption time of the surfactants, from the meniscus bulk at the reference concentration $c_0$ to the interface 2 at a concentration $\Gamma_2(s_m)$. For a purely diffusive case, this flux  would be $j_m= - r_m D \, (\Gamma_2(s_m) - \Gamma_{0})/(h_\Gamma\, \ell_{bl})$, with $\ell_{bl} \sim \sqrt{D t}$ the thickness of the mass boundary layer, of the order of 10 $\mu$m after one second.  This leads to  $\tau_{diff}\sim h_\Gamma \sqrt{t/D} \sim 1$ s and $U_m^{diff}=r_m/\tau^{diff} \sim 5 \cdot 10^{-4}$ m/s. However, convection and recirculations are important in the meniscus, and a faster transport can {\it a priori} be achieved.  
The comparison with the experimental results of Fig. \ref{fig:comp_exp_shear}  will evidence {\it a posteriori} that $U_m = r_m/\tau$  evolves during the dynamical process. It is  larger than $U^{diff}$ at short time and  becomes negligible afterwards.
 
The boundary condition at $s_m$  for the interface 2, which closes the model, is thus finally: 
\begin{equation}
   \Gamma_2(s_m) u_2(s_m) =  \frac{r_m}{\tau} \left( \Gamma_2(s_m) - \Gamma_{0} \right) \, . 
   \label{eq:cond_2}
\end{equation}

In order to perform a numerical resolution, the asymptotic conditions $u_1 = u_{\infty}$ and $\Gamma_1= \Gamma_{\infty}$ at $s \to - \infty$ need to be replaced by a condition at $s=0$. 
As shown below,  $\Gamma_1$ converges exponentially to  $\Gamma_{\infty}$ when $s \to -\infty$. We thus define the position origin $s=0$ as the point verifying 
\begin{equation}
 \Gamma_1 (0)  =\Gamma_{\infty} + \alpha ( \Gamma_0 - \Gamma_{\infty}) \, . 
  \label{eq:def_0}
\end{equation}
with $\alpha$ a small parameter. The corresponding value $u_1(0)$ is determined in section \ref{sec:lin} by linearization of the equation set. 
Numerically, the problem is solved  with  $\alpha =0.05$, without loss of generality. The length $s_m$, used in the numerical resolution, then depends on the arbitrary choice of $\alpha$. To correctly characterize the sheared film length we therefore introduced the characteristic length $\ell$ extracted {\it a posteriori} from an exponential fit of the numerical solution.

The system (\ref{eq:reduced_sys_1}-\ref{eq:reduced_sys_2}),  with the  boundary conditions both in the thin film and in the meniscus 
constitutes a closed problem, with $\Gamma_\infty$, $h_\infty$ and the physico-chemical constants as known parameters and $u_\infty$ and the length $s_m$ (or $\ell$) of the sheared film as solutions. 

\subsection{Scaling laws}
\label{sec:scallaw}
Before performing the whole numerical resolution, some scaling laws can be anticipated. In the following, we use  the notation $\delta$ for a difference  $X(s_m)- X(0)$ for any variable $X$, in order to estimate the spatial derivative of $X$. In contrast the notation $\Delta$ indicates a variation from the equilibrium value. 
Finally $\Delta^f$ indicates a difference between the interfaces 1 and 2 across the film, close to the meniscus.
We thus define $\delta \Gamma_1 =\Gamma_1(s_m) - \Gamma_\infty = \Gamma_0 - \Gamma_\infty$ and $\delta u_1 = u_1(s_m) - u_\infty$. Note that
Eq. \eqref{eq:gamma2} imposes that $\Gamma_1(s_m)  - \Gamma_2(s_m) = 2 \delta \Gamma_1$ so the same scaling and the same sign hold for both $\delta \Gamma_1$ and $\Delta^f \Gamma$, representing respectively  the concentration variation along the interface 1 and the concentration difference between both interfaces. Similarly, the eq. \eqref{eq:deltau} imposes that $\Delta^f u$ between both interfaces is of the same order as $\delta u_1$, as long as $\delta \Gamma_1 \ll \Gamma_0$. 

 With these definitions $u_\infty>0$, $\delta \Gamma_1 <0$, $\delta u_1 > 0$ for the pushing case, and  the opposite for the pulling case. The different scaling laws obtained below are built on three characteristic velocities: the capillary velocity $U_c= E/\eta \sim 3$ m/s, the diffusion velocity $U_d= D/h_\Gamma \sim 5.10^{-7}$m/s (as determined in section \ref{sec:comp_shear}), 
 and the  reservoir velocity $U_m= r_m/\tau$ associated to the meniscus. The film thickness always remains close to its asymptotic values and will be shown to be uniform in the linear regime. Scaling laws for  $\ell$ and for $U= - u_1(s_m)$ (given our conventions these two velocities are defined with an opposite sign),  are proposed below, in the three different regimes  that we identified. 
 
\subsubsection{Vanishing flux at the meniscus}
\label{sec:slowcomp}

\begin{figure}[htp!]
\begin{center}
\includegraphics[width=1\linewidth]{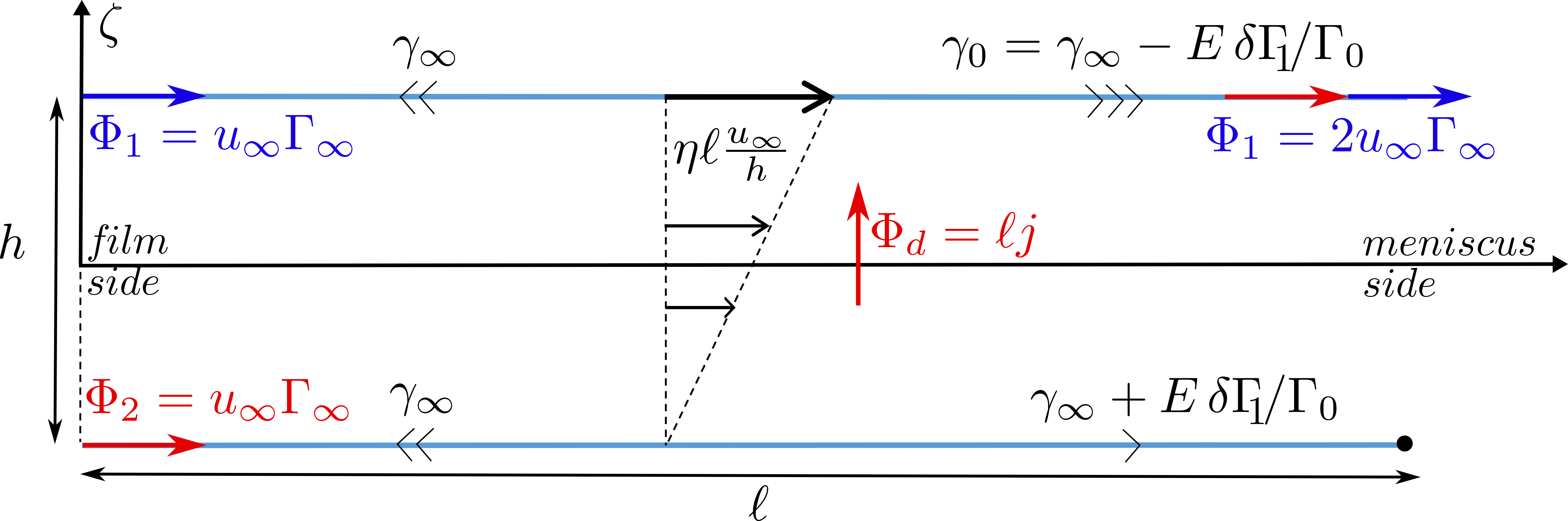}
\caption{Scheme of dynamics in the limit of vanishing velocity at the meniscus on interface 2. For the compression, the surfactant excess is lower in the central film than in the peripheral film, so $\delta \Gamma_1 <0$ and $u_\infty >0$. The signs are opposite for the stretching.  Surfactant fluxes initially coming from interface 1 and 2 are illustrated by the blue and red arrows respectively. Black arrows illustrate the tension and the viscous forces.   \label{fig:scaling1}}
\end{center}
\end{figure}

We first assume that the meniscus does not play any reservoir role, the flux $j_m$ being thus negligible in eq. \eqref{eq:mass_bal}. This limit, which can be reached either because of a vanishing velocity or because of a vanishing dodecanol surface excess at the meniscus on  interface 2, is explored first in the compression case, and then in the stretching case. 

In compression, the surface excess $\Gamma_2(s_m)$ is larger than its equilibrium value and can not vanish. It is however usually observed that the velocity may vanish in such cases, this effect being known as the stagnant cap limit \cite{cuenot97, cantat11, reichert19}.  In this limit, and in steady state, the whole flux $\Phi_2= u_\infty \Gamma_\infty$ advected on interface 2 must diffuse to interface 1, which imposes $\Gamma_\infty u_\infty \sim \ell j \sim  - D \ell \delta \Gamma_1 / (h_\infty \,  h_\Gamma)$ (as depicted in Fig. \ref{fig:scaling1}). The length $\ell$ of the sheared film can thus be seen as an exchange length, which must be large enough for the whole flux $\Phi_2$ to reach the interface 1 by diffusion, before reaching the stagnant cap at the meniscus. Then, on the interface 1, the flux at the meniscus must be twice the flux advected on the interface at the peripheral film side. As $\Gamma$ remains of the order of $\Gamma_0$, the velocity  $u_1(s_m)$ is of the order of $2  u_\infty$ (eq. \eqref{eq:uga1uga2}), which leads to $\delta u_1 \sim  u_\infty$. The viscous force between both interfaces is thus of the order of $\eta u_\infty \ell/h$. A second coupling between the unknown quantities $\ell$ and $u_\infty$ is given by the Marangoni law $-E \delta \Gamma_1/\Gamma_\infty \sim \eta u_\infty \ell/h_\infty$.

Combining both relationships, we get the following scalings laws
\begin{align} 
     & \frac{\delta \Gamma_1}{\Gamma_0}\sim -\frac{u_1(s_m)}{\sqrt{U_c \, U_d}}\quad  \mbox{so} \quad  \frac{\Delta\sigma}{2E} \sim \frac{U}{\sqrt{U_c \, U_d}} \label{eq:u1scalslow}\\
     & \ell \sim h_\infty \left(\frac{E h_\Gamma}{\eta D} \right)^{1/2} \sim h_\infty \sqrt{\frac{U_c}{U_d}}   \;.
     \label{eq:Lscalslow}
\end{align}

In this regime, the relevant velocity scale is  $\sqrt{U_c \, U_d} \sim 10^{-3}$ m/s. Using the experimental order of magnitude $\Delta \sigma/ E \sim 1$, the scaling eq. \eqref{eq:u1scalslow}  predicts a transfer velocity  of the order of $10^{-3}$ m/s, as expected.
The length $\ell$ of the sheared film is independent of the tension and its order of magnitude is $10^3 \,  h_\infty \sim 1$ mm. This validates the different assumptions made: $\ell \ll d$ ensuring that the film is not entirely sheared and $\ell \gg h_\infty$ ensuring that the lubrication approximation can be used. Moreover $\ell > \ell_m \sim 100\ \mu$m, the extension of the dynamical meniscus, so that Laplace is negligible in the sheared film. However, these two length scales may become similar for different solutions and the coupling between the sheared film and the dynamical meniscus should probably be considered in a more refined model. 

In the stretching case, we need to consider two situations: the flux at the meniscus may vanish because (i) the velocity vanishes or because  (ii) the concentration vanishes. In the limit of small tension in the film, the surface excess verifies  $\delta \Gamma_1 \ll \Gamma$, and only the first case needs to be considered. It leads to the same scaling as in the compression case: eqs. \eqref{eq:u1scalslow} and \eqref{eq:Lscalslow}, with  $\delta \Gamma_1 > 0$ and $ u_1(s_m)< 0$. 

However, at some critical tension, the surface excess vanishes on interface 2, at the meniscus, and the stretching dynamics strongly differs from the compression dynamics: the interface is not able to resist extension anymore and the velocity diverges. From eq. \eqref{eq:gamma2}  we deduce,  as the surface excess $\Gamma_2(s_m)$ must remain positive,  that this critical case occur for  $\Gamma_1 (s_m) = 2 \Gamma_\infty$, {\it i. e. } $\Gamma_\infty  = 0.5 \Gamma_0$. This corresponds to  $\Delta \sigma/(2E)=0.5$, which  constitutes an upper limit for this control parameter. The divergences associated to this particular regime are investigated in section \ref{sec:div}.

In summary, when the meniscus cannot supply nor adsorb any surfactant flux (small $U_m$) the dynamics induced by a meniscus can be described by a well defined  constitutive law: the tension difference $\Delta \sigma$ varies linearly with the transfer velocity ;  in the extension case, this tension saturates at   $\Delta \sigma=E$, and does not depend on the velocity at larger velocity values.

\subsubsection{Fast meniscus transport}
\label{sec:fast}

Here we consider the limit  $U_m \gg \sqrt{U_c U_d}$,
in which the meniscus plays the role of a reservoir for the surfactants and almost entirely absorbs (or provides in the stretching case) the flux $\Phi = \Gamma_\infty u_\infty$ advected on the interface 2. In that case, $u_1(s_m) \sim u_2(s_m)$ and  the  eq. \eqref{eq:cond_2} directly  provides the relationship between the velocity  at the meniscus and the surface excess:  

\begin{equation}
u_1(s_m) \sim  u_\infty  \sim -  \frac{\delta \Gamma_1}{\Gamma_0} U_m \quad  \mbox{so} \quad  \frac{\Delta\sigma}{2E} \sim \frac{U}{U_m} \; . 
\label{eq:u1scalfast}
\end{equation}

The scaling for $\ell$ can be deduced from the flux conservation eq. \eqref{eq:reduced_sys_2} leading to  $u_\infty \delta \Gamma_1 + \Gamma_0 \delta u_1 \sim - \ell  D \delta  \Gamma_1 /(h_\infty \, h_\Gamma)$. Coupled to the Marangoni law  $ \eta \ell \delta u_1 / h_\infty \sim  - E \delta \Gamma_1 /\Gamma_0$ it simplifies into
\begin{equation}
 u_\infty - U_c \frac{h_\infty}{\ell} \sim  - U_d \frac{\ell  }{h_\infty} \; .
\label{eq:bal_fast}
\end{equation}

Here two cases must be investigated, depending on the sign of $u_\infty$. 
In the pushing case,  $u_\infty>0$, and both terms of the left hand side of eq. \eqref{eq:bal_fast} contribute oppositely and can thus  balance each other. Physically, it comes from the fact that (i)  the interface flows from the peripheral  film at large surface excess to the central film at equilibrium coverage so $\delta \Gamma_1< 0$, (ii)  on the other hand,  $\delta u_1> 0$, leading to an increase of the area on interface 1, and thus a decrease of the surfactant excess $\Gamma_1$  along the shear film : a consistent solution can thus be reached without any transport between both interfaces, the two terms on the left of eq. (50)  ensuring the surfactant mass balance. 
The scaling obtained in the limit of large $U_m$, in compression, is thus  
\begin{equation}
\ell \sim  h_\infty \; \left( \frac{-\delta \Gamma_1}{\Gamma_\infty} \right)^{-1}  \; \frac{U_c}{U_m} \; .  
\label{eq:ellscalfastpush}
\end{equation}

The length $\ell$ diverges at small $\delta \Gamma_1$, which may seem surprising. However, in this limit  the velocity difference between both interfaces decreases, and the shear, even if spread over a large part of the film, is very small. In this limit, the interface can flow almost freely from one film to the other and the geometrical frustration induced by the meniscus becomes negligible. 

The dynamics is entirely different in the stretching case, as both terms of the left hand side of eq. \eqref{eq:bal_fast} have the same sign. Indeed, as $|u_\infty| \ll |u_1(s_m)|$, the interface 1 is surprisingly compressed during its transport through 
the sheared film. However, the positive tension associated to the film stretching imposes $\delta \Gamma_1 < 0$. This can only be achieved with a non negligible diffusion from the interface 2. In the limit $U_m \gg \sqrt{U_c U_d}$, the obtained scaling for the stretching case is 
\begin{equation}
\ell  \sim h_\infty \frac{\delta \Gamma_1}{\Gamma_\infty} \frac{U_m}{U_d} \; , 
\label{eq:ellscalfastpull}
   \end{equation}
whereas the scaling of eq. \eqref{eq:Lscalslow} is recovered in the other limit. 

It should be noted that $\ell$ becomes large for large $U_m$.  As for the small $\delta\Gamma_1$ limit in compression, the velocity difference between both interfaces decreases and the dissipation induced by the meniscus frustration becomes negligible.

In this second regime, governed by the meniscus, we obtain a linear relationship between the transfer velocity and the tension difference as in eq. \eqref{eq:u1scalslow}, but the tension is smaller for the same transfer velocity. The meniscus acts as a reservoir for the surfactants, and attenuates the film shear.  A strong asymmetry arises
for the exchange lengths: in the compression case, the surface excess gradients establish on each interface with a negligible diffusion from one interface to the other, and the sheared film length vanishes at large $U_m$. In contrast, these gradients requires a large exchange between interfaces in the stretching case, and $\ell$ increases with $U_m$. 

\subsubsection{Diverging behavior in the stretching case}
\label{sec:div}

In this last regime, observed at large $\delta \Gamma_1$, the tension
in the stretched film becomes independent of the velocity
and saturates at $\Delta \sigma = E$. The assumption $\delta \Gamma_1 \ll \Gamma_0$ used in the previous section is not valid anymore and a different scaling applies.

We  define $\Gamma_\infty = (\Gamma_0/2) (1 +\hat{\e})$ so that the concentration at the meniscus is $\Gamma_2(s_m)/\Gamma_0 = \hat{\e}$. At large $U_m$, the velocities in the stretched film are much higher than their gradients and $u_2(s_m) \sim u_\infty$. The condition eq. \eqref{eq:cond_2} thus becomes
\begin{equation}
u_\infty \sim - \frac{U_m}{\hat{\e}} \, .
\label{eq:divUdeG}
\end{equation}

The sheared film extension is obtained from eq. \eqref{eq:bal_fast}, in which the term proportional to $U_c$ can be neglected. This leads to 
\begin{equation}
\ell  \sim \frac{h_\infty}{\hat{\e}} \frac{U_m}{U_d} \; .
\label{eq:divLdeG}
\end{equation}

All the scalings obtained in this section are quantitatively verified  numerically in the following section.

\subsection{Numerical resolution}

\subsubsection{Non dimensionalization} 
In order to reduce the number of parameters to explore, we now build a dimensionless form of the problem using $u_1=|u_\infty|\bar{u}$, $\Gamma_1=\Gamma_{\infty}\bar{\Gamma}$, $h=h_{\infty}\bar{h}$ and $s=\frac{Eh_{\infty}}{\eta |u_\infty|}\bar{s}$. The scaling chosen for $s$ comes form equation (\ref{eq:reduced_sys_1}). In the dimensionless form, and combining equation (\ref{eq:reduced_sys_1}) and (\ref{eq:reduced_sys_2}), the new system writes :

\begin{gather}
    \partial_{\bar{s}}\bar{\Gamma}=-\frac{1}{\chi_{\infty}}\frac{\Delta^f \bar{u}}{\bar{h}} \, , \label{eq:sysadim1}\\
    \partial_{\bar{s}}\bar{u}=\frac{1}{\chi_{\infty}}\frac{\bar{u}\Delta^f \bar{u}}{\bar{\Gamma}\bar{h}}-2 A \frac{\bar{\Gamma}-1}{\bar{\Gamma}\bar{h}} \, ,\label{eq:sysadim2}
\end{gather}

with $\chi_{\infty}=\Gamma_{\infty}/\Gamma_0$, $A=DE/(\eta u_\infty^2h_{\Gamma})= U_c U_d / u_\infty^2 $ and :

\begin{gather}
    \Delta^f \bar{u}=\frac{2}{2-\bar{\Gamma}}  \left( \bar{u} -  \bar{u}_\infty   \right)  \, , \\
    \bar{h}=\bar{u}_\infty  \,   \frac{2-\bar{\Gamma}}{\bar{u}_\infty+  \bar{u}(1-\bar{\Gamma})}\;  .
\end{gather}

The asymptotic velocity in the film at small $s$ is $\bar{u}_\infty =1$ if the film is pushed toward the meniscus; in that case the condition $\chi_\infty >1$ must be fulfilled to ensure the existence of solutions.  If the film is pulled, we have $\bar{u}_\infty =-1$ and $\chi_\infty <1$. In both cases the asymptotic concentration in the film is $\bar{\Gamma} = 1$.

The conditions at the meniscus become 
\begin{gather}
    \bar{\Gamma}(\bar{s}_m) = \frac{1}{\chi_\infty}  \, ,\label{eq:condm1}  \\
     ( \bar{u} - \Delta^f\bar{u})(\bar{s}_m) = K \sqrt{A}\;  \frac{\chi_\infty -1}{2\chi_\infty -1} \, ,\label{eq:condm2}
\end{gather}
with 

\begin{equation}
  K= \frac{2 r_m}{\tau}\sqrt{\frac{\eta h_\Gamma}{DE}} = \frac{ 2 U_m}{\sqrt{U_c U_d}}   \, . \label{def_K}
\end{equation}

\subsubsection{Linearization and boundary conditions at $\bar{s}=0$}
\label{sec:lin}

To solve the system \eqref{eq:sysadim1}-\eqref{eq:sysadim2}, one needs to impose compatible boundary conditions at $\bar{s}=0$ which we obtain by linearizing the equations.
We introduce $\bar{u}=\bar{u}_\infty+\e_u$ and $\bar{\Gamma}=1+\e_\Gamma$ with $\e_u,\e_\Gamma\ll1$. At first order in these small parameters, we get $\Delta^f \bar{u}= 2 \e_u$, $\bar{h}=1$, $\partial_{\bar{s}} \e_\Gamma=- \frac{2}{\chi_{\infty}}\e_u$ and  $\partial_{\bar{s}}\e_u=\frac{2 \bar{u}_\infty}{\chi_{\infty}}\e_u - 2A\e_\Gamma$, having the solutions  $\e_\Gamma=a_\Gamma \text{e}^{ks}$ and $\e_u=a_u \text{e}^{ks}$. Using the convention chosen in eq. \eqref{eq:def_0} to define the origin of $s$, this imposes $a_\Gamma =\alpha (1 - \chi_\infty)/\chi_\infty$. Injecting these solutions in the linearized equations gives:
\begin{align}
&a_\Gamma \, k + a_u  \frac{2}{\chi_{\infty}}  = 0 \, , \\
& 2 A a_\Gamma +  a_u \left (k -\frac{2 \bar{u}_\infty}{\chi_{\infty}} \right)  = 0 \, ,
\end{align}
which characteristic equation is:
\begin{gather*}
    k^2 - 2\frac{\bar{u}_\infty}{\chi_{\infty}}k-\frac{4A}{\chi_{\infty}}=0 \, .
\end{gather*}
Since $A>0$,  $\chi_{\infty}>0$ and $\bar{u}_\infty = 1 $  for the pushing case and $\bar{u}_\infty = -1 $ for the pulling case, the only positive solution, compatible with the asymptotic behavior at  $s \to - \infty$, is in both cases

\begin{gather*}
    k=\frac{1}{\chi_{\infty}}(\sqrt{1+4A\chi_{\infty}}+ \bar{u}_\infty)\, , 
\end{gather*}
leading to  the initial conditions 
\begin{align}
&\bar{\Gamma}(0)=1+ \alpha \frac{1 - \chi_\infty}{\chi_\infty}\, , \label{eq:CInum1}\\
&\bar{u}(0)=\bar{u}_\infty - \alpha \frac{1 - \chi_\infty}{2 \chi_\infty} \left (\sqrt{1+4A\chi_{\infty}}+ \bar{u}_\infty \right)  \, .
\label{eq:CInum2}
\end{align}

\subsubsection{Resolution and relevant numerical quantities}

\begin{figure}
\begin{center}
\includegraphics[width=1\linewidth]{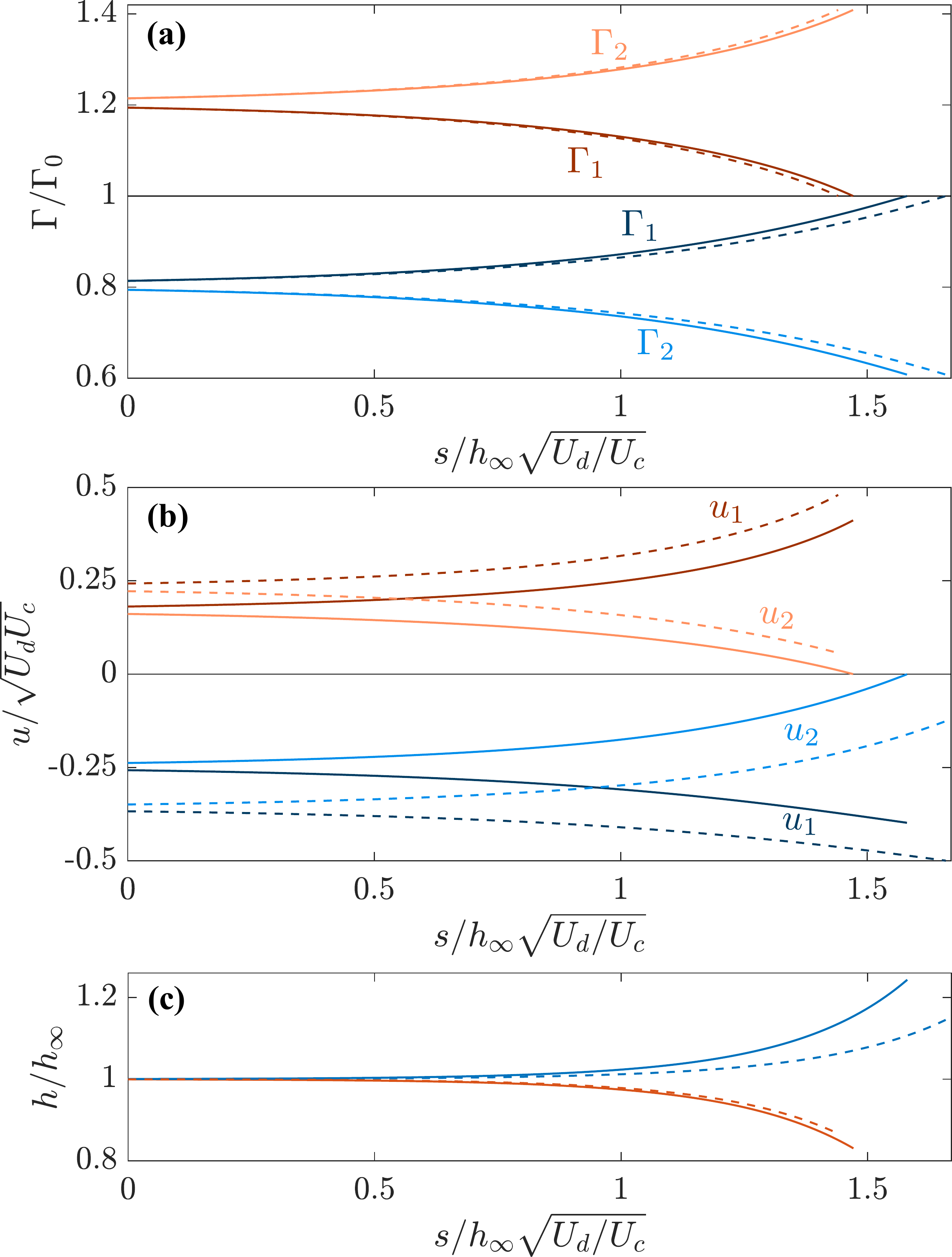} 
\caption{Evolution of the dimensionless surface concentrations (a), surface velocities (b), and film thicknesses (c)  in term of the dimensionless coordinate ($s/h_{\infty}\sqrt{U_d/U_c}$) for $|\delta\Gamma|/\Gamma_0=0.2$ and $K=10^{-3}$ (solid lines) and $K=0.39$ (dashed lines). The extension case is represented in blue and the compression in red. The curves stops at different $s=s_m$ since $s_m$ depends on $K$ and $\delta\Gamma$.
\label{fig:ex_res}}
\end{center}
\end{figure}

\begin{figure*}
\begin{center}
\includegraphics[width=1\linewidth]{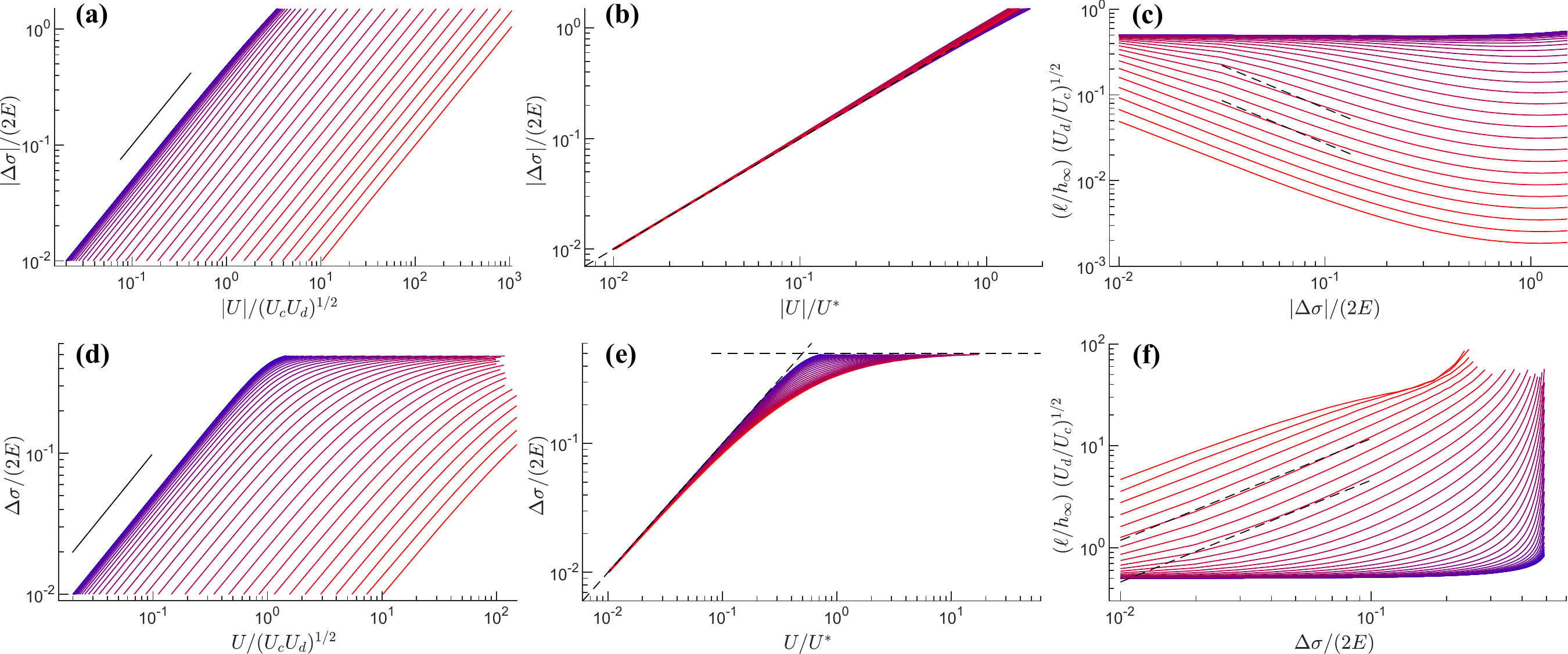} 
\caption{Results of the numerical resolution for the compression (top) and the extension (bottom). The control parameter $K=  2 U_m/\sqrt{U_c U_d}$ has been varied logaritmically from $10^{-3}$(blue curves) to $10^3$ (red curves). (a)-(d) Film tension difference $\Delta \sigma$ between the peripheral film and the central film, rescaled by the film elasticity $2 E$, as a function of the transfer velocity $U$ in the central film, rescaled by $\sqrt{U_c U_d}$. The slope of the  black lines correspond to a linear law. For the compression, $\Delta\sigma$ and $U$ are negative and we plotted their absolute values. (b)-(e) Same data, with the velocity rescaled by $U^*= 2(\sqrt{U_c U_d}+U_m)$.  The dashed line corresponds to $\Delta \sigma/(2 E)= U/U^*$ and the horizontal dashed line in (e) is  $\Delta \sigma/(2 E)= 0.5$. (c)-(f) Length  $\ell$ of the sheared  film, rescaled by $h_\infty \sqrt{U_c/U_d}$, as a function of the tension rescaled by $2 E$. The black dashed lines in (c) corresponds to  $\ell/(h_\infty \sqrt{U_c/U_d})= 0.57 /( K  \Delta \sigma/(2 E)) $, for two  values of $K$, corresponding to the closest  curve ($K = 81$ and $ 208$)(see the scaling of eq. \eqref{eq:ellscalfastpush}). Similarly, in (f) they correspond to $\ell/(h_\infty \sqrt{U_c/U_d})= 0.57 K  \Delta \sigma/(2 E)$, for the same $K$ values.
\label{fig:num_res}}
\end{center}
\end{figure*}

The non-linear coupled equations \eqref{eq:sysadim1}-\eqref{eq:sysadim2} are first solved with the Matlab solver ode45 with the initial conditions eqs. \eqref{eq:CInum1}, \eqref{eq:CInum2}, for a given value of $\chi_\infty$, $K$ and $A$. The upper $s$ value $s_m$ is determined with  \eqref{eq:condm1}. This resolution is performed with different values of the parameter $A$ until the condition \eqref{eq:condm2} is verified too, for the specific value $A^*( K,\chi_{\infty})$. The obtained parameter $A^*(K,\chi_\infty)$ eventually provides the film velocity as a function of its asymptotic film tension, by simply using the definition of $A$ in eq. \eqref{eq:sysadim2}:
\begin{equation}
   |u_\infty|
   =   \frac{\sqrt{U_c U_d}}{\sqrt{A^*(K,\Gamma_\infty/\Gamma_{eq})}} \, .
   \label{UdeAst}
\end{equation}
Note that the sign of $u_\infty$ must be prescribed {\it a priori}, as the pushing and pulling cases have a different initial  condition $\bar {u}_1(0) = \bar{u}_\infty= \pm 1$. Figure \ref{fig:ex_res} shows the typical spatial evolution in the sheared domain of the surface coverages, velocities of both interfaces and of  the film thickness for both the extension and compression cases.

From $A^*(K,\chi_\infty)$ we can now predict the quantities experimentally measured in Fig. \ref{fig:tension_velocity_men}. 
The film tension difference $\Delta \sigma$ between the peripheral film and the central film can be expressed as a function of the Gibbs elasticity and the numerical parameter $\chi_\infty$ with the relation  
\begin{equation}
    \frac{\Delta \sigma}{2 E}=\frac{\delta \Gamma_1}{\Gamma_0}= 1 - \chi_\infty \, . 
\end{equation}
 
The velocity in the central film $U$, defined as negative for the compression and positive for the extension, is identified with $-u_1(s_m)$, the velocity being assumed to be constant along the meniscus side (on interface 1). This velocity is thus expressed as
\begin{equation}
 \frac{U}{\sqrt{U_c U_d}} = -\frac{  |u_\infty|}{\sqrt{U_c U_d}}\; \bar{u}_1(\bar{s}_m) =- \frac{\bar{u}_1(\bar{s}_m)}{\sqrt{A^*}} \, . 
\end{equation}

The other important dynamical quantity is the length $\ell$ of the sheared film. Its non-dimensional value  $\bar{\ell}$ is obtained by fitting the evolution of $\bar{\Gamma}(\bar{s})$ with an exponential function. Then we get 
\begin{equation}
    \frac{\ell}{h_\infty}\sqrt{\frac{U_d}{U_c}}=\frac{\bar{\ell} \, h_\infty U_c}{h_\infty |u_\infty|}\sqrt{\frac{U_d}{U_c}}=\bar{\ell}\sqrt{A^*} \; .
\end{equation}

\subsubsection{Numerical results}
\label{sec:numres}
The numerical results for compression and extension are shown in figure \ref{fig:num_res}. The relationship between the tension difference and the transfer velocity is shown in (a) for the compression. For all the values of the parameter $K= 2 U_m/\sqrt{U_c U_d}$ the tension is proportional to the velocity. As expected from the scaling laws analysis, the prefactor is constant at small $K$ (eq. \eqref{eq:u1scalslow}) and decreases with $K$ at large $K$ (eq. \eqref{eq:u1scalfast}). As shown in Fig. \ref{fig:num_res} (b), all numerical data fall on a single master curve 
\begin{equation}
\frac{\Delta \sigma}{2 E } = \frac{U}{U^*}
\label{eq:master}
\end{equation}
if the velocity is rescaled by 
\begin{equation}
U^* = \sqrt{U_c U_d} (2+ K)= 2 (\sqrt{U_c U_d} + U_m)
\label{eq:ustar}
\end{equation}
which nicely interpolates between the asymptotic behaviors at small and large $K$. This is the sought constitutive relation for the meniscus, based on the  microscopic physico-chemical properties of the system.

The same behavior is recovered for the extension case (Fig. \ref{fig:num_res} (d)-(e)) at small tension. The master curve of eq. \eqref{eq:master} is still verified, which is consistent with a linear relationship between $U$ and $\Delta \sigma$, expected in the limit of vanishing $\Delta \sigma$. However at larger tension, the saturation at  $\Delta \sigma /(2 E) = 0.5$ breaks the symmetry between stretching and compression. It corresponds to a vanishing interface concentration at the meniscus, on the lateral side (interface 2). The tension reaches a plateau and the velocity diverges, as discussed in section \ref{sec:div}  and in Appendix \ref{app:divergence}.

The sheared film length $\ell$ is plotted in Fig. \ref{fig:num_res} (c) as a function of the film tension difference for the compression case. It is independent of the tension at small $K$ value and it decreases with $\Delta \sigma$ at large $K$,  as expected from eqs. \eqref{eq:Lscalslow} and \eqref{eq:ellscalfastpush}. Its upper bound is $ h_\infty \sqrt{U_c/U_d} \sim 10^{-3}$ m, much smaller than the film size, as {\it a priori} assumed by the model.  At large $U_m$, $\ell$  can becomes of the order of the dynamical meniscus length $\ell_m$. In that case, some corrections related to the Laplace pressure are expected in the sheared film domain. Finally,  for important tension variations the length reaches another constant values which depends on $U_m$ and is not captured by our scaling analysis.

For the extension case, figure \ref{fig:num_res} (f), the sheared length at small $U_m$ is constant and is the same as in compression. An important difference with the compression appears at large $U_m$ where $\ell$ increases with $\Delta \sigma$ as predicted by eq. \eqref{eq:ellscalfastpull}. Moreover, the sheared length diverges close the tension saturation $\Delta\sigma/(2E)=0.5$ with a scaling law predicted in eq. \eqref{eq:divLdeG} and verified in Appendix \ref{app:divergence}. When $\ell$ becomes important, the sheared film  invades the entire peripheral film and our model breaks down as the domain A represented in Fig. \ref{fig:men_frustration} entirely disappears.

On the basis of these numerical results, we can now refine our description of the transfer velocity. Its value $U$ is defined here as the velocity of the central part of the central film.
The velocity $U_l$ of the central part of the peripheral films has been assumed to be close to $U$  in section \ref{sec:def_kin}, on the basis of the observations reported in \cite{bussonniere20}. The corresponding numerical quantities are  $U= - u_1(s_m)$ and   $U_l= - u_\infty$, which actually differ from each other. Indeed, using   eqs. \eqref{eq:gamma2} and  \eqref{eq:cond_2}  at $s=s_m$ we get 
\begin{equation}
u_2(s_m)= 2 U_m \frac{\Gamma_\infty - \Gamma_0}{2 \Gamma_\infty - \Gamma_0}  \, , 
\end{equation}
so from  \eqref{eq:u2} and \eqref{eq:master} we deduce  
\begin{equation}
  U = \frac{-2 u_\infty}{1+ 2\frac{U_m}{U^*} - 2 \frac{u_\infty}{U^*} } =U_l \; \frac{ 2 \sqrt{U_c U_d} + 2 U_m}{U_l  + \sqrt{U_c U_d} + 2 U_m }\, . 
   \label{eq:scalingu1}
\end{equation}

At small tension ($U$ and $U_l$ much smaller than $U^*$), we thus find   $U \sim 2 U_l$ if   $U_m \ll \sqrt{U_c\,  U_d}$ and  $U \sim  U_l$ in the opposite limit. Note that the  Fig. 4 of \cite{bussonniere20} shows that both velocities $U$ and $U_l$ are of comparable values but does not allow us to make a  quantitative comparison.  

%%%%%%%%%%%%%%%%%%%%%%%%%%%%%%%%%%%%%%%%%%%%%%%%

\section{Comparison with experimental data}
\label{sec:comp_shear}

In this section, all the experiments performed at different deformation amplitudes ($\Delta d$), motor velocity ($V$) and mean deformation ($d_m$) are compared to the model developed in the previous section.

\subsection{Compared quantities}

The model predicts the film tension variation $\Delta \sigma$ associated to the sheared film only.  However, the tension jump in the dynamical meniscus, associated to the film extraction and  shown in Fig.   \ref{fig:tension_velocity_men}, is not entirely negligible and will be added  as a correction to the contribution associated to the shear. This additional tension jump $\Delta \gamma^{out}_-$ is given by eq. \eqref{eq:dgpull}.

Following the scheme of the surface tension distribution along the interfaces, we get the full theoretical prediction for the tension in the  compressed film as
\begin{equation}
    \frac{\Delta\sigma_{th}^-}{2E}=\frac{\Delta\sigma^-}{2E}-3.84\frac{\gamma_0}{E}\left(\frac{\eta |U|}{\gamma_0}\right)^{2/3} \, .
\label{eq:cor1}
\end{equation}
with $ \Delta\sigma^-$  the film tension associated to the sheared film  and predicted as a function of $U= -u_1 (s_m)$ by the model of section \ref{sec:const_exchange} (see Fig. \ref{fig:num_res}). 
It should be noted that the  tension of the meniscus interface connected to the central film is $\gamma_{mc}^-=\gamma_0-\Delta\gamma_-^{out}$ as shown in figure \ref{fig:men_frustration}, and thus differs from the boundary condition $\gamma_0$ imposed in the shear model. However, this correction would only provide a second order correction while greatly complicating the numerical resolution. 

The extension case is similar, but the tension jumps associated to the Frankel's films extractions are located in the peripheral films. The extraction velocity is thus not identical on both interfaces: it is $U= - u_1(s_m)$ on the top interface of the stretched film, and  $U_2^+ = -u_2(s_m)$, on the external interface. In that case, we show in appendix \ref{app:frankel_u} that eq. \eqref{eq:dgpull} remains valid if the averaged velocity is used. 

The total film tension difference is finally given by :
\begin{equation}
    \frac{\Delta\sigma_{th}^+}{2E}=\frac{\Delta\sigma^+}{2E}+3.84\frac{\gamma_0}{E}\left(\frac{\eta (U+U_2^+)}{2\gamma_0}\right)^{2/3} \, ,
\label{eq:cor2}
\end{equation}
with $ \Delta\sigma^+$  the film tension associated to the sheared film. 

Note that in both cases the absolute value of $\Delta\sigma$ is increased by the additional term.

\subsection{Time evolution and fitting procedure}

\begin{figure*}
\begin{center}
\includegraphics[width=1\linewidth]{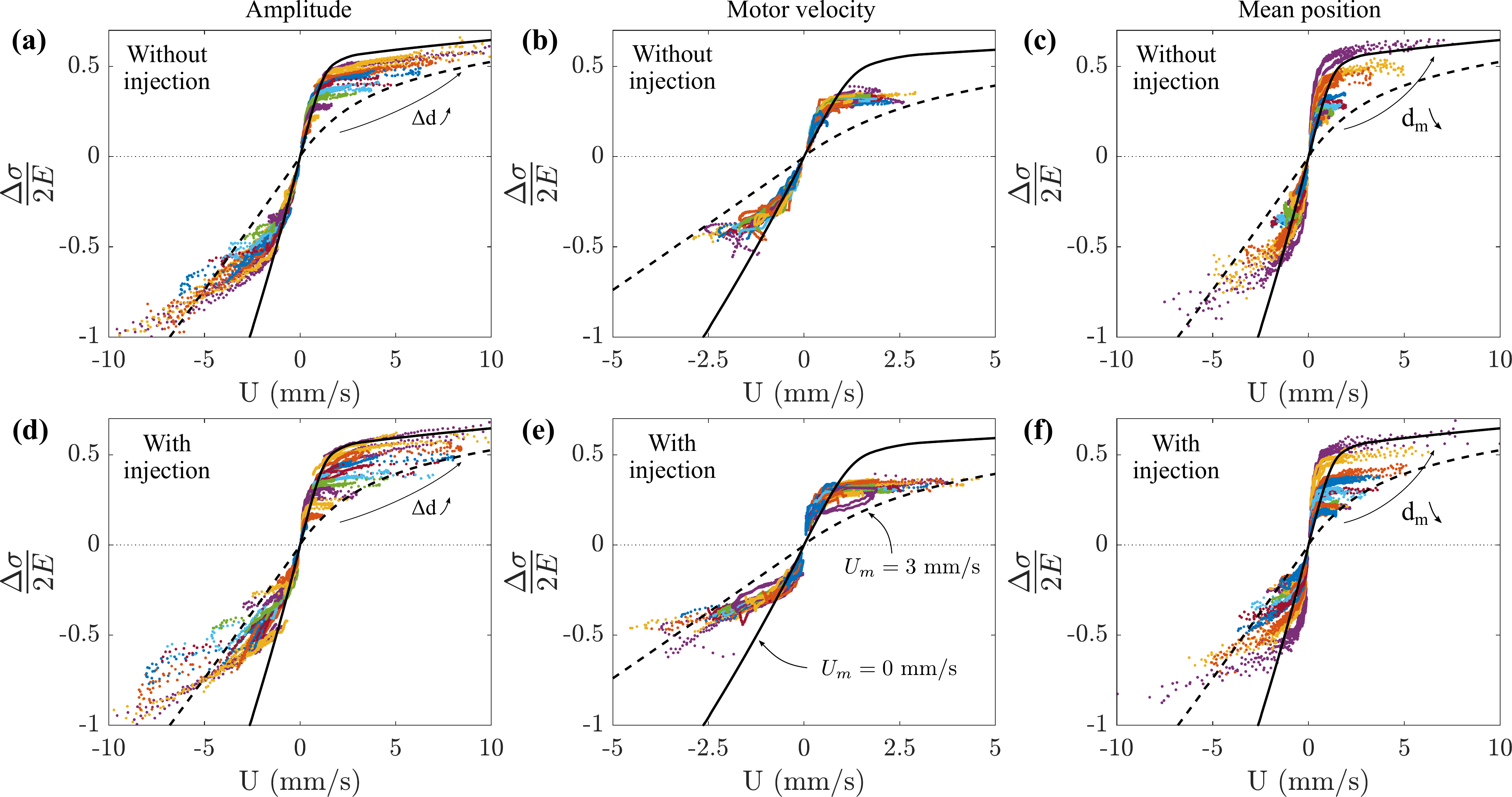} 
\caption{Comparison between the experimental and the theoretical viscous responses of an elementary foam. The solid lines and dashed lines are the predictions of eqs. \eqref{eq:cor1} and \eqref{eq:cor2}, based on the numerical results shown in Fig.  \ref{fig:num_res},  with  $U_c = 3$ m/s,  $U_d=5.10^{-7}$ m/s and, respectively  $U_m\rightarrow0$ mm/s and  $U_m=3$ mm/s. Experimental points correspond to the relaxation phase only (after the motor stop) and the large velocities thus correspond to the small times.  Three control parameters were independently varied: the amplitude $\Delta d$ in (a) and (d) with $d_m=12$ mm and $V=50$ mm/s ; the motor velocity $V$ in (b) and (e) with $\Delta d=6$ mm and $d_m=12$ mm ; and the mean position $d_m$ in (c) and (f) with $\Delta d=5$ mm and $V=50$ mm/s. Each color corresponds to one set of parameters. Experiments represented on the top (a-c) were performed without injection while liquid was supplied to the foam for the bottom experiments (d-f).
\label{fig:comp_exp_shear}}
\end{center}
\end{figure*}

The data associated to the experiments of the first campaign are plotted in figure \ref{fig:comp_exp_shear}. The time resolution of this  campaign is not good enough to compute the transfer velocity $U$ during the motor motion  and only the relaxation phase ({\it i. e.} after the motors stop) is shown.  A good  reproducibility is observed for each set of parameters (each color) but the relationship between $\Delta \sigma$ and $U$ differs from one parameter value to the other, especially  when $\Delta d$ and $d_m$ vary,  at early time ({\it i. e.} for important velocities).
Moreover, alimented experiments (with injection) consistently exhibit higher transfer velocities than non-alimented experiments at early time which suggest that the velocity $U_m$ plays an important role in this regime. At later time, \textit{i.e.} for smaller velocity, all the experiments remarkably converge toward a single master curve. 

The model is built on the three parameters $U_c= E/\eta$, $U_d= D/h_\Gamma$ and $U_m= r_m/\tau$ (see section \ref{sec:scallaw}). The capillary velocity $U_c = 3$ m/s has been precisely determined from the experimental results of section \ref{sec:const_film}. 
The diffusion velocity $U_d$ is also a well defined, constant, quantity. However, $h_\Gamma$ can not be deduced from our measures in section \ref{sec:const_film} and different theoretical definitions may be relevant, leading to different possible values,  as discussed in section \ref{sec:phychi}. It is thus kept as an adjustable parameter, assumed to be the same for all the data sets. Finally, the velocity $U_m$ has been introduced as a phenomenological parameter,  to quantify the ability of the meniscus to provide or absorb surfactants. This quantity can  vary with time, the meniscus being {\it a priori} a more efficient reservoir at the beginning of the deformation.  
As shown in Fig. \ref{fig:num_res}(a) and (d) all curves collapse on a single master curve at small $U_m$, this master curve being an upper limit for $\Delta \sigma$. 
 We thus interpret the superposition of the different curves in Fig. \ref{fig:comp_exp_shear}  at long time as the limit $U_m=0$. 

On the basis of this assumption,  we thus  fit the late relaxation phase, when menisci are potentially depleted/saturated, using $U_m=0$ mm/s and $U_d$ as a fitting parameter. We use the numerical predictions shown in Fig. \ref{fig:num_res}, with the corrections given by eq. \eqref{eq:cor1} and \eqref{eq:cor2}.  The best fit is obtained for $U_d = 5 \cdot 10^{-7}$m/s and the resulting  law is the solid line shown in the six graphs of Fig. \ref{fig:comp_exp_shear}. As expected from the model, all data points are close or  below this limiting case.   The dashed line corresponds to the  prediction obtained with   $U_m= 3$ mm/s while keeping $U_d$ and $U_c$ fixed.
This law is a lower boundary for all data points, thus indicating that  $U_m= 3$ mm/s is the maximal reservoir velocity reached by the system, when menisci can supply/adsorb a large quantity of surfactant.  A more detailed discussion of the agreement between theory and observations is made below. 

\subsection{Small velocity - long time regime}

First, experiments are compared to the model when velocity (and tension difference) is small ($|U|<1$ mm/s) which corresponds to the late time relaxation. In this regime, the results shown in figure \ref{fig:comp_exp_shear} are well captured by our model for all the motor parameters with $U_d=5.10^{-7}$ m/s and $U_m\rightarrow0$ m/s excepted the experiments with the shortest film lengths (small $d_m$ in figure \ref{fig:comp_exp_shear} (c)-(f)). For these extreme deformations we suspect that the shear length becomes of the same order as the film length ($d\sim 5$ mm) thus breaking down the assumptions of the model. For all other experiments, the velocity transfer is well captured by the simple law (from eqs. \eqref{eq:master} and \eqref{eq:ustar}, with $U_m=0$)
\begin{equation}
U= 2 \,  \sqrt{\frac{E D }{\eta h_\Gamma }}\, \frac{\Delta \sigma}{2 E} \, .
\label{eq:rellin}
\end{equation}
 Note that the corrections associated to the Frankel's film extractions (eqs. \eqref{eq:cor1} and \eqref{eq:cor2}) are negligible in this regime, and are thus omitted in this equation. 

The fitted parameter $U_d=D/h_\Gamma=5 \cdot 10^{-7}$ m/s can be discussed on the basis of the transport properties of the dodecanol given in section \ref{sec:phychi}. 
A part of the dodecanol is solubilized in SDS micelles and the other part is  in the monomeric form which leads two possible diffusion velocities: (i)  if the transport is dominated  by the  micelles of 1.8 nm \cite{duplatre96}, the Stokes-Einstein formula imposes $D_M\approx8\ 10^{-11}$ m$^2$/s and, using $h_\Gamma\approx 5.4$ $\mu$m, we get $U_d^M=D_M/h_\Gamma\approx1.5\ 10^{-5}$ m/s ; (ii) if only the monomers participate to the dynamics, $D_m=5\ 10^{-10}$ m$^2$/s, $h_\Gamma^m \approx 370$ $\mu$m and $U_d^m=D_m/h_\Gamma^m\approx 10^{-6}$ m/s, closer to the fitted value. This suggests that the Marangoni stress induced by the shear flow is controlled by the diffusion of dodecanol monomers only.

This result might seem surprising as the majority of DOH is solubilized in SDS micelles. However, micelle-assisted transport for important concentration variation (important film deformation) is limited by the micelle formation/disintegration step \cite{patist02,colegate05} which has a characteristic timescale of $\sim 200$ ms for SDS/DOH mixture \cite{patist98}. This time is much longer than the monomer diffusion timescale across the film: $\tau_d^m\sim h^2/D_m\sim 2$ ms, which may explain that only monomers are involved in the diffusion process across the films.

\subsection{High velocity - short time regime}

At early time, experiments systematically deviate from the prediction associated to $U_m\rightarrow0$ mm/s, and the tension observed are smaller than this prediction. These deviations are reproducible and depend on the imposed deformation. 
This  can be qualitatively rationalized if we consider that the  meniscus interface behaves as a reservoir at the beginning of the experiments and get saturated or depleted over time. Indeed, apart from extreme deformations (small $d_m$), all the experimental data are bounded by the model predictions obtained with $U_m\rightarrow0$ and $U_m=3$ mm/s (respectively the solid and dashed lines in Fig. \ref{fig:comp_exp_shear}). This suggests that the system response is governed by the surfactant transport in the  meniscus at early times and by the transport in the films  at later times. This scenario is corroborated by the difference between alimented and non-alimented experiments: for a given transfer velocity, a smaller tension is observed for alimented films, in which the menisci are potentially less depleted, thus having a larger $U_m$ value.

Importantly, a key experimental feature is captured by the model: a clear asymmetry is observed between the extension and compression  at high velocity: for a given $U$, the  tension difference reached in the compressed  films is higher than in the stretched films. As shown in figure \ref{fig:comp_exp_shear}, their ratio reaches a  factor around two at the highest velocities. This symmetry breaking is predicted by the shear model. Moreover the
highest tensions obtained in extension (at large deformation and at large film size) are in excellent agreement with the saturation predicted  in the sheared film. 
With the correction of the  equation \eqref{eq:cor1}, the upper bound $\Delta \sigma/(2E) = 0.5$ becomes
\begin{equation}
\Delta \sigma= E + 7.68 \gamma_0\left(\frac{\eta (U+U_2^+)}{2\gamma_0}\right)^{2/3}\! \! \approx  E + 7.68 \gamma_0 \left(\frac{\eta U}{\gamma_0}\right)^{2/3} \, . 
\label{eq:relsat}
\end{equation}
It corresponds to the high velocity limit of the numerical solution (the black solid line in Fig. \ref{fig:comp_exp_shear}). As this limit does not depend on $U_d$ nor on $U_m$, it  is predicted without any free parameter and its quantitative observation is thus an important validation  of the model.

\subsection{Influence of the reservoir velocity $U_m$ of the meniscus}

Predicting the evolution of $U_m$ with time would require a model of the flow inside the menisci (along the menisci and in the cross section), involving especially the uncontrolled drainage flow along the solid parts of the set-up,
and is out of the scope of this study. However, if we assume that resupplying (or discharge) of surfactants by flow along the menisci is slow compared to the experiment time, menisci depletion (saturation) depends on the amount of surfactants delivered (absorbed) since the experiment beginning. In the limit of large $U_m$ this latter quantity is characterized by the interface transfer length $L$, and $U_m$ should thus decreases with $L$.

\begin{figure}
\begin{center}
\includegraphics[width=.9\linewidth]{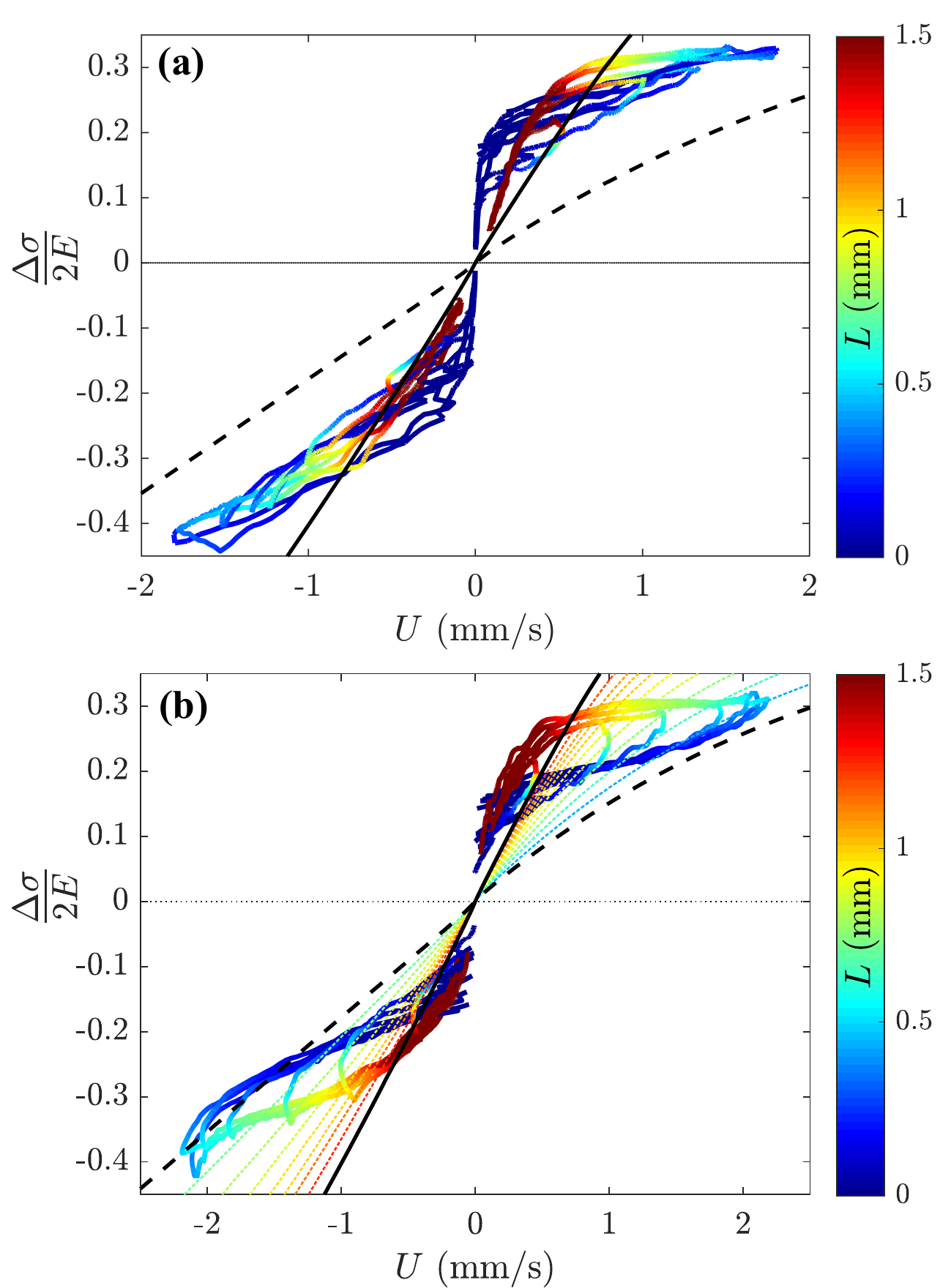} 
\caption{Evolution of the film tension variation, without (a) and with injection (b), as a function of the transfer velocity for $d_m=12$ mm, $\Delta d=6$ mm and $V$ ranging from $1$ to $12.5$ mm/s. The color represents the transfer length $L$ (between $0$ and $1.5$ mm). Curves are averaged over 3 experiments. The solid  and dashed black lines are the numerical predictions obtained  for $U_d=5.10^{-7}$ m/s,  and, respectively  $U\rightarrow0$ mm/s and $U_m=3$ mm/s. Color dashed lines in (b) correspond to model predictions for different $U_m$ ranging between $0$ and $3$ mm/s.
\label{fig:comp_exp_shear_ar_V}}
\end{center}
\end{figure}

In order to test qualitatively the correlation between this length $L$ and the reservoir velocity $U_m$, the data of Fig. \ref{fig:comp_exp_shear}(b) and (e) have been replotted in  figure \ref{fig:comp_exp_shear_ar_V} using a different color code: for each data point, the color represents the actual value of the  transfer length. 
We only kept the data series with $V<20$ mm/s to have  enough data points during motor motion, so that both the behavior during motor motion and during the relaxation are measurable. 
No definitive conclusion can be deduced from this representation, but it nevertheless provides some hints, that may serve as a basis for  future improvements of the model. 

The data shown in this figure are far from the saturation regime in extension, so the model predicts a linear relationship between $U$ and $\Delta \sigma$, with a slope controlled by $U_m$ only.  
If $U_m$ were a function of $L$ only, all points sharing the same color (so same $L$) should be on the same line (passing through the origin). These lines are represented in Fig. \ref{fig:comp_exp_shear_ar_V}(b). 

Some correlation between $U_m$ and $L$ appears for  $L> 0.4$: for a given $L$, experiment points are well captured by the model with a single $U_m$.
This is more visible for the alimented experiments (Fig. \ref{fig:comp_exp_shear_ar_V} (b)), for which  couples of data $(U, \Delta \sigma)$ differing by a factor 2 but sharing the same $L$, fall on the same line, and are thus associated to  the same $U_m$. 
For $L = 0.4$, it corresponds to $U_m\approx 3$ mm/s  and the saturation at $U_m=0$ is reached for $L \sim 1.5$ mm.  These observations consolidate our hypothesis that $U_m$ decreases with $L$, as the meniscus is less and less able to play its role of reservoir. 

The extraction velocity observed at the beginning of the experiment (dark blue part of the curves) is however in contradiction with this interpretation: the extraction velocity is lower than predicted, and even the expected hysteretic loop shown in Fig. \ref{fig:tension_velocity_men}, observed for most series,  is not observed at the shorter times for the series shown in figure \ref{fig:comp_exp_shear_ar_V} (b). 
 It seems to be a  time-delay between tension variation and transfer length, which is  more important when meniscus radius are smaller (non-alimented foams). This phenomena might be ascribed to unsteady effects either in the  film extraction dynamics, or in the sheared film dynamics, which have been modeled in a steady regime. Such effects  prevent a proper comparison with our model at early time. 

To summarize, the decreasing reservoir role of the menisci is able to rationalize the largest part of  our experiments. The hysteretic loop shown in Fig. \ref{fig:tension_velocity_men}, and observed for most series,  is captured by this decrease of $U_m$. The reservoir  velocity $U_m$ shows some  correlation with the transfer length at large transfer length.

\subsection{Non-linearities in compression}

\begin{figure}
\begin{center}
\includegraphics[width=.9\linewidth]{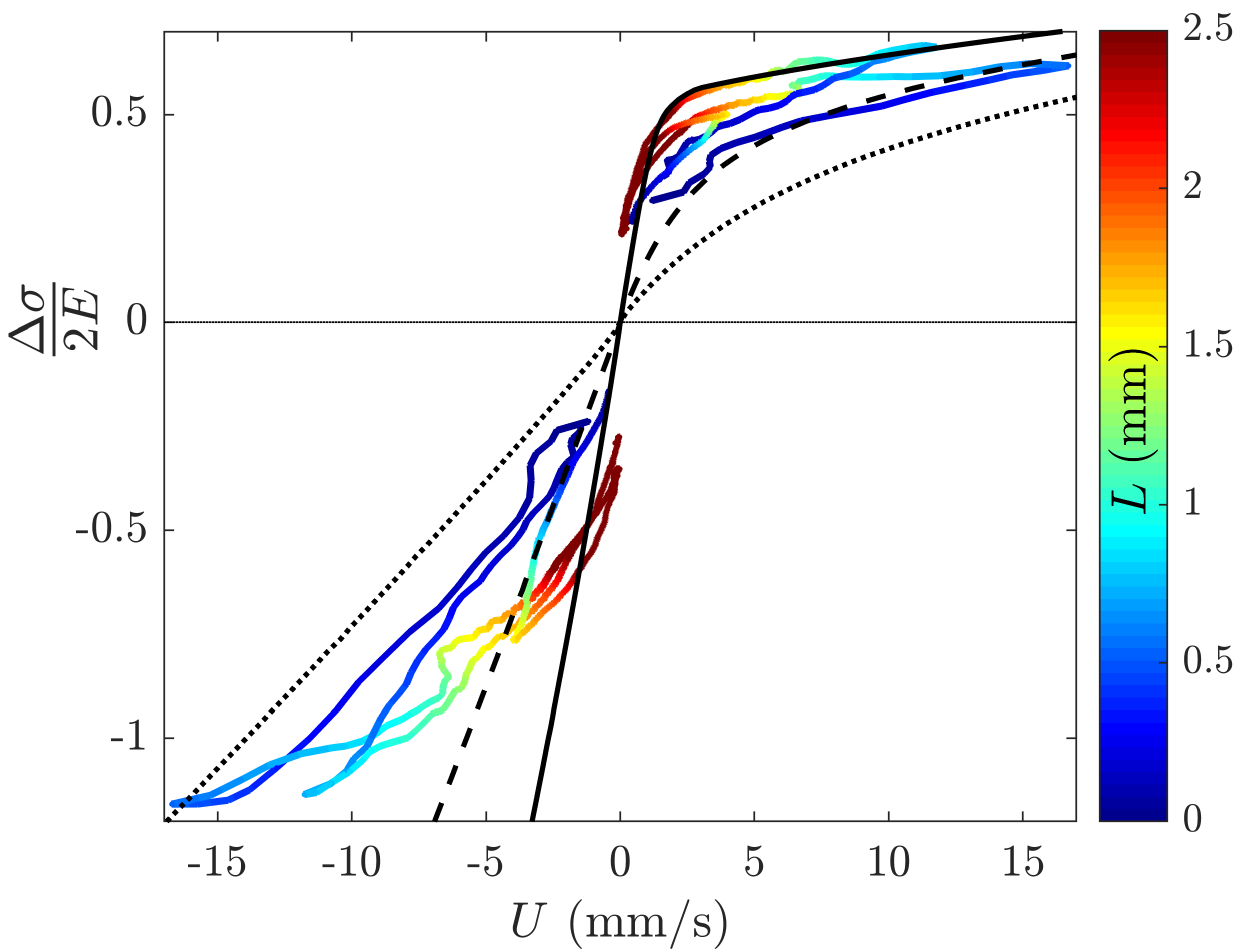} 
\caption{Evolution of the tension variation in term of the transfer velocity for experiments with $d_m=12$ mm, $\Delta d=10$ mm and different motor velocities : $V=10$ mm/s, $V=50$ mm/s and $V=100$ mm/s. Colors dots  correspond to the average  over 50 experiments. The color represents the transfer length $L$ (between $0$ and $2.5$ mm). The model is represented in black for $U_m\rightarrow0$ mm/s (solid line), $U_m=3$ mm/s (dashed line) and $U_m=8$ mm/s (doted line). 
\label{fig:comp_exp_shear_ar}}
\end{center}
\end{figure}

As shown in figure \ref{fig:comp_exp_shear_ar}, under important deformations the  viscous response in compression of the films assembly greatly deviates from the model with $U_m\rightarrow0$ mm/s. We previously discussed the possible influence of $U_m$ on these behavior. However, for the more extreme deformations, $U_m$ needs to increase from $3$ to $8$ mm/s in order to capture the early dynamic which is inconsistent with surfactant transport and accumulation in the menisci, and with the observations in extension.

Similar discrepancies are observed in figure \ref{fig:comp_exp_shear} where experiments consistently deviate from the linear behavior in compression at high tension difference ($\Delta\sigma/(2E) > 0.5$). These deviations may arise from the important compaction of the surfactant monolayer which is limited by the maximum surface coverage. This limit corresponds to the parameter $\Gamma_m$ of the Langmuir adsorption isotherm in appendix \ref{app:phys_chem} and is not captured by the model, as the physico-chemistry equations have been linearized. Some trends on the influence of these non linearities  can be anticipated from the evaluation of $E$ and $h_\Gamma^m$  at larger $\Gamma$ using the (non-linearized) equations \eqref{eq:est_E} and \eqref{eq:est_hGm}: one can see that  the elasticity diverges close to the maximum surface coverage. However,   the reservoir length decreases faster, and the product $E h^m_\Gamma$ tends to 0. 
Simply substituting these quantities $E(\Gamma)$ and $h^m_\Gamma(\Gamma)$ in the law  $\Delta\sigma\sim (E/\sqrt{U_cU_d})\ U=(\sqrt{\eta E h_\Gamma^m/D_m})\ U$ obtained with the linearized model shows that the  slope of the viscous relation $\Delta\sigma(U)$ should decreases when $\Delta\sigma$ increases which is consistent with the experimental trend. Note that in extension a decrease of the slope is also expected as $E\rightarrow0$ while $h_\Gamma^m$ tends to a constant value. However, this effect is hidden by the saturation predicted in the linear case. Incorporation of the full non-linear physico-chemistry in the model is needed to better capture viscous behavior at high tension variations.

Another possible failure of the model in compression at large velocity is the marginal regeneration instability, which has been shown to be triggered by a compression in \cite{gros20}. This instability breaks the invariance along the meniscus and may modify the relationship between the tension difference and the velocity. 

%%%%%%%%%%%%%%%%%%%%%%%%%%%%%%%%%%%%%%%%%%%%%%%%%%%%%%%%%%
\section{Relevance for the foam rheology}
\label{sec:foam_rev}

The aim of this last section is  to discuss the relevancy of the  local constitutive laws, obtained  in section \ref{sec:const_film} for the film and in section \ref{sec:const_exchange} for the meniscus,  to set the bases of a consistent dissipative model for a foam, which can be seen as a complex network of films and menisci. The geometry of the deformations in a sheared 3D foam obviously differs from the specific one we impose to our five films sample; characteristic time and length scales are also different. For these reasons, only scaling properties are discussed at the foam sample scale and, for the sake of the simplicity, we will  only keep the most salient features of our model and discuss their robustness with regard to  scale modifications.

\subsection{A closed dynamical model for the film assembly}

Before extrapolating our conclusions to 3D foam samples, a first step is to show that the two constitutive laws are sufficient to build a closed set of equations governing the dynamics of the 5 films assembly, able to predict, for example, the time behavior shown in Fig. \ref{fig:ex_tension}.

The kinematic variables of the problem are the actual film lengths $L^-$, $L^c$ and $L^+$, of the compressed, central and stretched films,  directly controlled by the motor position, and their lengths  at rest   $L_0^-$, $L_0^c$ and $L_0^+$, governed by the dynamical process. 
The evolution of the rest lengths is governed by the   transfer velocities $U^-$ and $U^+$, respectively at the compressed and stretched meniscus :
\begin{align}
&\frac{dL_0^-}{dt} = -U^- \, ,  \\
&\frac{dL_0^c}{dt} = U^- - U^+  \, ,\\
&\frac{dL_0^-}{dt} = U^+  \, .
\end{align}  
and the evolution of total film lengths obeys, for each film
\begin{equation}
\frac{dL}{dt} = \frac{dL_0}{dt} + L_0 \dot{\varepsilon} \, . 
\end{equation}
with $\varepsilon= (L-L_0)/L_0$ and  $dL/dt = 0$, $-V_m(t)$ and $V_m(t)$ respectively for the central,  compressed and stretched films. The motor velocity $V_m$ is $V$ during the time $[0 \; t_m]$ and $0$ otherwise.

These kinematic laws are coupled to the constitutive equations. Restricting the model to small deformations, we only  keep terms of  first order in $\varepsilon$ and we get  $ \Delta \sigma= \sigma - \sigma_0 = 2 E \varepsilon$ from eq.  \eqref{eq:elastheo2}, and, from  eq. \eqref{eq:master},
\begin{align}
&U^- = \frac{U^*}{2 E} (\sigma^c-  \sigma^-) \, \\
&U^+ = \frac{U^*}{2 E} (\sigma^+-  \sigma^c) \, .
\label{eq:Uscal}
\end{align}  
 Building a quantitative prediction  would require to propose a phenomenological law for  the time evolution of $U^*$, between its long time value  $U^* = \sqrt{D E/(h_\Gamma \eta)} \sim 3 \,10^{-4}$m/s and $U^*= U_m$,  which upper bound has been found experimentally to be of the order of $3 \,10^{-3}$m/s. 
In order to build a simple and illustrative analytical solution, we assume instead a constant value for $U^*$, and we obtain a  closed set of coupled equations, governing  the five films dynamics:

\begin{align}
&- V_m =  \frac{U^*}{2E} \sigma^-  -    \frac{U^*}{2E}  \sigma^c  + \frac{L_0^-}{2E}  \dot{\sigma}^- \, ,   \\
& 0 = -\frac{U^*}{2E} \sigma^-  -    \frac{U^*}{2E}  \sigma^+ +   2  \frac{U^*}{2E}  \sigma^c + \frac{L_0^c}{2E}  \dot{\sigma}^c \, ,  \\
&V_m = \frac{U^*}{2E} \sigma^+  -    \frac{U^*}{2E}  \sigma^c  + \frac{L_0^+}{2E}  \dot{\sigma}^+ \, . 
\end{align}  

Simplifying further the problem by assuming the same initial length $d$ for each film, and linearizing the last terms, we obtain a symmetrical situation in which $\Delta \sigma^c=0$ and $\Delta  \sigma^-  = - \Delta  \sigma^+$.  The system becomes, with $\eta_e= 2E/U^*$ and $k_e= 2 E/d$  
 \begin{align}
V_m =  \frac{\Delta \sigma^+}{\eta_e}  + \frac{\dot{\Delta \sigma}^+}{k_e}  \, .
\label{eq:eqdiff}
\end{align} 
which is the equation governing a Maxwell viscoselastic fluid. 
The solution is 
\begin{align}
&\Delta \sigma^+= \eta_e V \left(1 - e^{-t/\tau^*} \right)   \quad \mbox{for } t<t_m \, , \\
&\Delta \sigma^+= \eta_e V \left(e^{(t_m-t)/\tau^* }  - e^{-t/\tau^*} \right)   \quad \mbox{for } t> t_m \, ,\\
\end{align}
where  the characteristic time of the system is  $\tau^* = \eta_e/k_e= d/U^*$. 

This solution  captures the most important properties of the dynamical  behavior observed in Fig. \ref{fig:ex_tension} and it especially brings out the dissipative role of the parameter $\eta_e= 2E/U^*$, which has the dimension of a bulk viscosity. The possibility of an upscaling of the local laws at the scale of a foam sample will be  discussed on the  simple basis of eq. \eqref{eq:eqdiff}.

\subsection{From few films to a foam sample}

Most of the foam  effective viscosity measurements  are obtained either under steady shear at the rate $\dot{\varepsilon}$ or  in an oscillatory regime at the pulsation $\omega$.   The foam viscosity $\eta_e$ is defined from the expression of the stress $T$: in the first case,  $T = T_0 + \eta_e( \dot{\varepsilon}) \dot{\varepsilon}$ with $T_0$ the quasistatic plastic threshold ; in the second case $T = G'(\omega) \varepsilon+ \eta_e(\omega) \dot{\varepsilon}$ (with $\eta_e(\omega) = G''(\omega)/\omega$).

The link between the local and global scales is a central question in the rheology of complex systems. In steady regime, the relationship between the viscous stress $\eta_e \dot{\varepsilon}$ and the film tension differences in the foam involves both a direct influence of the tension difference, and a non-linear variation of the plastic threshold $T_0$ with this tension difference \cite{cantat11,grassia12}. This last contribution greatly complicates the up-scaling of local laws to steady experiments and therefore is out of the scope of this section.

In the oscillatory regime, on the other hand, the geometrical effects are {\it a priori} simpler, as the plastic threshold is never reached. At low frequency, the foam loss modulus  $G''(\omega)$ is governed by the coarsening-induced plasticity, which is entirely decoupled from the local dissipative modes  of the system and has its own time scale \cite{cohenaddad04}. At higher frequency (usually above few Hz), foam exhibits an anomalous dissipation with a loss modulus scaling as $\sqrt{\omega}$ \cite{krishan10,gopal03}. The origin of this peculiar behavior is still a matter of debate. As assumed in \cite{liu96} this scaling might arise from the effect of the disorder and the local scalings may be entirely screened at the sample scale. In this case, the internal time scale of the foam sample differs from the local internal time scale, and is governed by weak domains, close to the yield stress. However, this generic scaling predicted by the weak plane region model does not hold for solution with important interface "rigidity" \cite{costa13}. Moreover, experiments in ordered foam exhibit a similar anomalous dissipation suggesting a different origin \cite{costaphd}.

In the frame of this paper, we thus choose to restrict the discussion to the behavior under oscillatory stress of a foam without internal dynamics, and far enough, everywhere in the sample,  from its local yield stress. In that case, we deduce 3D scalings  from the local our scalings and we reveal a new  possible origin  for the $\sqrt{\omega}$ scaling of the foam dissipation.

During a simple shear deformation of the foam, thin films experience compression, extension and simple in-plane shear (at a constant film area). Although meniscus constitutive law has been built in the case of a specific imposed deformation, the meniscus frustration avoiding free film relaxation by interface transfer has been shown to be generic. The associated scaling laws eqs. \eqref{eq:Uscal} are thus expected to hold for any generic deformation leading to film area variations. Dissipation associated to the simple film shear has been studied in \cite{costa13b} but has been shown to be negligible in our case (see section \ref{sec:const_film}) and is thus neglected in the following.

During foam shearing, the characteristic extension rate of the films scales as $\dot{\varepsilon} \, d$, with $d$ the bubble diameter and $\dot{\varepsilon}$ the strain rate imposed at the sample scale \cite{livre_mousse_en}. In eq. \eqref{eq:eqdiff}, the motor velocity $V= dL^+/dt$  should thus be replaced by $\dot{\varepsilon} \, d$. Similarly, for dimensional reasons, the film tension difference is at the origin of a  stress $T$ scaling as $\Delta \sigma/d$. Once extrapolated at the foam scale, the equation \eqref{eq:eqdiff} governing the 5 films sample thus leads to a viscous stress $T_{visc} \sim \eta_e \dot{\varepsilon}$,  allowing us to identify  $\eta_e$ as the effective foam viscosity.

\subsection{Predicting 3D foam viscosity ? }

In our experiment, the effective viscosity is $\eta_e = 2E/U^*$.
The parameter $U^*$ depends on the solution physico-chemical properties and on the film length scales and should thus be reconsidered  using typical foam parameters. 
Especially, the bubble size is usually smaller in 3D foam than our millimetric films. The orders of magnitude of the various quantities will  be built using  a liquid fraction $\phi = 5 \%$, a bubble size $d \sim 10^{-4}$m, and a meniscus size  $r_m \sim d \sqrt{\phi} \approx 20 \,  \mu$m \cite{livre_mousse_en}. 

The average film thickness may also strongly differ. It is not precisely known in a 3D foam. However, assuming that the capillary suction is high enough to drain the films toward their equilibrium thickness faster than the dynamical time scales, the parameter $h_\infty$ would be governed by the disjoining pressure and be of the order of few dozen of nanometers. Importantly, this length scale does not appear in the constitutive relationship governing the meniscus, which should therefore not be modified significantly if the disjoining pressure effects were included in the model. The length of the sheared domain would in contrast be affected and will be discussed at the end of the section.   

The physico-chemical properties greatly differ from one experiment to another, and are not always easy to evaluate quantitatively. Here we first  consider the values obtained with our foaming solution to build and we then discuss qualitatively the influence of the different physico-chemical parameters.

The velocity $U^*$ is equal to the largest value between the two possible characteristic velocities $\sqrt{U_c U_d}$ and $U_m$. The first one, $\sqrt{U_c U_d} = \sqrt{D E/(h_\Gamma \eta)} \sim 10^{-3}$ m/s, depends on the foaming solution but not on the foam geometry. Using simple Langmuir adsorption, its value is expected to decrease for more insoluble surfactant.

The velocity $U_m$ is discussed in section  \ref{sec:boundcond}. In the purely diffusive case, it  scales as  $(r_m/h_\Gamma) \sqrt{D \omega}$ and thus depends on the bubble size through $r_m$. In the case of our millimetric foam, this prediction leads to a value for $U_m$ smaller than the observed one, the transport being potentially dominated by the convection. However, we expect the diffusive processes to become dominant at smaller scale. Using $U_m \sim  (d/h_\Gamma) \sqrt{\phi D \omega}$,  we get $U_m \sim 10^{-4}$m/s at  1 Hz and $U_m \sim 10^{-3}$m/s at 100 Hz. These velocities are of the same order than $\sqrt{U_c U_d}$, so the dominant term is difficult to determine {\it a priori}, and may vary from one foaming solution to the other. 

In the regime  $U_m <  \sqrt{U_c U_d}$, the scaling for the loss modulus $G''= \omega \eta_e$ is 
\begin{equation}
G''  \sim \omega \sqrt{\frac{\eta E h_\Gamma}{D}} \, , 
\end{equation}
whereas if $U_m >  \sqrt{U_c U_d}$, 
\begin{equation}
G''  \sim \omega \frac{E}{U_m} \sim    \frac{E  h_\Gamma}{d  \sqrt{\phi D}}   \sqrt{\omega}   \; .
\label{eq:gsec}
\end{equation}

Using our physico-chemical parameters $E= 5\, 10^{-3}$ N/m, $h_\Gamma = 5.4 \, 10^{-6}$m and $D= 8\, 10^{-11}$m$^2$/s, and  for a bubble size of 100 $\mu$m  at 100 Hz, the second law eq. \ref{eq:gsec} leads to $G'' \sim 1000$ Pa  which is the order of magnitude found for usual foams. More importantly, it recovers the  scaling  $G''  \propto \sqrt{\omega}/d$ found in \cite{krishan10, costa13}, which clearly indicates that  the  local laws  established in this paper may be at the origin of the foam anomalous dissipation scaling. 

Our dissipative model is based on the coexistence of a symmetrical part at the center of the films, and a sheared domain close to the meniscus, of size $\ell$. The local scalings we predict are thus valid only if $\ell < d$, with $d$ the bubble size. For small tension values, we have $\ell \sim   h_\infty (U_c/U_d)^{1/2} \sim 10^3  h_\infty$  (see Fig. \ref{fig:num_res}).
Assuming an average film thickness of 50 nm in the foam, our model should thus be relevant for bubbles larger than 50 $\mu$m, which corresponds to most of the usual foam samples. Note that if the Frankel's film extractions, induced by large deformations, increase the average film thickness at a larger value, despite the large capillary suction present in the foam, the lower bubble size limit would be more restrictive. For bubbles smaller than this lower limit, the whole film should be  sheared, and  leading to some  coupling between the neighboring menisci, as assumed in \cite{denkov08,berut19}. The induced dissipation is expected to be qualitatively different in this regime. 

Such a transition between film extension and film shear can also be observed at constant bubble size, when the foaming solution varies \cite{titta18}. Indeed, the product $E \, h_\Gamma$ appearing in $\ell$ can vary over several orders of magnitude with the solubility of the chemical species at the origin of the tension variations, leading to large $\ell$ values at poor solubilities. For a given bubble size, poorly soluble species would lead to entirely sheared films, whereas more soluble ones would obey the laws established in this paper.

\section{Conclusion}
In this paper, we demonstrate that, at the millimetric scale and for the investigated foaming solution, the main part of the foam films has as perfectly reversible elastic behavior, with a negligible viscous contribution.  The main part of the dissipation is in contrast localized in a small domain of the films, close to the meniscus, which is predicted to be sheared. This local shearing is the direct consequence of a generic geometrical frustration occurring at the meniscus, which forbid the free transfer of interface from one film to its neighbor. A model where Marangoni stress induced by the shear is coupled to surfactant transport across the film is developed to capture this shear dissipation. Numerical resolution reproduces the experimental relationship between the transfer velocity and the tension difference between the adjacent films and confirms the scaling laws that we establish.

Our model also predicts the length $\ell$ of the sheared region which is determined by the physico-chemical properties of the foaming solution. This length sets a transition between films shorter than $\ell$ where the whole film is sheared, as usually assumed in rheological models of foam, and films longer than $\ell$ where the sheared regions coexist with a film extension/compression, as observed in millimetric film experiments. This reconciles the different classes of model proposed in the literature and rationalizes the different experimental observations.

For foams with  a bubble size larger than $\ell$, our local constitutive laws are up-scaled and a new possible origin of the foam anomalous dissipation is proposed. The scaling obtained also qualitatively captures the order of magnitude and the dependency with the bubble size of the foam loss modulus. Nonetheless, much efforts are still needed to develop a complete model of foam dissipation. In particular, it will require (i) to extend of our shear model to bubble smaller than $\ell$; 
(ii) to take into account the foam disorder and the influence of the weak regions ; (iii) to build a more quantitative constitutive law for the meniscus surfactant exchange velocity ($U_m$). The resulting model will have to be quantitatively compared to  foam experiments with different well calibrated solutions, with known physico-chemical properties.

\section*{Acknowledgments}
This project has received funding from the European Research Council (ERC) under the European Union's Horizon 2020 research and innovation program (grant agreement No 725094). We thank A. Saint-Jalmes  for fruitful discussions, E. Schaub, M. Le Fur and P. Chasle for technical support, and A. Carr\`ere and X. Ah-Thon for experimental support.

\appendix

\section{Estimation of $E$ and $h_\Gamma$}
\label{app:phys_chem}

The co-adsorption of DOH and SDS in a micellar SDS/DOH solution has been  modeled using a generalized Langmuir-Von Szyszkowski equation in \cite{fang92}, leading to the following results :
\begin{equation}
    \gamma_{th}=\gamma_{\sds}-RT\Gamma_{m}\ln{\left(1+\frac{c^m}{a}\right)},
    \label{eq:VonS}
\end{equation}
with $\gamma_{\sds}=40.5$ mN/m the  surface tension of pure SDS, $R$ the ideal gas constant, $T$ the temperature, $\Gamma_{m}=6.10^{-6}$ mol/m$^2$ the maximum surface coverage, $c^m$ the dodecanol  concentration in the monomer form and $a=5.55\ 10^{-3}$ mol/m$^{3}$ the adsorption equilibrium constant. Here only the concentration of monomer influences the surface tension as DOH molecules solubilized in SDS micelles are not surface active. The partition coefficient $\beta=c^M/c^m$, with $c^M$ the concentration of DOH solubilized in micelles, depends linearly on the SDS micelles concentration. It has been measured  in \cite{fang92}, and for our solution, with $c_{\sds}=19.4$ mol/m$^3$, we obtained $\beta=67.2$ indicating that almost all the dodecanol is solubilized  in micelles. The monomer concentration can be expressed in term of the total dodecanol concentration $c$, $c^m=c/(1+\beta)$ and equation \eqref{eq:VonS} becomes :
\begin{equation}
    \gamma_{th}=\gamma_{\sds}-RT\Gamma_{m}\ln{\left(1+\frac{c}{(1+\beta)a}\right)}.
\end{equation}

The corresponding dodecanol surface coverage is given by a Langmuir-type adsorption :
\begin{equation}
    \Gamma=\Gamma_{m}\frac{c^m}{a+c^m}=\Gamma_{m}\frac{c}{a(1+\beta)+c}.
    \label{eq:langmuir_Gc}
\end{equation}

The surface tension can then be expressed in term of surface coverage :
\begin{equation}
    \gamma_{th}=\gamma_{\sds}+RT\Gamma_{m}\ln{\left(1-\frac{\Gamma}{\Gamma_{m}}\right)},
    \label{eq:tension_fang}
\end{equation}
and the Gibbs-Marangoni elasticity is estimated by linearizing equation \eqref{eq:tension_fang} around the equilibrium state, at $c_0=0.27$mol/m$^3$, the initial bulk dodecanol concentration 
\begin{equation}
    E=-\left.\frac{\partial\gamma_{th}}{\partial\Gamma}\right|_{\Gamma_0}\Gamma_0=\frac{RT\Gamma_0\Gamma_{m}}{\Gamma_{m}-\Gamma_0}=RT\Gamma_{m}\frac{c_0}{(1+\beta)a},
    \label{eq:est_E}
\end{equation}
which gives $E=10.6$ mN/m. The reservoir length $h_\Gamma = \partial\Gamma/\partial c$ characterizing the adsorption is given by :
\begin{equation}
    h_\Gamma=
\Gamma_{m}\frac{a(1+\beta)}{(a(1+\beta)+c_0)^2} = \frac{\Gamma_{m}}{a(1+\beta)} \left( \frac{\Gamma_{m}-\Gamma_0}{\Gamma_{m}} \right)^2 ,
\end{equation}
leading to $h_\Gamma=5.4$ $\mu$m. This length assumes an equilibrium between the monomer and the solubilized dodecanol concentrations. If the monomer transport is faster than the exchange time with SDS micelles \cite{patist98}, as discussed in section \ref{sec:comp_shear}, dodecanol transport only involves monomers and the reservoir length becomes :
\begin{equation}
    h_\Gamma^m=\left.\frac{\partial\Gamma}{\partial c^m}\right|_{c_0}=\Gamma_{m}\frac{a}{(a+c_0^m)^2}=\Gamma_{m}\frac{a(1+\beta)^2}{(a(1+\beta)+c_0)^2},
    \label{eq:est_hGm}
\end{equation}
and $h_\Gamma^m=(1+\beta)h_\Gamma\approx370$ $\mu$m.

\section{Film shape and calculation of the angles}
\label{app:angles}

\begin{figure}
\begin{center}
\includegraphics[width=0.7\linewidth]{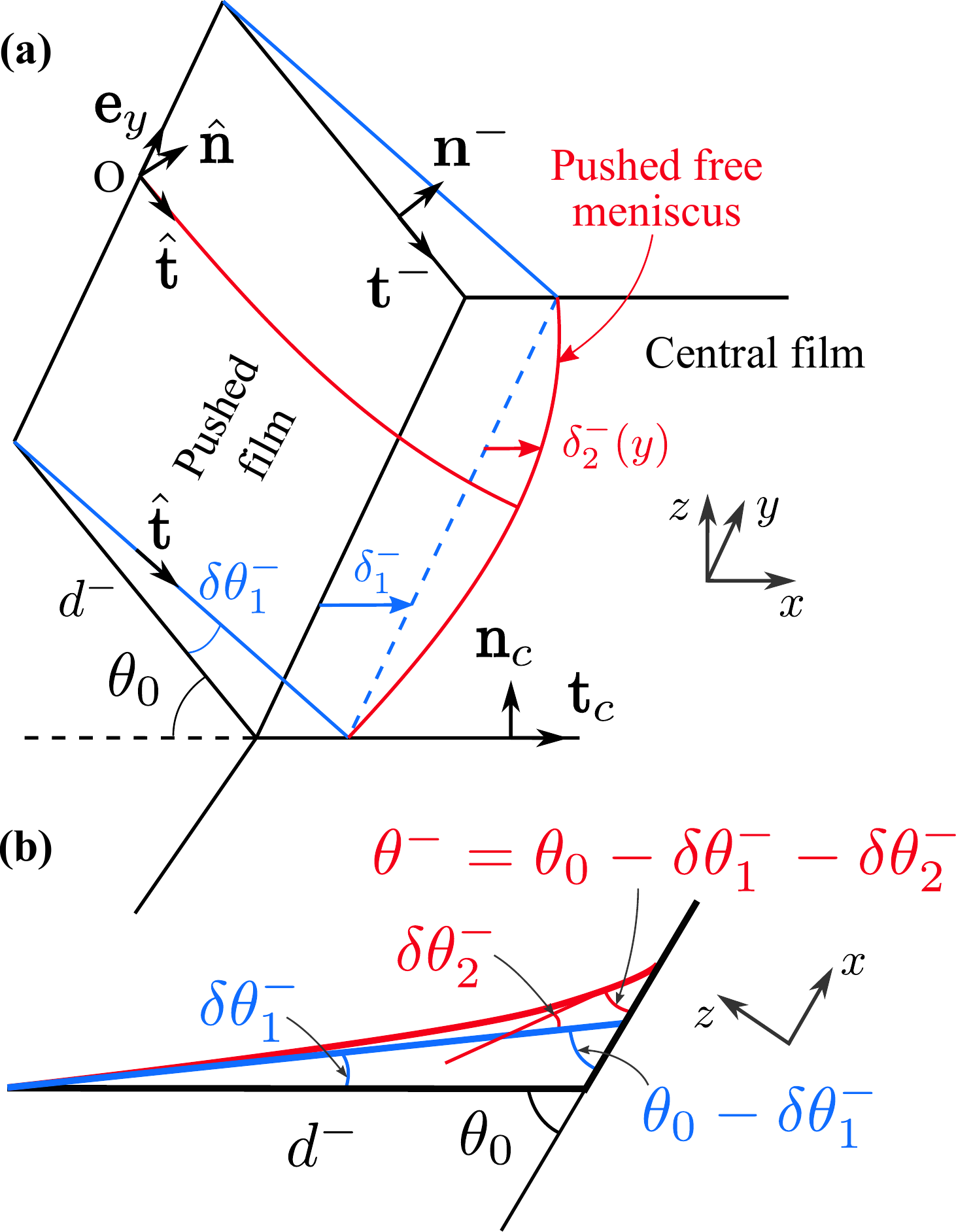}
\caption{Scheme of the pushed film deformation. (a) black lines: equilibrium plane of the pushed film (denoted by $^-$) and of the central film (denoted by $c$). blue lines: shape of the compressed film obtained for a uniform displacement $\delta_1^-$ of the free meniscus. The actual free meniscus displacement is $\delta_1^-$ only on the boundary (blue arrow) and an additional displacement $\delta_2(y)^-$ (red arrow) leads to a curved free meniscus (red line). (b) View in the $y=0$ plane of the same system showing the angle $\delta\theta_1^-$ and $\delta\theta_2^-$ induced by the respective displacements (same color code) and the resulting angle $\theta^-$. 
\label{fig:film_def}}
\end{center}
\end{figure}

The upper pushed film is initially in the plane (${\bf t}^- , {\bf e}_y$), and the pushed free meniscus is at the intersection between this plane and the plane  (${\bf t}_c, {\bf e}_y$) of the central film (see Fig. \ref{fig:film_def}).  After deformation,  the meniscus stays in the plane  (${\bf t}_c, {\bf e}_y$) (by top/bottom symmetry) and  is at the distance $s_c^m(y) = \delta_1^- + \delta_2^-(y)$ from its initial position, with $\delta_1^-$ the displacement of both ends of the meniscus,  $\delta_2^-(y)$ the variable part of the displacement and $s_c$ the coordinate along ${\bf t}_c$ (with an origin at the initial meniscus position).

We define the intermediate plane (${\bf \hat{t}},  {\bf e}_y$) as the plane containing the moving solid frame, at the position $s^- = d^-$ in the plane  (${\bf t}^- , {\bf e}_y$) and the line  $s_c = \delta_1^-$ in the (${\bf t}_c , {\bf e}_y$) plane (in blue in Fig. \ref{fig:film_def}). Without the variable part of the displacement, the upper left film would be in this plane after deformation.

The tilt angle $\delta \theta_1^-$ verifies, from simple geometry
\begin{equation}
\tan \left(\theta_0- \delta \theta_1^-\right) = \frac {d^- \sin \theta_0}{ d^- \cos \theta_0 + \delta_1^- }  \, , 
\label{dtheta1}
\end{equation}
with $\theta_0= \pi/3$. 

We now address the 3D shape of the deformed film, using the variables $(\hat{s}, y, \hat{z})$ in the (${\bf \hat{t}},  {\bf e}_y, {\bf \hat{n}}$) reference frame. For convenience, the origin of $\hat{s}$ is this time chosen at the external edge, with a positive direction toward the deformed film.  We characterize the film shape by the equation  $\hat{z} = f (\hat{s}, y)$. Its curvature, at the linear order is simply the Laplacian of $f$ so the vanishing curvature condition becomes 
\begin{equation}
 \frac{\partial^2 f }{\partial \hat{s}^2} + \frac{\partial^2 f }{\partial y^2} =0
\end{equation} 

The free meniscus is the  red line in Fig. \ref{fig:film_def} (a) parameterized by  $(\hat{s}^m(y),\hat{z}^m(y))$ with  $\hat{s}^m(y) = d^-$ (at order 0) and $\hat{z}^m(y)=\delta_2^-(y) \sin (\theta_0)$ (at order 1). 

At linear order in $\delta_1^-$ and $\delta_2^-$, the boundary condition at this meniscus is thus 
\begin{equation}
f(d^-,y) =  \delta_2^-(y) \sin (\theta_0) 
\end{equation}

The meniscus position along the solid frames can not be measured. The meniscus can slide on the frame, thus explaining the non-vanishing value of $\delta_1^-$, but this sliding motion is limited by uncontrolled viscous or geometrical stresses, which explain the free meniscus curvature and the non-vanishing value of $\delta_1^-$.
However, we know that the displacement of the menisci at $y=\pm w/2$ varies from 0 at the external edge to $\delta_1^-$ at the free meniscus (with respect to the undeformed shape in the  (${\bf t}^- , {\bf e}_y$) plane), with a smooth variation along the whole film. We thus simply assume a linear increase of this displacement, which corresponds to the condition 
\begin{equation}
f(\hat{s} ,\pm w/2) = 0   \quad \mbox{ and }  \quad  f(0, y) =0 \; . 
\end{equation}

The film shape is entirely determined by these boundary conditions. A simple analytical solution is obtained by fitting the free meniscus motion by 
$\delta_2^-(y)= \delta_2^-(0) \cos \left( \frac{\pi y}{w} \right)$ 
 and its equation is
\begin{equation}
f(\hat{s} , y) =  \delta_2^-(0) \sin (\theta_0)   \frac{\sinh\left( \frac{\pi \hat{s}}{w} \right)}{\sinh\left( \frac{\pi d^-}{w} \right)}  \cos \left( \frac{\pi y}{w} \right) \; . 
\end{equation}

The angle $\delta \theta_2^-$ verifies
\begin{equation}
\delta \theta_2^-=  \frac{\partial f}{\partial \hat{s}} (d^-, 0) =  \frac{ \delta_2^-(0)}{w}   \frac{\pi \sin (\theta_0)  }{\tanh\left( \frac{\pi d^-}{w} \right)}
\label{dtheta2}
\end{equation}

The angle $\theta^-$ needed to compute the tension differences is eventually 
\begin{equation}
\theta^- = \theta_0 - (\delta \theta_1^- +  \delta \theta_2^-)
\label{theta}
\end{equation}
with $\delta \theta_1^-$ and $\delta \theta_2^-$ functions of $\delta_1^-$ and $\delta_2^-(0)$  given by eqs. \eqref{dtheta1} and \eqref{dtheta2}.
The expressions are valid for positive or negative displacement of the meniscus and are thus identical for the right side.

\section{Non-linear Langmuir film elasticity}
\label{app:NL_elas}

The derivation of the non linear film elastic behavior relies on the same assumptions as the ones  made in section \ref{sec:gibbsmar}, but the Langmuir equation is used instead of its linearized form. The surfactant mass balance is 
\begin{equation}
    2\Gamma_0+c_0h_0=2\varepsilon\Gamma+ch_0,
\end{equation}
which  can be expressed in term of $\Gamma$ only, using eq. \eqref{eq:langmuir_Gc}:
\begin{equation}
    2\Gamma_0+\frac{(1+\beta)a\Gamma_0h_0}{\Gamma_m-\Gamma_0}=2\varepsilon\Gamma+\frac{(1+\beta)a\Gamma h_0}{\Gamma_m-\Gamma}.
\end{equation}
This equation can be reorganized into a second order polynomial using $x=\Gamma/\Gamma_m$ :
\begin{equation}
    2\varepsilon x^2-\left(2\varepsilon+2x_0+\frac{(1+\beta)ah_0}{\Gamma_m(1-x_0)}\right)x+2x_0+\frac{(1+\beta)ah_0}{\Gamma_m(1-x_0)}=0,
\end{equation}
and its roots provide the required relation between the surface coverage $\Gamma$ and the extension. Finally, equation \eqref{eq:tension_fang} gives the surface tension evolution with $\varepsilon$ shown in Fig. \ref{fig:elasticity} associated to a non-linear elasticity.

\section{Frankel law's for different interface velocities}
\label{app:frankel_u}

The extraction of a liquid film from a menisci at a velocity $U$ is a classical problem, and imposing different velocities $U_1$ and $U_2$ on both interfaces only leads to straightforward modifications of the usual equations (see the review \cite{cantat13}). In the lubrication regime ($\partial_sh\ll1$), the Stokes equation becomes:
\begin{equation}
    \frac{\partial^2v}{\partial\zeta^2}=\frac{1}{\eta}\frac{\partial p}{\partial s}=-\frac{\gamma_0}{2\eta}h''',
    \label{eq:stockes_u}
\end{equation}
with $h'=\partial_sh$. 
The interface velocity $v$ is assumed to be  constant over the dynamical meniscus (which is equivalent to $\ell\gg\ell_m$), and  $v(s,h/2)=U_1$ and $v(s,-h/2)=U_2$.
 Integrating two times the equation \eqref{eq:stockes_u} over $\zeta$ gives the velocity field :
\begin{equation}
    v(s,\zeta)=-\frac{\gamma_0}{4\eta}h'''\left(\zeta^2-\frac{h^2}{4}\right)+\frac{U_1-U_2}{h}\zeta+\frac{U_1+U_2}{2}.
\end{equation}
The flow rate in the dynamical meniscus is:
\begin{equation}
    Q=\int_{-h/2}^{h/2}v \, d\zeta=\frac{\gamma_0}{24\eta}h'''h^3+\frac{U_1+U_2}{2}h,
\end{equation}
and must equate the outgoing flux $Q=(U_1+U_2)h_\infty^f/2$ with $h_\infty^f$ the film thickness leaving the dynamical meniscus (that differs  from the thickness at the end of the sheared film $h_\infty$). The mass conservation leads to :
\begin{equation}
    h'''h^3=12\frac{\eta}{\gamma_0}(U_1+U_2)(h_\infty^f-h)=24Ca(h_\infty^f-h),
    \label{eq:landau}
\end{equation}
with $Ca=\eta(U_1+U_2)/(2\gamma_0)$ the capillary number based on the averaged velocity.
Equation \eqref{eq:landau} is  the well know Landau-Levich-Degardin (LDD) equation in which the interface velocity $U$ has been replaced by $(U_1+U_2)/2$. 
The surface tension difference between the meniscus and the film is calculated by integrating the Marangoni relation:
\begin{align}
    \Delta\gamma&=\int_{men}\frac{\partial\gamma}{\partial s}ds=\int_{men}\eta\frac{\partial v}{\partial\zeta}(s,h/2)ds\\
    &=-\int_{men}\frac{\gamma_0}{4}h'''hds+\int_{men}\eta\frac{U_1-U_2}{h}ds.
\end{align}
The second term on the right-hand side is the dissipation induced by the shear flow. It has been taken into account in our model of the sheared film and is disregarded here to avoid to count it twice. The first term is the classical term in the Frankel's problem which can be found in \cite{cantat13}.  Finally we obtained :
\begin{equation}
    \Delta\gamma^{out}=3.84\gamma_0\left(\frac{\eta(U_1+U_2)}{2\gamma_0}\right)^{2/3}.
\end{equation}

\section{Divergence}
\label{app:divergence}

\begin{figure}
\begin{center}
\includegraphics[width=.7\linewidth]{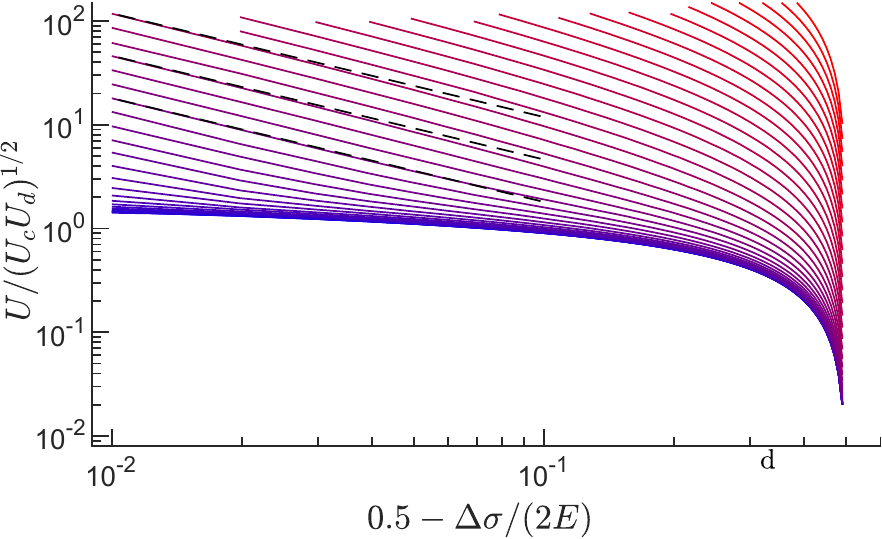} \\
\includegraphics[width=.7\linewidth]{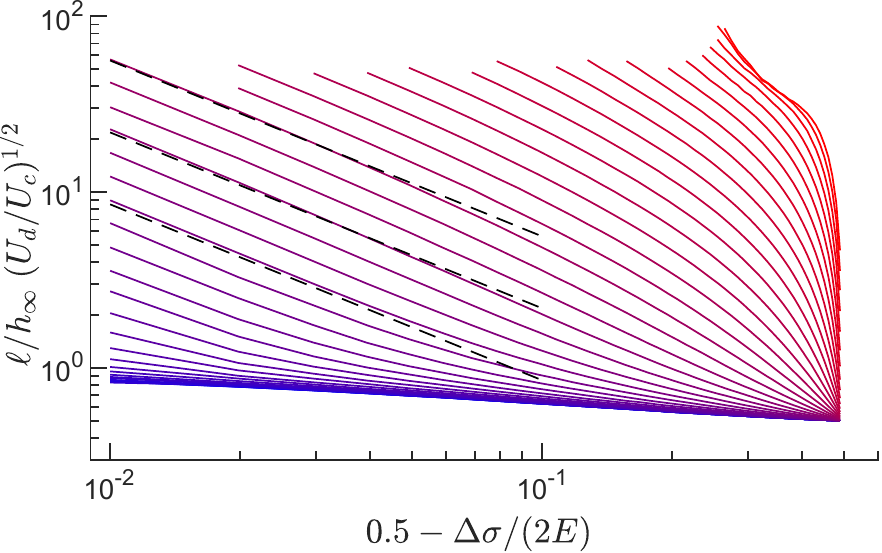} 
\caption{(Top) Velocity $U$ at the meniscus, rescaled by $\sqrt{U_c U_d}$, as a function of the concentration in the film, $\hat{\e} = 0.5 - \delta \Gamma/\Gamma_{0}= 0.5 - \Delta \sigma/(2 E)$, for  the pulling case. The parameter $K=  2U_m/\sqrt{U_c U_d}$ has been varied logaritmically from $10^{-3}$ (blue curves) to $10^3$ (red curves). The black lines corresponds to $U/\sqrt{U_c U_d} \sim 0.25 K/\hat{\e} $, for three values of $K$,  corresponding to the closed  curve ($K=0.73 ;  1.87 ;  4.8$ from bottom to top) (see the scaling of eq. \eqref{eq:divUdeG}). (Bottom) Similar plot for $\ell$, with the dashed black lines corresponding to  $\ell/(h_\infty \sqrt{U_c/U_d})= 0.12 K/\hat{\e}$, as predicted by eq. \eqref{eq:divLdeG}. 
\label{fig_div}}
\end{center}
\end{figure}

For the extension case,  the velocity diverges in the  limit $\Delta \sigma/(2 E)=0.5$, as shown in Fig. \ref{fig_div}. The scaling predicted in  eq. \eqref{eq:divUdeG} is recovered as emphasized by the dashed lines. Similar behavior is observed for the shear length $\ell$ which divergence follows the expected scaling law eq. \eqref{eq:divLdeG}.


%\bibliography{D:/ABI/Documents/bib/bib}

\end{document}